\documentclass[letterpaper]{JHEP3}
\usepackage{multicol}
\usepackage{amssymb,amsmath}

\usepackage{amsfonts}

\pdfoutput=1

\usepackage{slashed}
\usepackage{graphicx}

\def\raro{{\rightarrow}}

\newcommand{\roughly}[1]{\mathrel{\raise.3ex\hbox{$#1$\kern-0.85em
\lower1ex\hbox{$\sim$}}}}

\newcommand{\lsim}{\roughly<}
\newcommand{\gsim}{\roughly>}

\def\be{\begin{equation}}
\def\beq\begin{equation}
\def\ee{\end{equation}}
\def\bea{\begin{eqnarray}}
\def\eea{\end{eqnarray}}

\def\nn{\nonumber}

\def\pref#1{(\ref{#1})}

\def\ol#1{\overline{#1}}

\def\beq{\begin{equation}}
\def\eeq{\end{equation}}
\def\beqa{\begin{eqnarray}}
\def\eeqa{\end{eqnarray}}

\def\cA{{\cal A}}

\def\cL{{\cal L}}

\def\cO{{\cal O}}

\def\cV{{\cal V}}

\def\ssA{{\scriptscriptstyle A}}

\def\ssI{{\scriptscriptstyle I}}
\def\ssF{{\scriptscriptstyle F}}
\def\ssJ{{\scriptscriptstyle J}}

\def\ssL{{\scriptscriptstyle L}}
\def\ssM{{\scriptscriptstyle M}}
\def\ssN{{\scriptscriptstyle N}}

\def\ssR{{\scriptscriptstyle R}}
\def\ssT{{\scriptscriptstyle T}}
\def\ssV{{\scriptscriptstyle V}}
\def\ssW{{\scriptscriptstyle W}}
\def\ssX{{\scriptscriptstyle X}}
\def\ssY{{\scriptscriptstyle Y}}
\def\ssZ{{\scriptscriptstyle Z}}

\def\GUT{{\scriptscriptstyle GUT}}
\def\SM{{\scriptscriptstyle SM}}
\def\NC{{\scriptscriptstyle NC}}

\def\dsl{\hbox{/\kern-.5300em$\partial$}}
\def\sh{{\rm sh}}
\def\ch{{\rm ch}}

\newcommand{\bmat}{\left(\begin{array}}
\newcommand{\emat}{\end{array}\right)}

\def\-{\hphantom{-}}

\def\s2{\frac{1}{2}}

\def\IF{\relax{\rm I\kern-.18em F}}
\def\II{\relax{\rm I\kern-.18em I}}
\def\IP{\relax{\rm I\kern-.18em P}}
\def\IC{\relax{\rm I\kern-.48em C}}
\def\IR{\relax{\rm I\kern-.18em R}}
\def\IK{\relax{\rm I\kern-.20em K}}
\def\IM{\relax{\rm I\kern-.25em M}}

\def\Dsl{\,\raise.15ex\hbox{/}\mkern-13.5mu D} %this one can be subscripted
\def \one{\relax{\rm 1\kern-.26em I}}

\def\exd{{\rm d}}

\title{New Constraints (and Motivations) for Abelian\\
Gauge Bosons
in the MeV -- TeV Mass Range }

\author{ M.~Williams,${}^1$ C.P.~Burgess,${}^{1,2}$ Anshuman Maharana${}^3$ and F.~Quevedo${}^{3,4}$\\

${}^1$Department of Physics \& Astronomy, McMaster University\\ \qquad 1280 Main Street West, Hamilton ON, Canada.%\\

${}^2$Perimeter Institute for Theoretical Physics\\
\qquad 31 Caroline Street North, Waterloo ON, Canada.%\\

$^3$ DAMTP/CMS, University of Cambridge, %Wilberforce Road,\\
 Cambridge CB3 0WA, UK.%\\

$^4$ Abdus Salam ICTP, Strada Costiera 11, Trieste 34014, Italy.
}
\date{}
\abstract{ We survey the phenomenological constraints
on abelian gauge bosons having masses in the MeV to multi-GeV
mass range (using precision electroweak measurements, neutrino-electron and
neutrino-nucleon scattering, electron and muon anomalous magnetic
moments, upsilon decay, beam dump experiments,
atomic parity violation, low-energy neutron scattering
and primordial nucleosynthesis). We compute their implications for the
three parameters that in general describe the
low-energy properties of such bosons: their mass and their two
possible types of dimensionless couplings (direct couplings to
ordinary fermions and kinetic mixing with Standard Model
hypercharge).  We argue
that gauge bosons with very small couplings to
ordinary fermions in this mass range are natural in
string compactifications and are likely
to be generic in theories for which
the gravity scale is systematically smaller than
the Planck mass -- such as in extra-dimensional
models -- because of the necessity to suppress
proton decay. Furthermore, because its couplings are
weak, in the low-energy theory relevant to experiments
at and below TeV scales the charge gauged by the new boson
can appear to be broken, both by classical effects and by
anomalies. In particular, if the new gauge charge
appears to be anomalous,  anomaly cancellation
does {\em not} also require the introduction
of new light fermions in the low-energy
theory. Furthermore, the charge can appear to be conserved in the low-energy theory,
despite the corresponding gauge boson having a mass. Our
results reduce to those of other authors in the
special cases where there is no kinetic mixing or there is no direct coupling to ordinary
fermions, such as for recently proposed dark-matter scenarios.}

\begin{document}

\section{Introduction and summary of results}

New particles need not have very large masses in order to have evaded discovery; they can also be quite light provided they couple weakly enough to the other particles we {\em do} see. This unremarkable observation has been reinforced by recent dark matter models, many of which introduce new particles at GeV or lower scales in order to provide dark-matter interpretations for various astrophysical anomalies \cite{DarkSector}. This model-building exercise has emphasized how comparatively small experimental efforts might close off a wide range of at-present allowed couplings and masses for putative new light particles \cite{toroschuster, ovanesyan}.

\medskip\noindent{\em Light spin-one bosons}

\medskip\noindent
Spin-one gauge bosons are particularly natural kinds of particles to seek at low energies, since (unlike most scalars) these can have light masses in a technically natural way. Furthermore, their couplings are reasonably restrictive, allowing only two kinds of dimensionless interactions with ordinary Standard Model particles: direct gauge couplings to ordinary matter and kinetic mixing \cite{holdom2u1s} with Standard Model gauge bosons. Most extant surveys of constraints on particles of this type assume the existence of one or the other of these couplings, with older studies studying only direct gauge-fermion interactions \cite{carlson, bonly} and later studies (particularly for dark-matter motivated models) \cite{millichargebounds, ChangNgWu, HiddenU1Bounds, maximsec, cmb} usually allowing only kinetic mixing.

In this paper we have both motivational and phenomenological goals. On the phenomenological side, we analyze the constraints on new (abelian) gauge bosons, including both direct gauge-fermion couplings and gauge-boson kinetic mixing. In this way we include all of their dimensionless couplings, which (if all other things are equal) should dominate their behaviour at low energies. We can follow the interplay of these couplings with one another, and how this changes the bounds that can be inferred concerning the allowed parameter space. In particular we find in some cases (such as beam dump experiments) that bounds derived under the assumption of the absence of the other coupling can sometimes weaken, rather than strengthen, once the most general couplings are present.

Our motivational goal in this paper is twofold. First, we argue that the existence of gauge bosons directly coupled to ordinary fermions is very likely to be a generic and robust property of any phenomenologically successful theory for which the gravity scale is much smaller than the GUT scale \cite{StringLow,ADD,RS}. Next, we argue that these gauge bosons often very naturally have extremely weak gauge interactions within reasonable UV extensions of the low-energy theory, such as extra-dimensional models \cite{6Dflux} and low-energy string vacua \cite{IQ}. Besides motivating the otherwise potentially repulsive feature of having very small couplings, the smallness of these couplings (together with the low value for the fundamental gravity scale) also naturally tends to make the corresponding gauge bosons unusually light.

The remainder of this paper is organized as follows. The rest of this section, \S1, briefly summarizes the basic motivational arguments and phenomenological results. \S2\ then provides a more detailed theoretical background that motivates the sizes and kinds of couplings we consider, which may be skipped for those interested only in the bounds themselves. In \S3\ we briefly summarize the basic properties of the new gauge boson, with details given in an Appendix. By diagonalizing all kinetic terms and masses we identify the physical combination of couplings that are bounded in the subsequent sections. The next three sections, \S4, \S5\ and \S6, then explore the bounds on these couplings that are most restrictive for successively lighter bosons, starting at the weak scale and working down to MeV scales.

\subsection*{Motivational summary}

Why consider light gauge bosons that couple directly to ordinary fermions? And why should their couplings be so small? We here briefly summarize the more lengthy motivations given below, in \S2.

\medskip\noindent {\em Low-scale gravity and proton decay}

\medskip\noindent
Weakly coupled gauge bosons are likely to be generic features of any (phenomenologically viable) UV physics for which the fundamental gravity scale is systematically small relative to the GUT scale, $M_{\rm GUT} \sim 10^{15}$ GeV. Such bosons arise because of the difficulty of reconciling a low gravity scale with the observed stability of the proton. After all, higher-dimension baryon- and lepton-violating interactions that generically cause proton decay are not adequately suppressed if they arise accompanied by a gravity scale that is much smaller than $M_{\rm GUT}$. Similarly, global symmetries cannot themselves stop proton decay if the present lore about the absence of global symmetries in quantum gravity \cite{NoGlobinGrav} should prove to be true (as happens in string theory, in particular \cite{NoGlobSinST, hyper}).

This leaves low-energy gauge symmetries as the remaining generic mechanism for suppressing proton decay. Indeed, extra gauge bosons are often found in string vacua, and when the string scale is much smaller than the GUT scale, $M_s \ll M_{\rm GUT}$, these bosons typically play a crucial role in protecting protons from decaying. Furthermore, very weak gauge couplings appear naturally in such string compactifications, once modulus stabilization is included. In these systems the gauge couplings can be small because they are often inversely proportional to the volume of some higher-dimensional cycle, whose volume gets stabilized at very large values \cite{LV}. Similar things can also occur in non-stringy extra-dimensional models \cite{BraneBR}.

\medskip\noindent{\em Unbroken gauge symmetry without unbroken gauge symmetry}

\medskip\noindent
We believe there is a generic low-energy lesson to be drawn from how proton decay is avoided in phenomenological string constructions. This is because in these models, even though proton decay is forbidden by conservation of a gauged charge, the gauge boson that gauges this symmetry is not massless \cite{IQ}. This combines the virtues of an unbroken symmetry (no proton decay), with the virtues of a broken symmetry (no new forces mediated by a massless gauge boson).\footnote{Superconductors are similar in this regard: the photon acquires a mass without implying gross violations of charge conservation.} Usually this happy situation arises in the string examples because the gauge symmetry in question is anomalous, if judged solely by the light fermion content, with anomaly freedom restored through Green-Schwarz anomaly cancellation. But in four dimensions Green-Schwarz anomaly cancellation relies on the existence of a Goldstone boson, whose presence also ensures that the gauge boson acquires a nonzero mass.

For these constructions the effective lagrangian obtained just below the string scale from matching to the stringy UV completion is invariant under the symmetry apart from an anomaly-cancelling term that breaks the symmetry in just the way required to cancel the fermion loop anomalies. \S2\ argues that this property remains true (to all orders in perturbation theory) as one integrates out modes down to low energies. Leading symmetry breaking contributions arise non-perturbatively, exponentially suppressed by the relevant gauge couplings. Consequently they remain negligibly small provided only that the gauge groups involved in the anomalies are weakly coupled. Although supersymmetry also plays a role in the explicit string examples usually examined, our point here is that this is not required for the basic mechanism that allows massive gauge bosons to coexist with conservation of the corresponding gauge charge.

\FIGURE[tbh]{
\begin{tabular}{cc}
\includegraphics[scale=0.62]{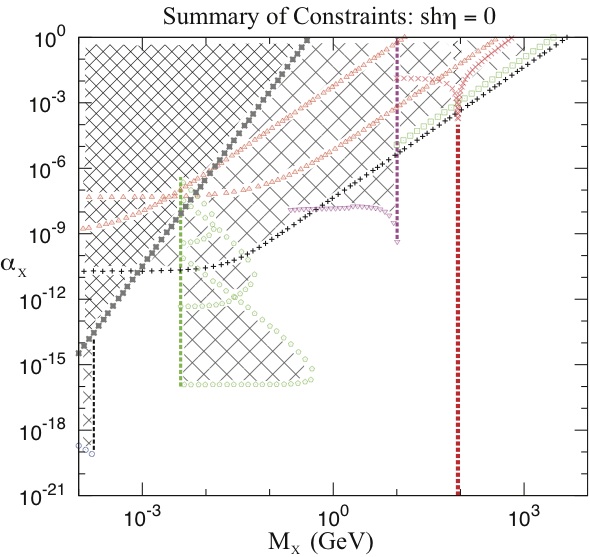} &
\includegraphics[scale=0.62]{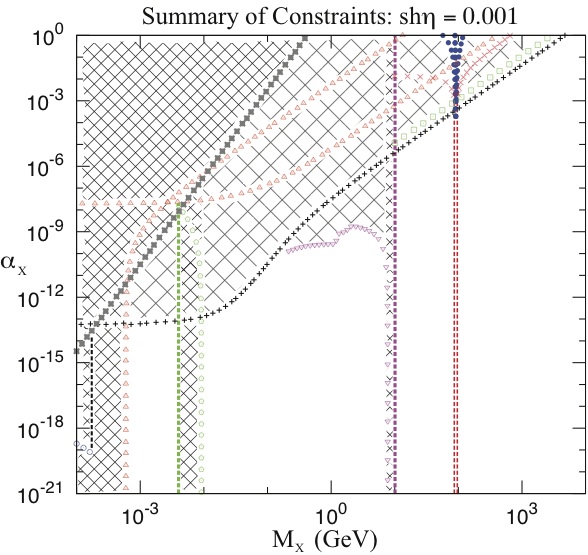} \\
\includegraphics[scale=0.62]{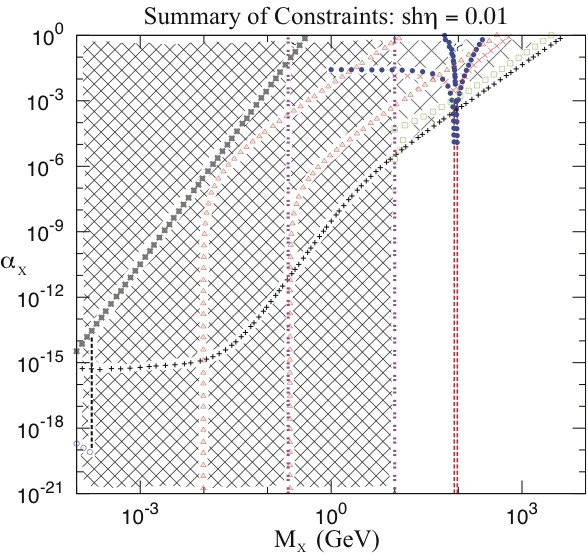} &
\includegraphics[scale=0.62]{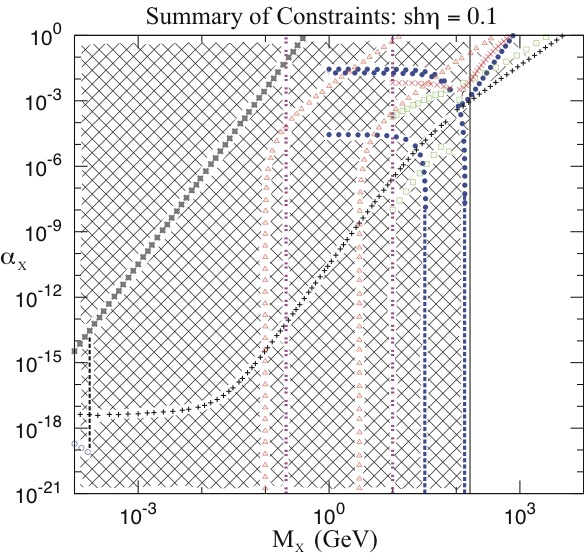} \\
\includegraphics[scale=0.62]{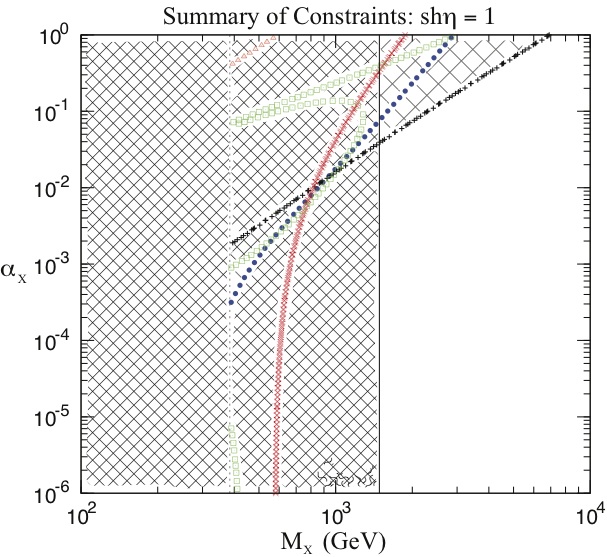} &
\includegraphics[scale=0.62]{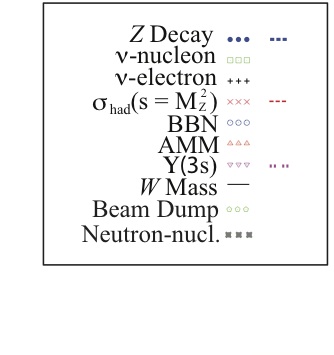}
\end{tabular}
\caption{Summary of the constraints presented herein. Each plot shows the bound on the new gauge coupling, $\alpha_\ssX$, as a function of $M_{\ssX}$ for various values of the kinetic-mixing parameter, $\sh\,\eta$, assuming a vector coupling $X_{f \ssL} = X_{f \ssR}:=X$, with $X=B-L$ ($X=B$) drawn as sparse (dense) cross-hatching.}
\label{allofem}
}

\subsection*{Phenomenological summary}

We next summarize, for convenience of reference, the combined bounds obtained from the constraints examined throughout the following sections.

\medskip\noindent {\em Mass vs coupling}

\medskip\noindent
Fig.~\ref{allofem} presents a series of exclusion plots in the $\alpha_\ssX-M_\ssX$ plane, where $\alpha_\ssX = g^2_\ssX/4\pi$ is the gauge-fermion coupling and $M_\ssX$ is the gauge boson mass. Each panel shows these bounds for different fixed values of the kinetic mixing parameter, $\sh\,\eta$ (for details on the definition of variables, see \S\ref{Xprops}). The figure shows the collective exclusion area of all of the different bounds considered in this paper. For concreteness they are calculated for a vector-like charge assignment, $X_{f\ssL} = X_{f\ssR}$, with the choice $X = B-L$ denoted by a lighter shading and the choice $X=B$ denoted with a heavier shading. Comparison of the cases $X = B$ and $X = B-L$ shows how much the bounds strengthen once direct couplings to leptons are allowed.

For $\eta =0$, the dominant bounds are from neutrino scattering, upsilon decay, anomalous magnetic moments, beam-dump experiments, neutron-nucleus scattering and nucleosynthesis. Once kinetic mixing is introduced, many of these bounds improve, with the exception of the beam-dump bounds. Once $\sh\,\eta\gtrsim0.06$, kinetic mixing becomes sufficiently strong that the $W$-mass bound prevails over any other bounds in the $M_\ssX \lsim M_\ssZ$ region. For $\sh\,\eta=1$, we discard the region where the oblique $T$ parameter is large (for details, see \S \ref{e+e-annihil}), and focus on the region where $M_\ssX>385$ GeV. In this region, it is the neutrino-electron scattering bound and the $W$-mass bound that dominate.

\medskip\noindent{\em Mass vs Mixing angle}

\medskip\noindent
It is useful to show these same bounds as exclusion plots in the mixing-angle/boson-mass plane, for fixed choices of the gauge-fermion coupling, $\alpha_\ssX$. This allows contact to be made with similar bounds obtained in the context of dark matter-inspired $U(1)$ models \cite{HiddenU1Bounds, toroschuster, ovanesyan, maximsec}, which correspond to the $\alpha_\ssX \to 0$ limit of the bounds we find here. This version of the plots is shown in Figure \ref{alphasummary}, restricted to the MeV-GeV mass range (in order to facilitate the comparison with earlier work).

\FIGURE[h!]{
\begin{tabular}{cc}
\includegraphics[scale=0.62]{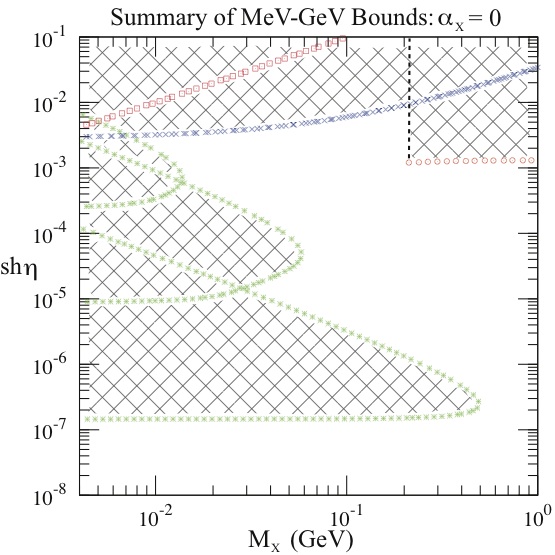} &
\includegraphics[scale=0.62]{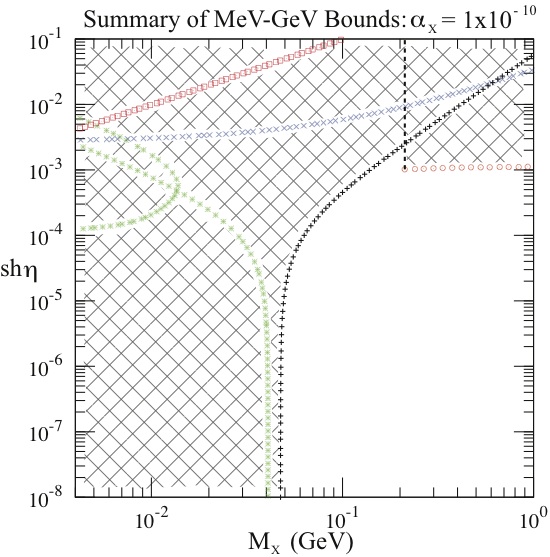} \\
\includegraphics[scale=0.62]{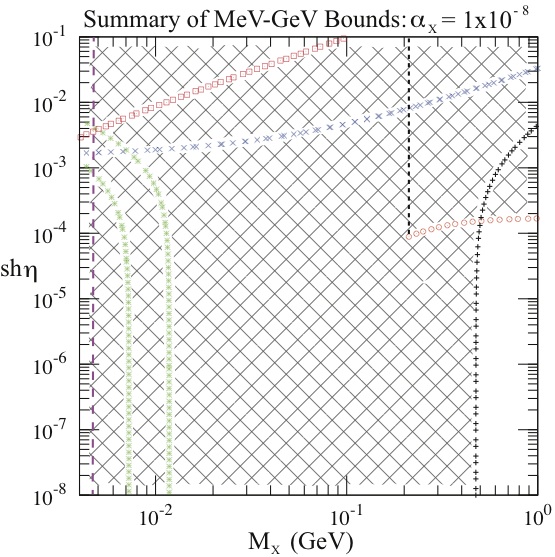} &
\includegraphics[scale=0.62]{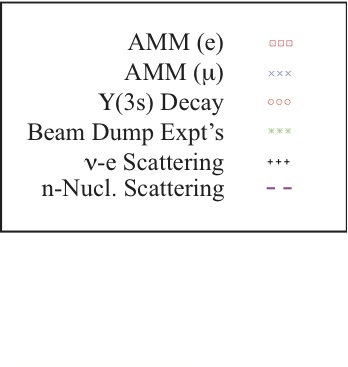}
\end{tabular}
\caption{Summary of the constraints on kinetic mixing relevant in the MeV-GeV mass range. Each plot shows the bound on the kinetic mixing parameter $\sh\,\eta$ as a function of $M_{\ssX}$, for $\alpha_\ssX=0$, $1\times10^{-10}$ and $1\times10^{-8}$. The plot assumes a coupling $X_{\ell \ssL} = X_{\ell \ssR} = -1$, such as would be true if $X = B-L$. Hatched regions are excluded.}
\label{alphasummary}
}

For small, but non-zero, gauge coupling ($\alpha_\ssX \sim 10^{-10}$) the bounds from beam dump experiments weaken significantly. However, another strong bound from neutrino-electron scattering also begins to take effect. This bound dominates for larger $\alpha_\ssX$, and once $\alpha_\ssX \gsim 10^{-7}$ the entire MeV$-$GeV mass range is excluded.

Since the bounds in Figure \ref{alphasummary} all rely on coupling to leptons, in the case where $X=B$ the constraints arise through the kinetic mixing and are independent of $\alpha_\ssX$. The resulting plot for $X=B$ is therefore the same as is shown in the figure for $\alpha_\ssX=0$. However, as the gauge coupling is increased the neutron-nucleus scattering bound --- discussed in \S\ref{neutnuclsec} --- eventually becomes important, first being visible as an exclusion in the $\sh\,\eta$ -- $M_\ssX$ plane in the panel for $\alpha_\ssX \simeq 10^{-8}$ in Fig.~\ref{alphasummary}.

\section{Theoretical motivation}

This section elaborates the motivations for weakly coupled, very light gauge bosons alluded to above. This is done both by summarizing the consistency conditions they must satisfy within the low-energy effective theory relevant to experiments, and by describing how such bosons actually arise from several representative UV completions in string theory and extra-dimensional models.

\subsection{Low-energy gauge symmetries, consistency and anomaly cancellation}

Very general arguments \cite{AnySpin,MasslessGauge} indicate that the interplay between unitarity and Lorentz invariance require massless gauge bosons only to couple to conserved charges that generate exact symmetries of the matter action. Consequently we normally expect the direct couplings of very light gauge bosons to be similarly restricted. This section reviews these arguments, emphasizing how they can break down \cite{GaugeUnitarity, AnomalyScale, NonlinIsBroken} if the energy scale, $\Lambda$, of any UV completion is sufficiently small compared with the gauge boson mass, $M$, and coupling, $g$: $\Lambda \lsim 4\pi M/g$. For the present purposes it suffices to restrict our attention to abelian gauge bosons (see however \cite{NonlinIsBroken} for some discussion of the nonabelian case).

The upshot of the arguments summarized here is that massive spin-one bosons can couple in an essentially arbitrary way if their mass, $M$, lies within a factor $g/4\pi$ of the scale of UV completion. But once $M$ becomes smaller than $g\Lambda/4\pi$, then the corresponding boson must gauge an honest-to-God, linearly realized exact symmetry. In particular this symmetry must be anomaly free. However any anomalies that Standard Model fermions give a putative new gauge charge needn't be cancelled by adding new, exotic low-energy fermions; they can instead be cancelled by the Goldstone boson whose presence is in any case required if the gauge boson has a mass. But this latter sort of cancellation also requires the UV completion scale to satisfy $\Lambda \lsim 4\pi M/g$.

Notice that for any given $M$ the condition $\Lambda \lsim 4\pi M/g$ need not require $\Lambda$ to lie below the TeV scale if the coupling $g$ is small enough. For instance, if $M \simeq 1$ MeV then $\Lambda$ lies above the TeV scale provided $g \lsim 10^{-5}$ (an upper limit often required in any case by the strong phenomenological bounds we find below). And, as subsequent sections argue, such small couplings can actually arise in a natural way from reasonable UV completions.

\subsubsection*{Massless spin-one bosons}

What goes wrong if a spin-one particle is not coupled to matter by gauging an exact symmetry of the matter action? If the spin-one particle is massless, then the problem is that one must give up either Lorentz invariance or unitarity (provided the particle has non-derivative, Coulomb-like couplings that survive in the far infrared). Lorentz invariance and unitarity fight one another because the basic field, $A_\mu(x)$, cannot transform as a Lorentz 4-vector if $A_\mu$ creates and destroys massless spin-one particles \cite{AnySpin,MasslessGauge}. Instead it transforms as a 4-vector {\em up to a gauge transformation}, $A_\mu \to A_\mu + \partial_\mu \omega$, and so interactions must be kept gauge invariant in order to be Lorentz invariant \cite{WbgBook}.

\subsection*{Massive spin-one bosons}

For massive spin-one particles the argument proceeds differently, as is now described. The difference arises because a 4-vector field, $A_\mu$, can represent a {\em massive} spin-one particle \cite{AnySpin}.

To examine the relevance of symmetries, it is worth first considering coupling massive spin-one particles to other matter fields, $\psi$, in some arbitrary non-gauge-invariant way, with lagrangian density $\cL(A_\mu,\psi)$. The first observation to make is that any such a lagrangian can be made gauge invariant for free, by introducing a St\"uckelberg field, $\phi$, according to the replacements $A_\mu \to \cA_\mu := A_\mu - \partial_\mu \phi$ and $\psi \to \Psi := \exp[-i \phi \, Q] \psi$, where $Q$ is a hermitian matrix acting on the fields $\psi$. With this replacement the lagrangian $\cL(\cA_\mu, \Psi)$ is automatically invariant under the symmetry $A_\mu \to A_\mu + \partial_\mu \omega$, $\phi \to \phi + \omega$ and $\psi \to \exp[i \omega \, Q] \psi$, since both $\cA_\mu$ and $\Psi$ are themselves invariant under these replacements. The original non-symmetric formulation corresponds to the specific gauge $\phi = 0$. For gauge symmetry, absence of gauge invariance is evidently equivalent to nonlinearly realized gauge invariance (similar arguments can also be made in the nonabelian case \cite{NonlinIsBroken}).

But this gauge invariance is obtained at the expense of introducing a new scale. Since $\phi$ is dimensionless, its kinetic term involves a scale, $v$,
\be
 \cL_{\rm kin} = - \frac1{4g^2} \, F_{\mu\nu} F^{\mu\nu} -
  \frac{v^2}2 \, (\partial_\mu \phi - A_\mu) \, (\partial^\mu \phi
  - A^\mu)  \,.
\ee
In $\phi = 0$ gauge the scale $v$ is seen to be related to the gauge boson mass by the relation $M = g v$. In a general gauge the scale $v$ controls the size of couplings between the canonically normalized field, $\varphi = \phi \, v$, and other particles. For instance the coupling
\be \label{psipsiphicoupling}
 \cL_{\rm coupling} = - i (\overline \psi \gamma^\mu Q\psi) \left(A_\mu - \frac{\partial_\mu \varphi}{v} \right) \,,
\ee
shows that the $(\overline\psi \gamma^\mu Q\psi) \partial_\mu\varphi$ coupling is dimension-five, being suppressed by the scale $v = M/g$.

Lagrangians with nonrenormalizable couplings like this must be interpreted as effective field theories, whose predictive power relies on performing a low-energy expansion in powers of $E/\Lambda$, for some UV scale $\Lambda$. The interpretation of the scale $v$ then generically depends on the how high $\Lambda$ is relative to $4\pi M/g = 4\pi v$. We consider each case in turn.

\medskip\noindent{\em Light spin-one bosons: $M \ll g \Lambda/4\pi$}

\medskip\noindent
If the gauge boson is very light compared with the UV scale, then its low-energy interactions should be describable by some renormalizable theory. But renormalizability is only consistent with a dimension-five interaction\footnote{The careful reader will recognize that this argument assumes negligible anomalous dimensions, and so needs re-examination for strongly coupled theories.} like the $(\overline\psi \gamma^\mu Q\psi) \partial_\mu \varphi$ coupling of eq.~\pref{psipsiphicoupling} if this coupling is a {\em redundant} interaction, such as it would be if it could be removed by a field redefinition. A sufficient condition for an interaction of the form $J^\mu \partial_\mu \varphi$ to be redundant in this way is if the field equations for $\psi$ were to imply the quantity $J^\mu(\psi)$ satisfies $\partial_\mu J^\mu = 0$ \cite{EFTrev}. This shows that if the gauge boson is to be arbitrarily light relative to $\Lambda$, its low-energy, renormalizable couplings must be to a (dimension-three) conserved current. This is the usual prescription for obtaining these couplings by gauging a linearly realized matter symmetry, for which $J^\mu$ is the usual Noether current.

\medskip\noindent{\em More generic massive spin-one bosons: $M \gsim g \Lambda/4\pi$}

\medskip\noindent
If, on the other hand, the dimension-five coupling $(\overline\psi \gamma^\mu Q\psi) \partial_\mu\varphi$ is not redundant, then there must be an upper bound on the UV scale: $\Lambda \lsim 4\pi v = 4\pi M/g$. Sometimes this may be seen from the energy-dependence predicted for the cross section of reactions in the low energy theory: if $\sigma(E) \propto 1/(4 \pi v)^2$ then this would be larger than the unitarity bound $\sigma \lsim 1/E^2$ for energies $E \gsim \Lambda \simeq 4 \pi v$, indicating the failure at these energies of the low-energy approximation. If so, the full UV completion must intervene at or below these energies to keep the theory unitary.

The upshot is that spin-one particles can couple fairly arbitrarily to matter provided they are massive, and provided the energy scale, $\Lambda$, of any UV completion satisfies $\Lambda \lsim 4\pi M/g$, where $M$ is the gauge boson mass and $g$ is its coupling strength. (Everyday examples of spin-one particles of this type include the $\rho$ meson or spin-one nuclei.) It is only spin-one particles with $M < g\Lambda/4\pi$ that must gauge linearly realized symmetries.

\subsection{Anomaly cancellation}

Any new gauge symmetry --- henceforth denoted $U(1)_\ssX$ --- must be an exact symmetry (though possibly spontaneously broken), and in particular must be anomaly free. This is true regardless of whether the symmetry is the linearly realized symmetry of a light gauge boson, or the nonlinearly realized symmetry of a massive gauge boson.

Of particular interest in this paper are models where the new symmetry acts on ordinary fermions, because a robust motivation for thinking about light gauge bosons is the avoidance of proton decay in models with a low gravity scale (more about which below). In this case these ordinary fermions usually contribute gauge anomalies for the new symmetry, and an important issue is how these anomalies are ultimately cancelled. The two main anomaly-cancellation scenarios then divide according to whether or not anomalies cancel among the SM fields themselves, or require the addition of new particles.

\subsubsection*{Anomaly cancellation using only SM fields}

The simplest situation is where the new gauge symmetry is simply a linear combination of one or more of the SM's four classical global symmetries --- baryon number $B$, electron number $L_e$, muon number $L_\mu$ and tau number $L_\tau$. In this situation there are only two independent combinations of these symmetries that are anomaly free\footnote{Notice that $B-L$ carries a Standard Model anomaly in the absence of sterile right-handed neutrinos (see below).} \cite{SM}, corresponding to arbitrary linear combinations of the anomaly-free symmetries $L_e - L_\mu$ and $L_\mu - L_\tau$:
\be
 X = a (L_e - L_\mu) + b (L_\mu - L_\tau) \,.
\ee
Of course, evidence for neutrino oscillations \cite{Numassev} make it unlikely that these symmetries are unbroken in whatever replaces the Standard Model in our ultimate understanding of Nature.

\subsubsection*{Anomaly cancellation using the Green-Schwarz mechanism}

If more general combinations of $B$, $L_e$, $L_\mu$ and $L_\tau$ are to be gauged, it is necessary to introduce new particles that can cancel their Standard Model anomalies. For a new $U(1)_\ssX$ symmetry the minimal way to do this is to add only the Goldstone boson, which must in any case be present if the corresponding gauge boson has a mass (as it typically must to avoid mediating a macroscopic, long-range new force, whose presence is strongly disfavoured by observations \cite{NewForce}). For a $U(1)_\ssX$ symmetry this can always be done using the 4D version \cite{4DGS} of the Green-Schwarz mechanism \cite{GS}. Besides its intrinsic interest, this is a way of cancelling anomalies that actually arises from plausible UV physics, such as low energy string models.

In principle, there are four types of new anomalies that can arise in 4D once the SM is supplemented by a new gauge symmetry, $U(1)_\ssX$. These are proportional to Tr[$XXX$], Tr[$XXY$], Tr[$XYY$] and Tr[$X G^a G^a$], where the trace is over all left-handed fermions and $X$ denotes the new symmetry generator, $Y$ is Standard Model hypercharge, and $G^a$ represents the generators of the Standard Model nonabelian gauge groups, $SU(2)_\ssL \times SU(3)_c$, as well as the generators of Lorentz transformations. In four dimensions CPT invariance implies the absence of pure gravitational anomalies, and anomaly cancellation within the Standard Model ensures the absence of anomalies of the form Tr[$YYY$] and Tr[$G^a G^b G^c$].

It is always possible to redefine the new symmetry generator, $V := X + \zeta \, Y$, to remove one of the two mixed anomalies. For instance, Tr$[VVY] = \hbox{Tr}[XXY] + 2 \,\zeta \, \hbox{Tr}[XYY]$ can be made to vanish by choosing $\zeta$ appropriately (provided Tr[$XYY$] does not vanish). It suffices then to consider only the case of nonzero anomalies of the form Tr[$VVV$] and Tr[$V G^a G^a$], where $G^a$ now includes also the generator $Y$. The anomaly then can be written in the $G^a$- and Lorentz-invariant form\footnote{A similar formulation can be made using the anomaly in its `consistent' form, rather than the `covariant' form used in the text.}
\bea \label{anomaly}
 \delta \Gamma &=& -  \int \exd^4 x \; \omega \Bigl\{ c_\ssX F_\ssV \wedge F_\ssV  + c_a  \hbox{Tr} [ F_a \wedge F_a ] - c_\ssL \hbox{Tr}[ R \wedge R ]  \Bigr\} \\
 &=& - \int \exd^4 x \; \omega \Bigl\{ c_\ssX \Bigl( F_\ssX + \zeta \, F_\ssY) \wedge (F_\ssX + \zeta \, F_\ssY) + c_a \hbox{Tr} [ G^a \wedge G^a ]
   - c_\ssL\hbox{Tr} [ R \wedge R ]\Bigr\}  \,,\nn
\eea
where $\Gamma$ is the `quantum action' (generator of 1PI correlations), the symmetry parameter is normalized by $\delta X_\mu = \partial_\mu \omega$ and the coefficients, $c_\ssX$, $c_a$ and $c_\ssL$, are calculable. Here $F_\ssV = \exd V = F_\ssX + \zeta \, F_\ssY$ is the gauge-boson field strength for the generator $X + \zeta \, Y$, while $F_a$ is the same for the Standard Model gauge bosons and $R$ is the gravitational curvature 2-form.

Given the coefficients $c_\ssX$, $c_a$ and $c_\ssL$, here is how 4D Green-Schwarz anomaly cancellation works \cite{4DGS}. Consider the gauge kinetic lagrangian, including the St\"uckelberg field $\phi$,
\bea
\label{stract}
   {\cal L} &=& \cL_{\rm inv} - {1 \over 4g^{2} } F^\ssX_{\mu\nu} F_\ssX^{\mu\nu} - {1 \over 4 g_{a}^{2} } {\rm Tr} [G^{a}_{\mu\nu} G_{a}^{\mu\nu} ] - {v^2 \over 2} ( \partial^{\mu} \phi - X^{\mu} )  (\partial_\mu \phi -  X_\mu) \nn\\
   && \qquad + \phi \Bigl\{ c_\ssX \Bigl( F_\ssX + \zeta \, F_\ssY) \wedge (F_\ssX + \zeta \, F_\ssY) + c_a \hbox{Tr} [ G^a \wedge G^a ]
   - c_\ssL\hbox{Tr} [ R \wedge R ]\Bigr\} \,.
\eea
Here $\cL_{\rm inv}$ denotes those parts of the lagrangian that are invariant under all of the gauge symmetries that are not written explicitly. The second line is not invariant under gauge transformations because $\phi$ is not; its variation precisely cancels the fermion anomaly, eq.~\pref{anomaly}.

An important observation is that the anomaly cancelling term is dimension-five, and so is not renormalizable. For instance, in terms of the canonically normalized field, $\varphi = \phi v$, the first anomaly cancelling term is $\cL_{\rm anom} = (\varphi/f) F_\ssX \wedge F_\ssX + \cdots$, where $f = v/c_\ssX$. As before, this implies the existence of a UV-completion scale, $\Lambda$, above which the low-energy effective description breaks down \cite{AnomalyScale}. For weakly coupled theories typically $\Lambda  \lsim 4\pi v \simeq 4\pi M/g \simeq 4 \pi c_\ssX f$ marks the scale where the fields arise that are required to extend the Goldstone boson to a linear representation of the symmetry.

Perhaps the most interesting feature of cancelling anomalies with the Green-Schwarz mechanism in this way is that the lagrangian remains invariant under the $U(1)$ symmetry, apart from the anomaly-cancelling term. This is interesting because it means that the corresponding charge still appears to be conserved in the low-energy theory, {\em despite the gauge field being massive}. This opens up interesting phenomenological possibilities for the gauging of symmetries like $U(1)_{B}$ and $U(1)_{B-L}$, which appear to be conserved in Nature but which are also ruled out as sources of the new long-range force that a massless gauge boson would imply.

One might worry that arbitrary symmetry-breaking interactions might be generated by embedding the anomaly cancelling interactions (or the fermion triangle anomaly graph) into a quantum fluctuation. For instance if $X = B$, so the new gauge boson couples to baryon number, then why can't some complicated loop generate a $\Delta B = \pm 1$ interaction, $\cO_{\pm 1}$, that can mediate proton decay? After all, this can be $U(1)_B$ invariant if it arises multiplied by a factor $e^{\mp i\phi}$, which carries baryon number $\Delta B = \mp 1$.

The difficulty with generating this kind of interaction is that it must involve $\phi$ undifferentiated. But if we restrict $\cL_{\rm anom}$ to constant $\phi$ configurations, it becomes a total derivative. For constant $\phi$, the dependence of observables on $\phi$ is similar to the dependence of observables on the vacuum angle, $\theta$. Consequently it arises at best only non-perturbatively, proportional at weak coupling to a power of $\sim \exp[- 8\pi^2/g^2]$, where $g$ is the anomalous gauge coupling. As a result the only potentially dangerous contribution of this type comes from the mixed $X$-QCD-QCD anomaly, which can generate nontrivial $\phi$-dependence once we integrate down to scales $\lsim \Lambda_{\scriptscriptstyle QCD}$. This is not dangerous in particular for the classical symmetries, $B$, $L_e$, $L_\mu$ and $L_\tau$, since these do not have mixed QCD anomalies \cite{SM}.

\subsubsection*{Anomaly cancellation using new fermions}

More complicated possibilities for new gauge bosons emerge if new, light exotic fermions are allowed that also carry the new $X$ charge (and so can also take part in the anomaly cancellation). We briefly describe some features involving such new exotic particles, although they do not play any role in our later phenomenological studies.

The simplest example along these lines is $X = B-L$, which is anomaly-free provided only that the SM spectrum is supplemented by three right-handed neutrinos (one for each generation). Furthermore, conservation of $L$ is consistent with all evidence for neutrino oscillations, although it would be ruled out should neutrinoless double-beta decay ever be witnessed.

A practical way in which such new fermions can arise at TeV scales is if the UV theory at these scales is supersymmetric. In this case the plethora of new superpartners can change anomaly cancellation in one of two ways (or both). They can either directly contribute to the anomalies themselves, and possibly help anomalies cancel without recourse to the Green-Schwarz mechanism. Alternatively, they can modify the details of how the Green-Schwarz mechanism operates if the UV scale, $v$, associated with it is larger than the supersymmetry breaking scale, $M_{\rm susy}$.

In particular, supersymmetry typically relates the kinetic term for the St\"uckelberg field, ({\ref{stract}}), with a Fayet-Iliopoulos term in the scalar potential \cite{GS4Dhet},
\be
  S_{FI} = - {1 \over g^{2} } \int d^4 x\bigg( \tau - \sum_{i} q_{i} \phi_{i}^{\dagger} \phi_{i} \bigg)^{2} \,,
\ee
where $\tau$ is a dynamical field whose vev acts as the low-energy Fayet-Iliopoulos parameter; the $q_{i}$ are the charges of the fields $\phi_{i}$ under the $U(1)$ in question. In string examples the field $\tau$ corresponds to a modulus of the compactification, which controls the size of a cycle in the internal geometry on which some branes wrap. We note that the vanishing of the D-term is consistent with vanishing vevs of the charged fields if $\tau = 0$, {\em i.e.} the symmetry survives as an {\em exact} global symmetry when the cycle size vanishes (the singular locus). Small values of the vev are obtained if the cycle size is small.

\subsection{Motivations from UV physics}

The above summary outlines some of the theoretical constraints on coupling ordinary fermions to very light gauge bosons. This section shows how very small couplings can naturally appear in well-motivated ultraviolet physics, such as extra-dimensional models or string vacua. In particular, they often arise due to considerations of proton stability in constructions for which the gravity scale is small compared with the Planck scale, as we now explain.

\subsubsection*{Proton decay in low-scale gravity models}

One of the surprises of the late 20th century was the discovery that the scale, $M_g$, of quantum gravity could be much smaller than the Planck scale, $M_p = (8\pi G)^{-1/2} \simeq 10^{18}$ GeV \cite{StringLow}. From the point of view of particle physics this possibility is remarkable for several reasons. Most obvious is the potential it allows for experimental detection if it should happen that $M_g$ is in the vicinity of the TeV scale \cite{ADD, RS}.

But there is a potentially more wide-reaching consequence that $M_g \ll M_p$ has for the low-energy sector: the suppression by powers of $M_g/M_p$ it allows for otherwise UV-sensitive radiative corrections \cite{Ubernat}. This suppression arises because the contribution of short-wavelength degrees of freedom can saturate at $M_g$, allowing their effects to be suppressed by powers of the gravitational coupling.

The most precise examples of this are provided by string theory, in the regime where the string scale is low, $M_g := M_s \ll M_p$ \cite{StringLow}. String theory makes the suppression of UV-sensitive contributions precise by providing an explicit stringy ultraviolet completion within which the effects of the full UV sector can be explored. Large-volume (LV) models \cite{LV} are particularly useful laboratories for these purposes, since these systematically exploit the expansion of inverse powers of the extra-dimensional volume (in string units), $\cV := (\hbox{Vol})/\ell_s^6 \gg 1$, and it is ultimately these kinds of powers that enforce the suppressions of interest since $M_s/M_p \propto \cV^{-1/2}$.

Proton decay --- that is, its experimental absence --- turns out to impose a very general constraint on any fundamental theory of this type, with $M_g \ll M_p$. It does so because having $M_g$ very small removes two of the standard ways of keeping the proton stable in specific models. On one hand quantum gravity, and string theory in particular \cite{NoGlobSinST, hyper}, seems to preclude the existence of global symmetries, and this forbids ensuring proton stability by simply using a conserved global charge (such as baryon number).

If $M_g$ is too small then it is also unlikely that such a symmetry simply emerges by accident for the lowest-dimension interactions in the low-energy effective theory. The problem in this case is that we know that generic higher-dimensional interactions,
\be
 \cL_{\rm eff} = \sum_i \frac{c_i \cO_i}{M_g^{d_i-4}}  \,,
\ee
eventually do arise in the low-energy effective theory, such as the standard baryon-number violating 4-quark operators arising at dimension $d_i = 6$ \cite{BVdim6} in the low-energy limit of grand-unified theories (GUTs) \cite{GUTFirst, GUTRevs}. But a dimension-six interaction of the form $\cO/M^2$ generically contributes a proton-decay rate of order $\Gamma \simeq m_p^5/M^4$, where $m_p$ is the proton mass, which is too large to agree with observations once $M$ falls below $M_\GUT \simeq 10^{16}$ GeV.

The way theories with $M_g \ll M_p$ usually evade proton decay is through the appearance of a {\em gauged} $U(1)$, whose conservation forbids the decay. Of course, to be useful the gauged $U(1)$ that appears must couple to the proton or its decay products in order to forbid its decay. But because this means ordinary particles couple to the new gauge boson, it potentially introduces other phenomenological issues. If the gauge symmetry is conserved, why isn't the gauge boson massless? If the gauge boson is light, why isn't the new boson seen in low-energy observations? If the gauge boson is heavy, the corresponding symmetry must be badly broken and so how can it help with proton decay? Interestingly, extant models can naturally address both of these issues, and often the low-energy mechanism that is used is Green-Schwarz anomaly cancellation with gauge boson mass generated through the St\"uckelberg mechanism described above. Sometimes this mechanism is also combined with supersymmetry to suppress the dangerous decays.

The existence of these gauge bosons, their properties, and the way they evade the above issues, may be among the few generic low-energy consequences of viable theories with a low gravity scale: $M_g \ll M_\GUT$.

\medskip \noindent {\em Sample symmetries:}

\medskip\noindent
The simplest proposals for new low-energy gauge groups that forbid proton decay are either baryon or lepton number, $X = B$ or $X = L$. If the anomalies for these symmetries due to Standard Model fermions are cancelled through the Green-Schwarz mechanism, then no new light particles are required besides the massive gauge boson itself.

More complicated examples are possible if the low-energy theory at TeV scales is supersymmetric. In this case symmetries like $B-L$, that in themselves cannot forbid proton decay, can help suppress proton decay if taken together with supersymmetry \cite{IQ}. (For instance, the parity $R = (-)^{F + 3(B-L)}$ that is usually used to suppress proton decay in the MSSM is a combination of fermion number and $B-L$.)

More general combinations of $B$ and $L$ can also suppress proton decay in supersymmetric theories. Ref.~\cite{IQ} provides a list of the kinds of symmetries of this type that can be relevant to proton decay, as well as the conditions they must satisfy in order to have their anomalies be cancelled through the Green-Schwarz mechanism. The general form for the low-energy charge may be written
\be
 X = m T_\ssR + n A + p L \,,
\ee
where $T_\ssR$ is right-handed isospin; $A$ is an axionic PQ symmetry; and $L$ is lepton number, with the charge assignments given in the Table. The coefficients $m$, $n$ and $p$ are subject to (but not over-constrained by) several anomaly cancellation conditions \cite{IQ}. In particular $B$ and $L$ violating interactions can be forbidden up to and including dimension six for some choices of these symmetries in the supersymmetric limit, as can the $\mu$-terms of the superpotential -- $W \simeq \mu_\ssL L \ol{H}$ and $W \simeq \mu H \ol{H}$ -- if $n \ne 0$.

\begin{center}
\begin{tabular}{c|ccccccc}
    & $Q$ & $U$ & $D$ & $L$ & $E$ & $H$ & $\ol{H}$ \\
  \hline
  $T_\ssR$ & 0 & 1 & $-1$ & 0 & $-1$ & 1 & $-1$ \\
  $A$ & 0 & 0 & 1 & 1 & 0 & $-1$ & 0 \\
  $L$ & 0 & 0 & 0 & 1 & $-1$ & 0 & 0 \\
  $X$ & 0 & $m$ & $n-m$ & $n+p$ & $-m-p$ & $m-n$ & $-m$ \\
\end{tabular}
\end{center}

\subsubsection*{Very light and weakly coupled gauge bosons from extra-dimensional models}

For the phenomenological discussions of later sections we consider gauge bosons in the MeV to TeV mass range, whose direct couplings to Standard Model fermions are much smaller than those arising within the Standard Model itself. This section and the next one describe several way that very light and weakly coupled bosons can arise from reasonable UV physics.

Extra-dimensional supergravity provides a simple way to obtain very light gauge bosons that are very weakly coupled. A concrete example is six-dimensional chiral gauged supergravity \cite{NS}, for which the bosonic part of the gravity multiplet contains the metric, $g_{\ssM\ssN}$, a Kalb-Ramond 2-form potential, $B_{\ssM\ssN}$, and a scalar, $\phi$. Because it is chiral this supergravity potentially has anomalies, whose cancellation imposes demands on the matter content. In six dimensions Green-Schwarz anomaly cancellation is not automatic, because cancellation of the pure gravitational anomalies requires the existence of a specific number of gauge multiplets \cite{6Danom}. Given these multiplets, mixed gauge-gravity anomalies can be cancelled through the Green-Schwarz mechanism using the couplings of the field $B_{\ssM\ssN}$.

The resulting supergravity admits simple solutions for which the extra dimensions are a sphere \cite{SSsoln}, whose moduli can be stabilized by a combination of background fluxes in the extra dimensions \cite{6Dflux}, and branes coupling to the 6D dilaton \cite{6DGWmech,BraneBR}. An important feature of this stabilization is that the value of the dilaton field becomes related by the field equations to the size of the extra dimensions:
\be \label{6Dmodstab}
 e^\phi = \frac{1}{(M_6 r)^{2}} \,,
\ee
where $M_6$ denotes the 6D Planck scale. This ensures these models are a rich source of $U(1)$ gauge bosons, some of whom can have massless modes that survive to low energies below the Kaluza-Klein scale. Some of these gauge modes also naturally acquire masses through the St\"uckelberg mechanism \cite{6Dflux} (with the St\"uckelberg field arising as a component of the Kalb-Ramond field, $B_{\ssM\ssN}$).

Besides having light gauge bosons, these models also naturally furnish them with very small coupling constants. This is because the loop-counting parameter for all bulk interactions turns out to be the value of the 6D dilaton, $\phi$, with $g^2 \simeq e^\phi$. But modulus stabilization, eq.~\pref{6Dmodstab}, ensures that this coupling can be extremely small because it scales inversely with the size of the extra dimensions (measured in 6D Planck units).

\subsubsection*{Very light and weakly coupled gauge bosons from low-energy string vacua}

A related mechanism also often arises in low-scale string models. In early heterotic models the role of the Goldstone boson is played by a member of the dilaton super-multiplet: $a \simeq \hbox{Im} \, S$ \cite{GS4Dhet}, while in later Type I and Type II models it is twisted closed string  multiplets that instead play this role \cite{GS4DII,IQ}. Although the universal couplings of the dilaton restrict the kinds of symmetries that can arise in heterotic constructions of this type, the same is not true for Type I and II models.

There is a simple reason why additional $U(1)$ gauge groups often arise. The basic building blocks for constructing models of particle physics in type IIB and IIA string theory are D-branes. Generically, the gauge group associated with a stack of $N$ D-branes is $U(N)$, but the Standard Model gauge group involves special unitary groups, $SU(3) \times SU(2) \times U(1)$. Typical GUT models also involve special unitary groups, like $SU(5)$, $SU(3) \times SU(2) \times SU(2) \times U(1)$ (Left-Right symmetric models) or $SU(4) \times SU(2) \times SU(2)$ (Pati-Salam models). It is the additional $U(1)$s that distinguish the Standard Model $SU(N)$ factors from the $U(N)$ factors arising from the
D-branes, that give new low-energy gauge symmetries.

Furthermore, anomaly cancellation in string theory typically demands the presence of additional D-brane stacks, in addition to those providing the Standard Model gauge group factors. These stacks also lead to extra $U(1)$s under which
Standard Model particles are charged. Extra $U(1)$s also appear naturally in F-theory models (for a recent discussion see \cite{timo}). In many concrete examples these additional gauge fields correspond to $U(1)_{B}$ or $U(1)_{B-L}$, hence can be relevant for the stability of the proton \cite{fqib, IQ} (see also \cite{cmq} for a recent discussion).

\medskip\noindent{\em Masses and couplings}

\medskip\noindent
For string vacua the masses and couplings of any gauged $U(1)$s can be computed, as we now briefly describe.

Consider first the $U(1)$s associated with the same stack of D-branes as gives rise to the Standard Model gauge group. As discussed earlier, such gauge bosons often acquire masses from the St\"uckelberg mechanism. The size of the mass generated in this way is the string scale when the $U(1)$ is anomalous \cite{wijn, cmq}, but it is the smaller Kaluza-Klein scale for non-anomalous $U(1)$s.

For models with the compactification volume not too much larger than the string scale, these $U(1)$ gauge bosons are very heavy. On the other hand, for large-volume models the string scale can be quite low, leading to additional $U(1)$s potentially as light as the TeV scale. The latter can have interesting low energy phenomenology (see for instance \cite{fqib, david, zwo, zwt, kr}). In these models the strength of the gauge coupling for the additional $U(1)$s is roughly the same as for the Standard Model gauge couplings (evaluated at the string scale), because both have the same origin: the world-volume theory of the stack. Hence they cannot be extremely small.

The masses and couplings of the extra $U(1)$ gauge bosons vary more widely when they arise from D-brane stacks whose $SU(N)$ factors are not part of the Standard Model gauge group. For instance, the case of additional $U(1)$s associated with D7 branes wrapping bulk four cycles of the compactification is discussed in detail in \cite{hyper}. The value of the gauge coupling in this case is inversely proportional to the volume (in string units) of the cycle, $\Sigma$, that the D7 brane wraps,
\be
    g^{2} \approx { 4 \pi \over {{\cal{V}}_{\Sigma} }} \,.
\ee
In the context of the large volume scenario (LVS) of modulus stabilization {\cite{LV}}, the size of the bulk cycle associated with the overall volume of compactification can easily be approximately $\cV_\Sigma \gsim 10^{9}$ in string units, set by the requirement that one generate TeV-scale soft terms. Thus one can obtain gauge couplings as low as $g \lsim 2 \times 10^{-4}$ \cite{hyper, LVSU1} (couplings larger than this can be obtained if the D7 brane wraps a cycle different from the one associated with the overall volume). With couplings this small, the gauge boson mass can be $M_\ssX \simeq g v \lsim 100$ MeV even if $v$ is a TeV.

\section{Gauge boson properties} \label{Xprops}

With the above motivation, our goal in the remainder of the paper is to work out various constraints on the parameters of a massive (yet comparatively light) gauge boson, the $X$ boson, that couples to a new $U(1)_\ssX$ symmetry. Since the lowest dimension interactions dominate in principle at low energies, we include in our analysis all of the dimensionless couplings that such a boson could have with Standard Model particles: {\em i.e.} both direct fermion-gauge couplings and gauge kinetic mixing. We see how these are constrained by present data as a function of the gauge boson mass.

More specifically, we consider an effective lagrangian density below the supersymmetry breaking scale of the form
\begin{equation}
 \cL = \cL_\SM + \cL_\ssX +
 \cL_{\rm mix}
\end{equation}
where $\cL_\SM$ is the usual Standard Model lagrangian; $\cL_\ssX$ describes the $X$ boson, including its couplings to the SM fermions; and $\cL_{\rm mix}$ is the kinetic-mixing interaction between the $X$ boson and that of the SM gauge factor $U(1)_\ssY$ \cite{holdom2u1s}. Explicitly
\be
 \cL_\ssX = - \frac14 \, X_{\mu\nu} X^{\mu\nu} - \frac{m_\ssX^2}{2} \,
 X_\mu X^\mu +  i J^\mu_\ssX X_\mu \,,
\ee
where $X_{\mu \nu } := \partial_\mu X_\nu - \partial_\nu X_\mu$ is the curl of the appropriate gauge potential, $X_\mu$, and $J^\mu_\ssX$ is the current for the $U(1)_\ssX$ gauge symmetry involving the SM fermions. Similarly, $\cL_{\rm mix}$ has the form,
\be
 \cL_{\rm mix} = \chi \, B_{\mu \nu } X^{\mu \nu }
\ee
where $B_\mu$ is the SM gauge boson for the gauge factor $U(1)_\ssY$.

The analysis we provide complements and extends earlier studies of extra gauge boson phenomenology. In the lower part of the mass range we may compare with \cite{carlson}, who some time ago considered the special cases $X = B-L$ and $\chi = 0$. Contact is also possible in this mass range with more recent Dark Matter models \cite{HiddenU1Bounds, toroschuster, ovanesyan} in the absence of direct matter couplings, $g_\ssX J^\mu = 0$. At masses much lower than those considered here other constraints on kinetic mixing have also been studied, from the cosmic microwave background \cite{cmb}, and from the absence of new long-range forces \cite{NewForce} or milli-charged particles \cite{holdom2u1s, millichargebounds}.

There is also a broad literature on the phenomenology of gauge bosons at the upper end of the mass range, largely done in the context of a $Z^{\prime}$ field and often motivated by GUTs \cite{GUTFirst, so10, e6, hewettrizzo, leftright, atthetev}. Until recently, most did not include the kinetic mixing term. Constraints including kinetic mixing arising from precision electroweak experiments are considered in \cite{preseweakmix, BKM, Leike}; more recent bounds are found in \cite{HIW,langacker,wells,ChangNgWu}. Many of these analyses overlap parts of our parameter space. For instance $Z^\prime$ searches, such as \cite{d0}, give bounds on the mass of the $Z^\prime$ that apply in the regime that the couplings to fermions are identical to that of the $Z$. Others \cite{bonly,CMbarnum}\ derive bounds for a $Z^{\prime }$ coupled only to baryon number.

One difference between the models examined here and those usually considered for $Z'$ phenomenology at the weak scale, such as those of ref.~\cite{massmixing}, is the absence in $\cL$ of mixing between the $X$ and the $Z$ bosons in the mass matrix ({\em i.e.} a term of the form $\cL_{\rm mix} = \delta m^2 Z_\mu X^\mu$). We do not consider this type of mixing because we imagine the models of interest here to break the $X$ symmetry with a SM singlet. Notice that because the SM Higgs is uncharged under the $X$ symmetry, the strong bounds as found, for example, in \cite{CGR} don't apply.

\subsection{The mixed lagrangian}

In this section we diagonalize the gauge boson kinetic mixing terms (and SM mass terms) and identify the physical combination of parameters relevant for phenomenology within the accuracy to which we work. Our goal in so doing is to follow ref.~\cite{stuvwx, bigfit} and identify once and for all how the gauge boson mixing contributes to fermion couplings and to oblique parameters \cite{oblique} modified by the gauge-boson mixing. This allows an efficient identification of how observables depend on the mixing parameters.

We begin by writing the lagrangian of interest more explicitly, after spontaneous symmetry breaking. Because it is the $Z$ and photon that potentially mix with the $X$ boson, we also focus on these sectors of the SM lagrangian. In order to distinguish the fields before and after mixing, where appropriate we denote the still-mixed fields with carets, {\em e.g.} $\hat X_\mu$, reserving variables like $X_\mu$ for the final, diagonalized fields.

With this notation, the lagrangian of interest is
\begin{equation}
 \cL = \cL_{\rm gauge} + \cL_{f}
 +\cL_{\rm int} \,,
\end{equation}
where
\be
\cL_{\rm gauge} = \cL_{\rm kin} + \cL_{\rm mass}
\ee
with
\begin{eqnarray}
 \cL_{\rm kin} &=& - \frac14 \hat{W}_{\mu\nu}^3
 \hat{W}_3^{\mu\nu} - \frac14 \hat{B}_{\mu\nu}
 \hat{B}^{\mu\nu} - \frac14 \hat{X}_{\mu\nu}
 \hat{X}^{\mu\nu}  + \frac{\chi}{2} \hat{B}_{\mu\nu}
 \hat{X}^{\mu \nu } \\
 \cL_{\rm mass} &=& -\frac12 \left( m_3 \hat{W}_\mu^3
 -m_0 \hat{B}_\mu \right) \left( m_3 \hat{W}_3^\mu
 -m_{0} \hat{B}^\mu \right) - \frac{m_\ssX^2}{2}
 \, \hat{X}_\mu \hat{X}^\mu \,,
\end{eqnarray}
and
\begin{eqnarray}
 \cL_{f} &=& - \sum_f \ol{f} \left( \dsl
 + m_f \right) f \\
 \cL_{\rm int} &=& i \sum_f \left\{ g_2 \left( \ol{f}
 \gamma^\mu T_{3f} \gamma_\ssL f \right) \hat{W}_\mu^3
 + g_1 \left[ \ol{f} \gamma^\mu
 \left( Y_{f \ssL} \gamma_\ssL + Y_{f \ssR} \gamma_\ssR \right)
 f \right] \hat{B}_\mu \right. \\
 && \quad \left. + g_\ssX \left[ \ol{f} \gamma^\mu
 \left( X_{f \ssL} \gamma_\ssL + X_{f \ssR} \gamma_\ssR \right)
 f \right]
 \hat{X}_\mu \right\} \,.
\end{eqnarray}
Here $T_{3f}$, $Y_{f\ssL}$ and $Y_{f\ssR}$ denote the usual SM charge assignments, while $X_{f\ssL}$ and $X_{f\ssR}$ are the fermion charges under the new $U(1)_\ssX$ symmetry. The SM masses, $m_3$ and $m_0$, are defined as usual \cite{SM} in terms of the standard model gauge couplings, $g_1$ and $g_2$, and the Higgs VEV, $v$: $m_3 = \frac12 \, g_2 v$ and $m_0 = \frac12 \, g_1 v$. $\gamma_\ssL$ and $\gamma_\ssR$ are the usual left- and right-handed Dirac projectors.

Defining the gauge-field-valued vector $\hat{\bf V}$ to be
\begin{equation}
 \hat{\bf V}=\begin{bmatrix}
 \hat{W}^{3} \\
 \hat{B} \\
 \hat{X}
\end{bmatrix} \,,
\end{equation}
the above lagrangian can be written in matrix form
\begin{equation} \label{startLeq}
 \cL_{\rm gauge} + \cL_{\rm int} = -\frac14 \hat{\bf V}_{\mu\nu}^\ssT \hat K
 \hat{\bf V}^{\mu\nu} - \frac12 \hat{\bf V}_\mu^\ssT
 \hat M \hat{\bf V}^\mu + i \hat{\bf J}_\mu^\ssT
 \hat{\bf V}^\mu
\end{equation}
where
\begin{equation} \label{startKMeq}
 \hat K :=
 \begin{bmatrix}
 1 & 0 & 0 \\
 0 & 1 & -\chi \\
 0 & -\chi & 1
 \end{bmatrix}
 \quad \hbox{and} \quad
 \hat M :=
 \begin{bmatrix}
 m_{3}^2 & -m_{3}m_{0} & 0 \\
 -m_{3}m_{0} & m_{0}^2 & 0 \\
 0 & 0 & m_\ssX^2
 \end{bmatrix} \,,
\end{equation}
and
\begin{equation}
 \hat{\bf J}_\mu :=
 \begin{bmatrix}
 J_\mu^3 \\
 J_\mu^\ssY \\
 \hat{J}_\mu^\ssX
 \end{bmatrix}
 = \sum_f \begin{bmatrix}
 g_2 \left[ \ol{f} \gamma_\mu T_{3f} \gamma_\ssL f \right] \\
 g_1 \left[ \ol{f} \gamma_\mu \left( Y_{f\ssL} \gamma_\ssL
 + Y_{f\ssR} \gamma_\ssR \right) f \right] \\
 g_\ssX \left[ \ol{f} \gamma_\mu \left( X_{f\ssL}
 \gamma_\ssL + X_{f\ssR} \gamma_\ssR \right) f \right]
 \end{bmatrix} \,.
\end{equation}
The off-diagonal elements of $\hat M$ ensure it has a zero eigenvalue, and the condition that the matrix $\hat K$ be positive definite requires $\chi^2 < 1$.

\subsection{Physical couplings} \label{physcouplings}

In order to put this lagrangian into a more useful form we must diagonalize the kinetic and mass terms, and then eliminate the SM electroweak parameters in terms of physically measured input quantities like the $Z$ mass, $M_\ssZ$, the fine-structure constant, $\alpha = e^2/4\pi$, and Fermi's constant, $G_\ssF$, as measured in muon decay.

The diagonalization is performed explicitly in the Appendix, leading to the diagonalized form
\begin{equation}
 \cL = -\frac14 \, {\bf V}_{\mu\nu}^\ssT
 {\bf V}^{\mu\nu} - \frac{M_\ssZ^2}{2} \, Z_\mu Z^\mu
 -\frac{M_\ssX^2}{2} \, X_\mu X^\mu +
  i {\bf J}_\mu^\ssT {\bf V}^\mu  \,,
\end{equation}
where the physical masses are
\be
 M_\ssX^2 = \frac{m_\ssZ^2}{2}
 \left( 1 + \hat s_\ssW^2 \sh^2 \eta + r_\ssX^2 \ch^2 \eta
 + \vartheta_\ssX \sqrt{ \left( 1+ \hat s_\ssW^2 \sh^2 \eta
 + r_\ssX^2 \ch^2 \eta \right)^2
 - 4 r_\ssX^2 \ch^2 \eta} \right)
\label{masseigs1}
\ee
and
\be
 M_\ssZ^2 = \frac{m_\ssZ^2}{2}
 \left( 1 + \hat s_\ssW^2 \sh^2 \eta + r_\ssX^2 \ch^2 \eta
 - \vartheta_\ssX \sqrt{ \left( 1+ \hat s_\ssW^2 \sh^2 \eta
 + r_\ssX^2 \ch^2 \eta \right)^2
 - 4 r_\ssX^2 \ch^2 \eta} \right) \,.
\label{masseigs}
\ee
In these expressions $m_\ssZ^2 := \frac14 \left( g_1^2 + g_2^2 \right) v^2$,
\be
 \hat{c}_\ssW := \cos \hat\theta_\ssW :=
 \frac{g_2}{\sqrt{g_1^2 + g_2^2}}
 \quad \hbox{and} \quad
  \hat{s}_\ssW := \sin\hat \theta_\ssW
 := \frac{g_1}{\sqrt{g_1^2 + g_2^2}} \,,
\ee
while
\be
 \sh \, \eta := \sinh \eta := \frac{\chi}{\sqrt{1 - \chi^2}}
 \quad \hbox{and} \quad
 \ch \, \eta := \cosh \eta := \frac{1}{\sqrt{1 - \chi^2}} \,.
\ee
Finally, the quantities $r_\ssX$ and $\vartheta_\ssX$ are defined by
\be
 r_\ssX := \frac{m_\ssX}{m_\ssZ}
 \quad \hbox{and} \quad
 \vartheta_{\ssX} :=
 \left\{ \begin{array}{c} +1 \quad \hbox{if} \quad  r_{\ssX} > 1 \\
 -1 \quad\hbox{if} \quad r_{\ssX} < 1 \end{array} \right. \,,
\ee
which ensures $M_\ssZ \rightarrow m_\ssZ$ and $M_\ssX \rightarrow m_\ssX$ as $\eta \rightarrow 0$.

The currents in the physical basis are similarly read off as
\begin{equation}
 {\bf J}_\mu
 :=
 \begin{bmatrix}
 J_\mu^\ssZ \\
 J_\mu^\ssA \\
 J_\mu^\ssX
 \end{bmatrix} =
 \begin{bmatrix}
 \check J_\mu^\ssZ c_\xi +
 \left( -\check J_\mu^\ssZ \hat s_\ssW \sh \, \eta
 + \check J_\mu^\ssA \hat c_\ssW \sh \, \eta
 + \check J_\mu^\ssX \ch \, \eta \right) s_\xi \\
 \check J_\mu^\ssA \\
 - \check J_\mu^\ssZ s_\xi
 + \left( - \check J_\mu^\ssZ \hat s_\ssW \sh \, \eta
 + \check J_\mu^\ssA \hat c_\ssW \sh \, \eta
 + \check J_\mu^\ssX \ch \, \eta \right) c_\xi
 \end{bmatrix}  \,,
\end{equation}
where
\be
  \begin{bmatrix}
 \check J_\mu^\ssZ \\
 \check J_\mu^\ssA \\
 \check J_\mu^\ssX
 \end{bmatrix}  :=
 \begin{bmatrix}
 \hat J_\mu^3 \,\hat c_\ssW - \hat J_\mu^\ssY \,\hat s_\ssW \\
 \hat J_\mu^3 \,\hat s_\ssW + \hat J_\mu^\ssY \,\hat c_\ssW \\
 \hat{J}_\mu^\ssX
 \end{bmatrix}
 = \sum_f
 \begin{bmatrix}
 i \hat e_\ssZ \ol{f} \gamma_\mu \left[ T_{3f} \gamma_\ssL
 - Q_f \hat s_\ssW^2 \right] f \\
 ie \, \ol{f} \gamma_\mu Q_f f \\
 ig_\ssX \ol{f} \gamma_\mu \left[ X_{f\ssL} \gamma_\ssL
 + X_{f \ssR} \gamma_\ssR \right] f
 \end{bmatrix} \,, \notag
\ee
and $e := g_2 \hat s_\ssW = g_1 \hat c_\ssW$, $\hat e_\ssZ := e/(\hat s_\ssW \hat c_\ssW)$ and $Q_{f} = T_{3f} + Y_{f\ssL} = Y_{f\ssR}$. Finally, $c_\xi := \cos \xi$ and $s_\xi := \sin \xi$ with the angle $\xi $ given by
\begin{equation}
 \tan 2\xi =\frac{-2 \hat s_\ssW \sh \eta }{1-\hat s_\ssW^2 \sh^2 \eta-r_\ssX^2 \ch^2\eta} \,.
 \label{tan2alpha}
\end{equation}

Writing the resulting lagrangian as
\begin{equation}
 \cL = \cL_\SM + \delta \cL_\SM +
 \cL_\ssX \,, \label{Leff1}
\end{equation}
shows that the $X$ boson has two kinds of physical implications: $(i)$ direct new couplings between the $X$ boson and SM particles; $(ii)$ modifications (due to mixing) of the couplings among the SM particles themselves.

\subsubsection*{Modification of SM couplings}

The modification to the SM self-couplings caused by $Z-X$ mixing are given by
\be \label{Leff2}
 \delta \cL_\SM = - \frac{z}{2} \, m_\ssZ^2 Z_\mu Z^\mu
 + i \hat e_\ssZ \sum_f \left[ \ol{f} \gamma^\mu
 \left( \delta g_{f\ssL} \gamma_{\ssL} + \delta g_{f \ssR}
 \gamma_{\ssR}\right) f \right] Z_\mu \,,
\ee
with \cite{bigfit} $z := (M_\ssZ^2 - m_\ssZ^2)/m_\ssZ^2$ and
\be
 \delta g_{f\ssL(\ssR)} = \left(c_\xi-1\right) \hat g_{f\ssL(\ssR)}+ s_\xi \left(\sh\,\eta \, \hat s_\ssW (Q_f \hat c_\ssW^2 -\hat g_{f\ssL(\ssR)}) + \ch \,\eta \, \frac{g_\ssX}{\hat e_\ssZ} X_{f\ssL(\ssR)} \right) \,.
\ee

The last step before comparing these expressions with observations is to eliminate the parameters $\hat s_\ssW$ and $m_\ssZ$ (the second of which enters the interactions through $r_\ssX$) from the lagrangian in favour of a physically defined weak mixing angle, $s_\ssW$, and the physical mass, $M_\ssZ$. This process reveals the physical combination of new-physics parameters that is relevant to observables, and thereby provides a derivation \cite{bigfit} of the $X$-boson contributions to the oblique electroweak parameters \cite{oblique}.

To this end define the physical weak mixing angle, $s_\ssW$, so that the Fermi constant, $G_\ssF$, measured in muon decay is given by the SM formula,
\begin{equation}
 \frac{G_\ssF}{\sqrt{2}} := \frac{e^2}{8 s_\ssW^2 c_\ssW^2 M_\ssZ^2} \,.
 \label{G_fdef}
\end{equation}
But this can be compared with the tree-level calculation of the Fermi constant obtained from $W$-exchange using the above lagrangian, giving (see Appendix)
\begin{equation}
 \hat s_\ssW^2 = s_\ssW^2 \left[
 1 + \frac{z \, c_\ssW^2}{c_\ssW^2 - s_\ssW^2}
 \right] \,,
\end{equation}
to linear order in $z$ (which we assume is small --- as is justified shortly by the phenomenological bounds).

Eliminating $\hat s_\ssW$ in favour of $s_\ssW$ in the fermionic weak interactions introduces a further shift in these couplings, leading to our final form for the neutral-current lagrangian:
\be
 \cL_{\NC} =  i e_\ssZ \sum_f \overline{f}
 \gamma^\mu \left[ \left( g_{f\ssL}^\SM + \Delta g_{f\ssL}
 \right) \gamma_\ssL + \left( g_{f\ssR}^\SM + \Delta g_{f\ssR}
  \right) \gamma_\ssR \right] f \, Z_\mu \,, \label{dLNC}
\ee
where $e_\ssZ := e/s_\ssW c_\ssW$ and
\begin{eqnarray} \label{DgfLR}
  \Delta g_{f \ssL(\ssR)} &=& - \frac{z}{2} g_{f \ssL(\ssR)}^{\SM} - z \left(\frac{s_\ssW^2 c_\ssW^2}{c_\ssW^2-s_\ssW^2}\right) Q_f
  +\delta g_{f\ssL(\ssR)} \nn\\
  &=& \frac{\alpha T}{2} \, g_{f\ssL(\ssR)}^{\SM}
  +\alpha T \left( \frac{s_\ssW^2 c_\ssW^2}{c_\ssW^2
  -s_\ssW^2} \right) Q_f + \delta g_{f\ssL(\ssR)} \,.
\end{eqnarray}
The SM couplings are (as usual) $g_{f\ssL}^\SM := T_{3f} - Q_f s_\ssW^2$ and $g_{f\ssR}^\SM := - Q_f s_\ssW^2$, while the oblique parameters \cite{oblique} $S$, $T$ and $U$ are given by
\be \label{obliqueresult0}
 \alpha S = \alpha U = 0 \,,
\ee
and
\begin{eqnarray} \label{obliqueresult}
 \alpha T = - z \,.
\end{eqnarray}

\subsubsection*{Direct $X$-boson couplings}

The terms explicitly involving the $X$ boson similarly are
\be
 \cL_\ssX = - \frac14 \, X_{\mu \nu } X^{\mu \nu } -
 \frac{M_\ssX^2}{2} \, X_\mu X^\mu
  + i \sum_f \ol{f} \gamma_\mu \left( k_{f\ssL} \gamma_{\ssL}
 + k_{f\ssR} \gamma_{\ssR} \right) f  X^\mu \,, \notag
\ee
with
\be
  k_{f\ssL(\ssR)} = c_\xi \,\ch\,\eta \, g_\ssX X_{f\ssL(\ssR)} + c_\xi \, \sh\,\eta \, \frac{e}{c_\ssW} (Q_f c_\ssW^2-g_{f\ssL(\ssR)}^\SM)-s_\xi \,e_\ssZ g_{f\ssL(\ssR)}^\SM \,. \label{DkfLR}
\ee

We are now in a position to compute how observables depend on the underlying parameters, and so bound their size. When doing so we follow \cite{bigfit} and work to linear order in the deviations, $\Delta g_{f\ssL(\ssR)}$, of the SM couplings, since we know these are observationally constrained to be small.

\section{High-energy constraints\label{e+e-annihil}}

This section considers the constraints on the $X$ boson coming from its influence on various precision electroweak observables measured at high-energy colliders. There are two main types of observables to consider: those that test the changes that $X$-boson mixing induces in SM couplings; and those sensitive to the direct couplings of the $X$ boson to SM fermions. We consider each type in turn.

We begin with two well-measured observables that are sensitive only to changes to the SM self-couplings: the $W$ boson mass, $M_\ssW$, and the $Z$-boson branching fraction into leptons, $\Gamma(Z\rightarrow \ell^+ \ell^-)$. Later subsections then consider reactions to which direct $X$ exchange can contribute, such as the cross section, $\sigma_{\rm res}(e^+ e^- \to h)$, for electron-positron annihilation into hadrons evaluated at the $Z$ resonance.

\subsubsection*{Consistency limits on accessible parameter space} %\label{smalleta}}

Since the SM is in such good agreement with experiment \cite{ewwg}, it is useful to linearize corrections to the SM parameters as we have done in the previous section. To be consistent, we limit ourselves to considering the subset of parameter space which is consistent with this linearization procedure. In practice, we require that the following two conditions of $z$ be satisfied:
\begin{enumerate}
\item $z$ must be real (see the discussion in the Appendix), which amounts to demanding that it is obtained by a physically allowed choice for the initial parameters $m_\ssX$ and $\chi$. This implies that
\be
\Delta_\ssX^2 - R_\ssX^2 s_\ssW^2 \sh^2\eta \geq 0 \,,
\ee
where $\Delta_\ssX$ is defined in eq.~\pref{AppDeltaeqn}. This simplifies to
\be
\left|\Delta_\ssX-\kappa\right| \geq \sqrt{\kappa(\kappa+1)} \label{zrealconstr}
\ee
where
\be
\kappa:=s_\ssW^2\sh^2\eta \,.
\ee
\item $z$ must be small: $z\ll1$. To quantify this statement, we assume that $z$ (or, equivalently, $\alpha T$) will be at most within $2\sigma$ from its global fit value \cite{bigfit}:
\be
|z|\leq0.014 \,.
\ee
This bound has been considered in \cite{CPS} in the context of hidden sector dark matter models.
\end{enumerate}

In figure \ref{zbound}, we show the regions in the $M_\ssX-\sh\,\eta$ parameter space that are excluded by each of these bounds. From this, we see that the first condition is dominant when $\sh\,\eta < 3\times10^{-2}$, whereas for greater values of $\sh\,\eta$, it is the second condition that is dominant.

\FIGURE[h!]{
\includegraphics[scale=1]{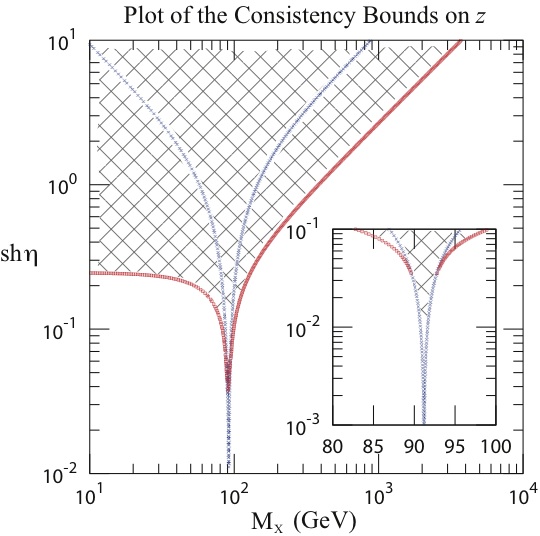}
\caption{Plot of the bounds on $z$ as a function of $M_{\ssX}$ and $\sh \, \eta$. The blue crosses limit the region in which $z$ is real, and the red squares limit the region in which $z\ll1$. The hatched regions are excluded.}
\label{zbound} }

\subsection{Effects due to modified $W,Z$ couplings}

We start with several examples of constraints that probe the induced changes to the SM self-couplings.

\subsubsection*{The $W$ mass\label{wmasssect}}

Mixing with the $X$ boson modifies the SM prediction for the $W$ mass due to its contribution to the electroweak oblique parameter $T$, as follows \cite{bigfit, oblique}:
\begin{eqnarray}
 M_\ssW^2 &=& m_\ssW^2 = m_\ssZ^2 \left( 1- \hat s_\ssW^2\right) \\
 &=& \Bigl[ M_\ssZ^2 \left( 1+\alpha T\right) \Bigr]
 \left[ 1-s_\ssW^2 \left( 1-\frac{c_\ssW^2 \alpha T }{c_\ssW^2
 -s_\ssW^2} \right) \right] \\
 &\simeq & (M_\ssW^2 )_\SM \left[
 1 + \alpha T \left( 1+\frac{s_\ssW^2}{c_\ssW^2
 -s_\ssW^2}\right) \right] \,,
\end{eqnarray}
where $(M_\ssW^2)_\SM$ is the full SM prediction, including radiative corrections: $(M_\ssW^2)_\SM = M_\ssZ^2 (1 - s_\ssW^2) +$ loops. Because both the SM radiative corrections and the oblique corrections are known to be small, we can neglect their product in the above expression.

At this point one might ask why bother examine the $W$ mass correction separately, since the $W$ mass is one of the observables included in the global fits to oblique parameters, and we have already assumed that $z$ must be small enough to ensure that the oblique parameter $T$ lies within its 2-$\sigma$ range obtained from global electroweak fits (as in Figure E.2 of \cite{ewwg}). The reason we re-examine the $W$ mass is that it leads to a slightly stronger constraint, because the mixing between the $Z$ and $X$ bosons does not contribute to the $S$ parameter, and this prior information leads to a slightly stronger limit on $T$ (as is shown in Figure \ref{STellipse}).
\FIGURE[h!]{ \includegraphics[scale=0.65]{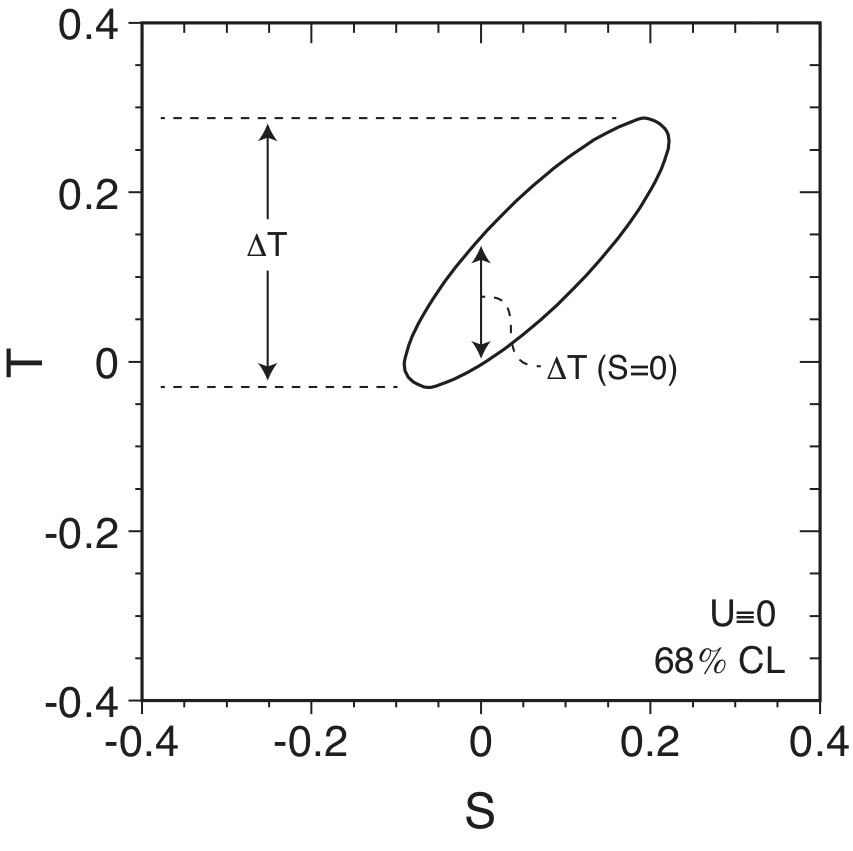}
\caption{Plot of the EWWG bound on the $S$ and $T$ oblique parameters, showing how $T$ is more tightly constrained given prior knowledge that $S=0$.}
\label{STellipse} }

Using the result, eq.~\pref{obliqueresult}, $\alpha T = - z$ together with eq.~\pref{Appzeqn} for $z$ as a function of $\eta$ and $M_\ssX$ gives the desired expression for $\Delta M_\ssW$ as a function of $\eta$ and $M_\ssX$. In the limit when the $Z$ and $X$ masses are very close to one another --- {\em i.e.} when $\Delta_\ssX$ is such that the equality in eq.~\pref{zrealconstr} holds --- the expression for $z$ becomes
\be
z = \frac{\kappa -\Delta_\ssX}{1+\Delta_\ssX} =  - \vartheta_\ssX s_\ssW |\eta| + s_\ssW^2 \eta^2 + \cO(\eta^3) \qquad \hbox{(near-degenerate $Z$ and $X$ masses)} \,,
\ee
and so
\be
 |\Delta M_\ssW| \simeq M_\ssZ c_\ssW
 \left[ \frac{c_\ssW^2}{c_\ssW^2-s_\ssW^2}
 \left( \frac{s_\ssW |\eta|}{2} \right)
 \right] \simeq
 \,2.75  ~\text{GeV} \left( \frac{\eta}{0.1}
 \right) \,.  \label{deltamw1}
\ee

Moving away from degeneracy, we find the expression for $z$ can be simplified as follows:
\be
 z = \frac{\kappa - \Delta_\ssX\left(1-\sqrt{1 - \frac{\kappa R_\ssX^2}{\Delta_\ssX^2}}\right)}
 {1 + \Delta_\ssX\left(1-\sqrt{1 - \frac{\kappa R_\ssX^2}{\Delta_\ssX^2}}\right)}
 = \kappa-\frac{\kappa R_\ssX^2}{2\Delta_\ssX} + \cO(\kappa^2)
 = \frac{s_\ssW^2 \eta^2}{1-R_\ssX^2} + \cO(\eta^4) \,,
\ee
where $R_\ssX := M_\ssX/M_\ssZ$ ({\em c.f.} eq.~\pref{AppRXeqn}). So when $M_\ssX$ and $M_\ssZ$ are very different,
\be
 \Delta M_\ssW \simeq  \frac{s_\ssW^2  c_\ssW^3}{2(c_\ssW^2-s_\ssW^2)}
 \left( \frac{\eta^2 M_\ssZ^3}{M_\ssX^2 - M_\ssZ^2}
 \right) \simeq 1.10\times 10^{5}\left( \frac{\eta^2}{M_\ssX^2
 -M_\ssZ^2}\right) ~\text{GeV}^{3} \,. \label{deltamw}
\ee
The large-$M_\ssX$ limit of eq.~\pref{deltamw} agrees with the result given in \cite{wells}, which finds
\begin{equation}
 \Delta M_\ssW \simeq \left( 17\ \text{MeV}\right)
 \left( \frac{\eta }{0.1}\right)^2
 \left( \frac{250\text{ GeV}}{M_{\ssX}} \right)^2 \,.
\end{equation}

\FIGURE[h!]{ \includegraphics[scale=1]{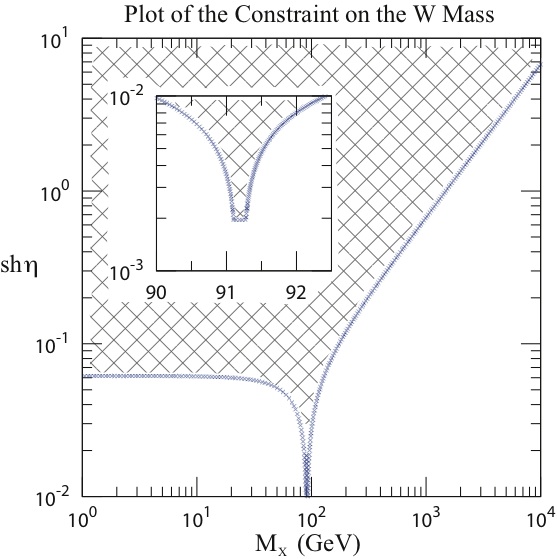}
\caption{Constraint obtained from limiting the influence of kinetic mixing
on the SM value of the $W$ mass. The hatched regions are excluded.}
\label{wmass} }

The experimental agreement of the measured $W$ mass with the SM prediction implies $\Delta M_\ssW \leq 0.05$ GeV \cite{pdg} ($2\sigma$ uncertainty), and the constraint this imposes on $\sh\,\eta$ as a function of $M_\ssX$ is shown in
Figure \ref{wmass}. Several points about the comparison given in the figure are of note:
\begin{itemize}
 \item The $W$-mass bound on $\eta$ is model-independent inasmuch as it relies only on the kinetic mixing and does not depend at all on the fermion quantum numbers to which $X$ couples;

\item The strongest constraints on $\eta$ occur for $M_\ssX$ nearest to the $Z$ pole, where $|\eta_{\rm pole}| \leq 1.8 \times 10^{-3}$;

\item When $M_\ssX \ll M_\ssZ,$ the bound on $\eta$ becomes approximately $M_\ssX$-independent: $|\eta| \leq 6.2\times 10^{-2}$. This behaviour is also visible in the analytic expression, eq.~\pref{deltamw};

\item When $M_\ssX \gg M_\ssZ,$ the $W$ mass bounds the ratio $\eta/M_\ssX$: giving $M_\ssX/\eta \gsim 1.5$ TeV.
\end{itemize}

\FIGURE[h!]{
\begin{tabular}{cc}
\includegraphics[scale=0.69]{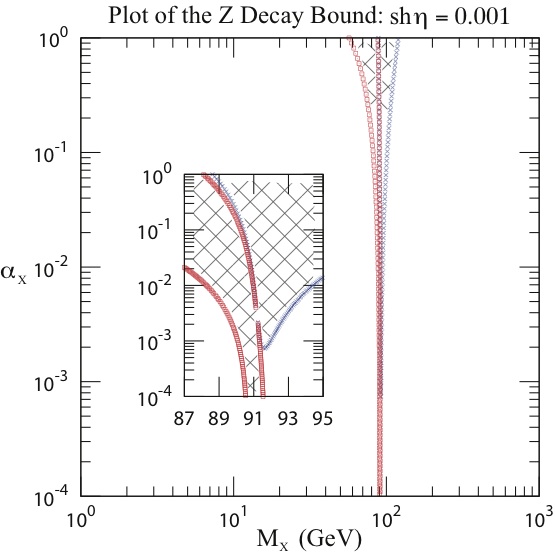} &
\includegraphics[scale=0.69]{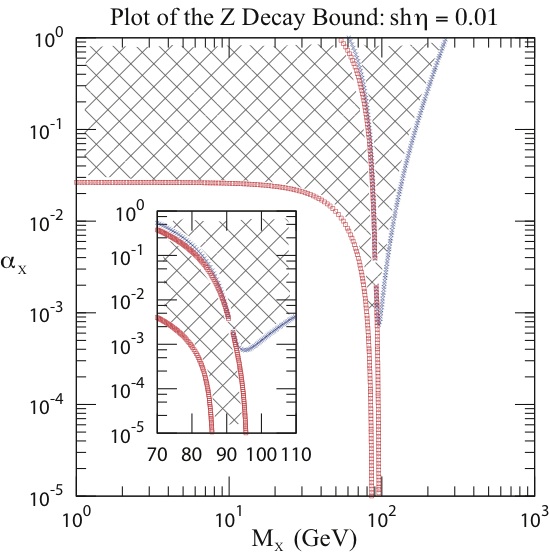} \\
\includegraphics[scale=0.69]{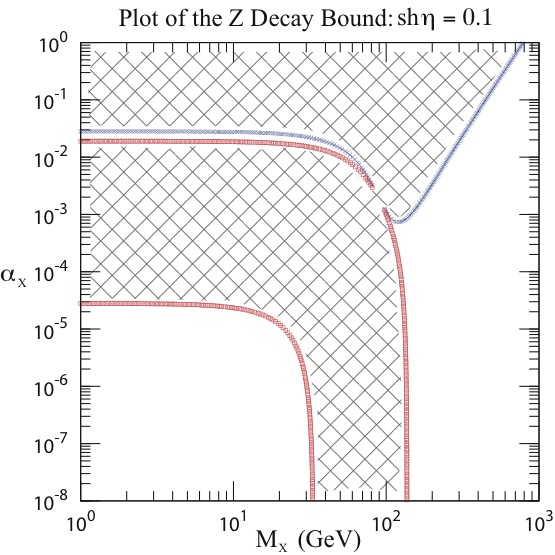} &
\includegraphics[scale=0.69]{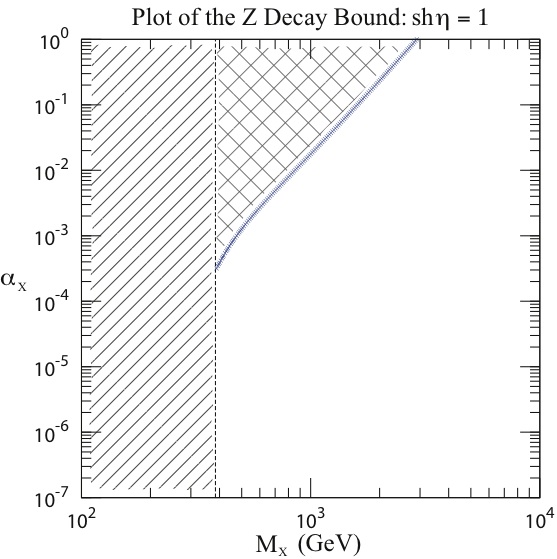} \\
\end{tabular}
\caption{The constraint arising from $Z \to \ell^+ \ell^-$ decay on the coupling $\alpha_\ssX = g^2_\ssX/4\pi$ as a function of $M_\ssX$, for various values of $\sh\,\eta$. The parameters agreeing with the positive bound ($\Delta \Gamma = + \Delta \Gamma_{\exp})$ are marked with blue crosses, while those agreeing with the negative bound ($\Delta \Gamma =-\Delta \Gamma_{\exp})$ are marked with red squares. The plot assumes a coupling $X_{\ell \ssL} = X_{\ell \ssR} = -1$, such as would be true if $X = B-L$. Hatched regions are excluded.}
\label{zdecay}
}

\subsubsection*{$Z$ Decay}

The $Z$ decay rate has been measured with great accuracy at LEP and SLC (for
details regarding their analysis, see \cite{ewwg}). The experimental value \cite{pdg} for the decay $Z\rightarrow \ell^+ \ell^-$, where $\ell$ can be any of the charged leptons, is $\Gamma_{\ell^+ \ell^-} = 83.984 \pm 0.086$ MeV (1 $\sigma$), and agrees well with the SM result \cite{pdg} $83.988\pm 0.016$ MeV. The modified $Z$-fermion couplings change the tree-level decay rate,
\begin{equation}
 \Gamma_{\ell^+ \ell^-} = \frac{M_\ssZ e_\ssZ^2}{24\pi }
 \left( g_{\ell \ssL}^2 + g_{\ell \ssR}^2 \right) \,,
\end{equation}
where the couplings $g_{\ell \ssI} = g_{\ell \ssI}^\SM + \Delta g_{\ell \ssI}$ (with $I = L, R$) are defined by the interaction \pref{dLNC}. The deviation from the SM prediction therefore is
\begin{equation}
 \Delta \Gamma_{\ell^+ \ell^-} :=
 \Gamma_{\ell^+ \ell^-} - \Gamma_{\ell^+ \ell^-}^\SM
 \simeq \frac{M_\ssZ e_\ssZ^2}{24\pi} \sum_{\ssI = \ssL,\ssR}
 \left[ 2 g_{\ell \ssI}^\SM + \Delta g_{\ell \ssI} \right]
  \Delta g_{\ell \ssI} \,.
\end{equation}
Notice that this vanishes if $\Delta g_{\ell \ssI} = 0$ or when $\Delta g_{\ell \ssI} = - 2 g_{\ell \ssI}^\SM$. It can therefore happen that $\Delta \Gamma_{\ell^+ \ell^-}$ vanishes for two separate regions as one varies through parameter space.

\FIGURE[h!]{ \includegraphics[scale=1]{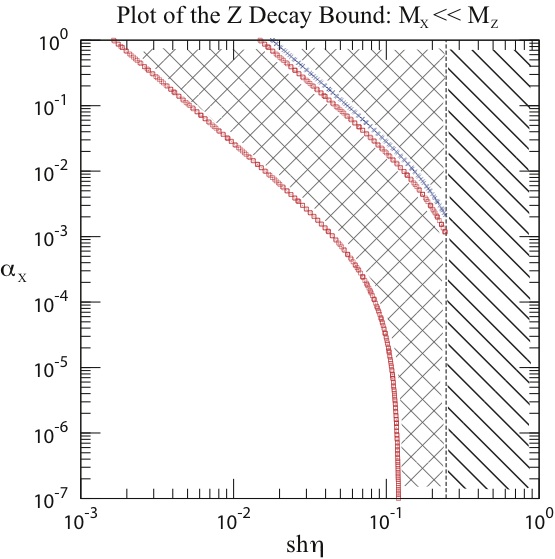}
\caption{Plot of the constraint arising from considering $Z$ decay into
leptons in the limit where $M_\ssX\ll M_\ssZ$. The upper bound ($\Delta \Gamma =+\Delta \Gamma_{\exp})$ is marked with blue crosses; the lower bound ($\Delta \Gamma =-\Delta \Gamma _{\exp})$ is marked with red squares. Hatched regions are excluded.}
\label{zdecaylowmcut} }

To obtain bounds on $\eta$ and $M_\ssX$ we use eq.~\pref{DgfLR} to eliminate $\Delta g_{f\ssL(\ssR)}$, giving
\be \label{Dgell}
 g_{\ell \ssI} = \left(c_\xi-\frac{z}{2}\right)\left(-\frac12 \, \delta _{\ssI\ssL} + s_\ssW^2\right) +\frac{z c_\ssW^2 s_\ssW^2}{c_\ssW^2-s_\ssW^2}  -s_\xi\left[\left(-\frac12\, \delta _{\ssI\ssL}+1\right)
 s_\ssW \sh\,\eta - X_{\ell \ssI} \frac{g_\ssX}{e_\ssZ} \ch\,\eta\right]
\ee
Here $-\frac12 \, \delta_{\ssI \ssL} + s_\ssW^2$ is the SM contribution, $g_{\ell \ssI}^\SM$, where $\delta_{\ssI \ssL}$ denotes a Kronecker delta function. Requiring $\Delta \Gamma_{\ell^+ \ell^-}$ to be smaller than the experimental (2 $\sigma$) experimental error gives the desired bound on the parameters $g_\ssX$, $\eta$ and $M_\ssX$. Figure \ref{zdecay} shows the excluded values in the $\alpha_\ssX = g_\ssX^2/4\pi$ vs $M_\ssX$ plane, with the leptonic $X$-boson charge assumed to be $X_{\ell \ssL} = X_{\ell \ssR} = -1$ (such as would apply if $X = B-L$). Each panel of the figure corresponds to a different choice for $\sh\,\eta$. For the panel in which $\sh\,\eta=1$ bounds at lower mass scales than roughly $385$ GeV are not plotted, since these would conflict with a $z=-\alpha T$ satisfying the global electroweak fit, as outlined in figure \ref{zbound}.

In order to understand the features present in the plots it is useful to consider the small-$\eta$ limit of $z$ and $\xi$. As discussed above for the $W$ mass bound, the small-$\eta$ limit when $M_\ssX$ and $M_\ssZ$ are very similar or very different must be considered separately. The expressions when $M_\ssX$ and $M_\ssZ$ are very different are
\be
z \simeq \frac{s_\ssW^2\eta^2}{1-R_\ssX^2} \quad \hbox{and} \quad
 \xi \simeq \frac{s_\ssW \eta}{R_\ssX^2-1} \,.
\ee
As might be expected, all terms in $\Delta g_{\ell \ssI}$ are suppressed by a factor of $1/R_\ssX^2 \simeq M_\ssZ^2/M_\ssX^2$ and so go to zero when $M_\ssX \gg M_\ssZ$. In the opposite limit, $R_\ssX \to 0$, $\Delta g_{\ell \ssI} \simeq X_{\ell \ssI} \eta s_\ssW (g_\ssX/e_\ssZ) + \eta^2 s_\ssW^2 c_\ssW^4/(c_\ssW^2 - s_\ssW^2)$, which can pass through zero (if $X_{\ell \ssI}\eta < 0$) when $|X_{\ell \ssI}| (g_\ssX/e_\ssZ) \simeq \cO(\eta)$.

Several features of these plots should be highlighted:
\begin{itemize}
 \item The best bounds come for $M_\ssX \simeq M_\ssZ$, even for small couplings $g_\ssX$, because in this limit the $Z-X$ mixing parameter $\xi$ becomes maximal ($\tan 2 \xi \, \raro \, \infty$), leading to strong constraints.
\item For a similar reason, once $\eta$ is sufficiently large ($\sh\,\eta \simeq 0.1$ --- see also figure \ref{zdecaylowmcut}) the regime of vanishingly small $\alpha_\ssX$ remains excluded because $\Delta g_{\ell \ssL(\ssR)}$ is dominated by the oblique corrections to the weak mixing angle.
 \item For $M_\ssX \gg M_\ssZ$ the excluded area approaches a straight line, corresponding to a bound on the ratio $g_\ssX/M_\ssX^2$, as expected from the form of $\Delta g_{\ell \ssL(\ssR)}$.
 \item The graph is more intricate for $M_\ssX \ll M_\ssZ$, with slivers of allowed parameter space emerging for a narrow, $\eta$-dependent but $M_\ssX$-independent, value of $\alpha_\ssX$. This happens (for sufficiently large $\eta$) because $\Delta \Gamma = 0$ is a multiple-valued condition on the parameters, as discussed above.
\end{itemize}

Figure \ref{zdecaylowmcut} provides a view of the bounds taken on a different slice through the three-dimensional parameter space ($\eta$, $\alpha_\ssX$, $M_\ssX$). This figure plots the constraints on $\alpha_\ssX$ vs $\sh\,\eta$, in the regime where $M_\ssX \ll M_\ssZ$, showing how a wider range of $\alpha_\ssX$ is allowed as $\sh\,\eta$ shrinks. Note that bounds are only shown for the region where $z \ll 1$.

\subsection{Processes involving $X$-boson exchange}

In this section we consider precision electroweak observables, like the resonant cross section for $e^+ e^- \to$ hadrons, that receive direct contributions from $X$-boson exchange, in addition to the modifications to SM $Z$-boson couplings.

\subsubsection*{The annihilation cross section}

We again proceed by computing the leading change to the tree-level cross section for $e^+ e^- \to f \ol{f}$ at leading order in the new interactions. Interference terms between SM loops and $X$-boson contributions may be neglected under the assumption that their product is negligible \cite{bigfit}. The relevant Feynman diagrams are shown in Figure \ref{schannel}, where the exchanged boson is either a photon, $Z$ or $X$ boson.

\FIGURE[h!]{
\includegraphics[scale=0.7]{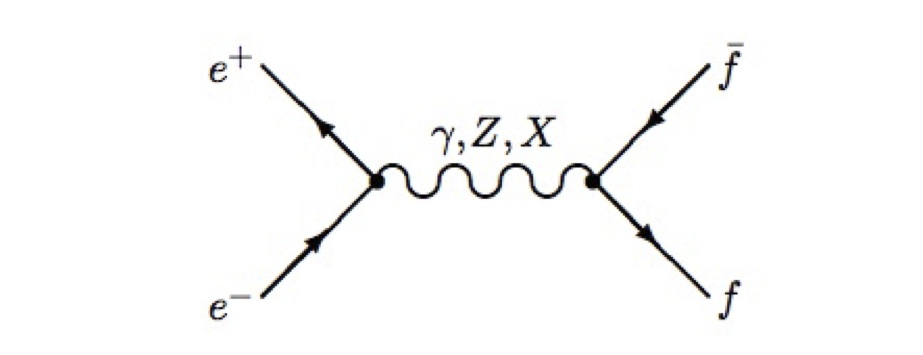}
\caption{Relevant tree-level Feynman diagrams corresponding to electron-positron annihilation into fermion-antifermion pairs.
\label{schannel}}
}

Neglecting fermion masses the relevant spin-averaged squared matrix element for this process is (see, {\em e.g.} \cite{SM} for a treatment of SM scatterings using similar conventions)
\begin{equation}
 \frac14 \sum \left\vert \mathcal{M} \right\vert^2
 = N_c \left[ \left( \left\vert A_{\ssL\ssL}
 \left( s\right) \right\vert ^2 + \left\vert A_{\ssR\ssR}
 \left( s\right) \right\vert ^2\right) u^2
 +\left( \left\vert A_{\ssL \ssR} \left(
 s\right) \right\vert ^2 + \left\vert A_{\ssR \ssL}
 \left( s\right) \right\vert^2\right) t^2\right] \,,
\end{equation}
where $s$, $t$ and $u$ are the usual Mandelstam variables and
\begin{equation}
 A_{\ssI\ssJ} \left( s\right) := \frac{e^2 Q_e Q_f }{s}
 + \frac{e_\ssZ^2 g_{e\ssI}g_{f\ssJ}}{s - M_\ssZ^2
 +i\Gamma_\ssZ M_\ssZ} + \frac{k_{e\ssI} k_{f\ssJ}}{s
 -M_\ssX^2 + i\Gamma_\ssX M_\ssX} \,.
\end{equation}
The total unpolarized cross section that follows from this is
\begin{equation}
 \sigma \left( e^+ e^- \rightarrow f \overline{f} \right)
 = \frac{N_c s}{48\pi } \left( \left\vert A_{\ssL\ssL}
 \right\vert^2 + \left\vert A_{\ssR\ssR} \right\vert^2
 +\left\vert A_{\ssL\ssR} \right\vert^2
 +\left\vert A_{\ssR\ssL} \right\vert^2\right) \,.
\end{equation}
The couplings $g_{f\ssI}$ and $k_{f\ssI}$ in these expressions are defined in terms of $\eta$, $g_\ssX$ and $M_\ssX$ by eqs.~\pref{DgfLR} and \pref{DkfLR}. The quantities $\Gamma_\ssZ$ and $\Gamma_\ssX$ are only important near resonance, and denote the full decay widths for the $Z$ and $X$ boson, respectively:
\begin{eqnarray}
 \Gamma_\ssZ &=& \frac{e_\ssZ^2 M_\ssZ}{24\pi }
 \sum_{2m_f \leq M_\ssZ}
 \left[ g_{f\ssL}^2 + g_{f\ssR}^2 \right] N_{c} \\
 \hbox{and} \quad
 \Gamma_\ssX &=& \frac{M_\ssX}{24\pi } \sum_{2m_f \leq M_\ssX}
 \left[  k_{f\ssL}^2 + k_{f\ssR}^2 \right] N_c \,,
\end{eqnarray}
where $N_c$ is the colour degeneracy for fermion $f$.

\FIGURE[h!]{
 \begin{tabular}{cc}
\includegraphics[scale=0.63]{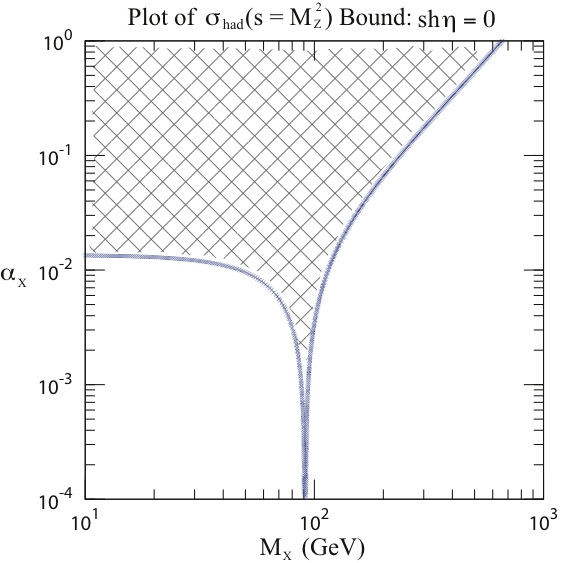} &
\includegraphics[scale=0.63]{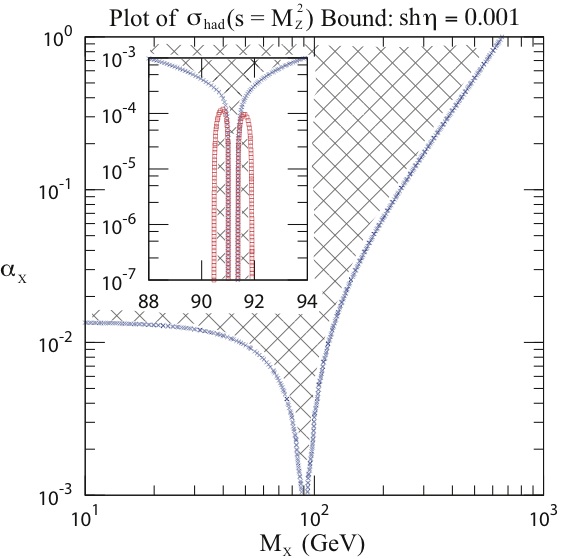} \\
\includegraphics[scale=0.63]{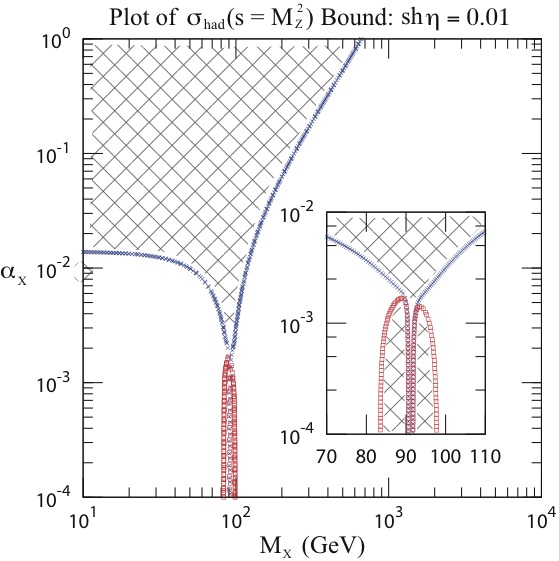} &
\includegraphics[scale=0.63]{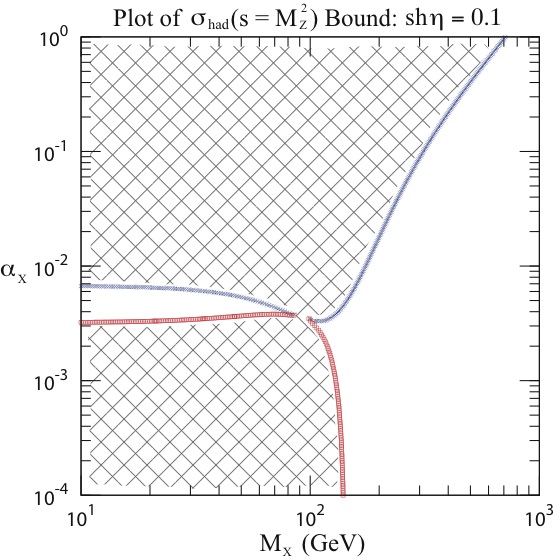} \\
 \multicolumn{2}{c}{
\includegraphics[scale=0.63]{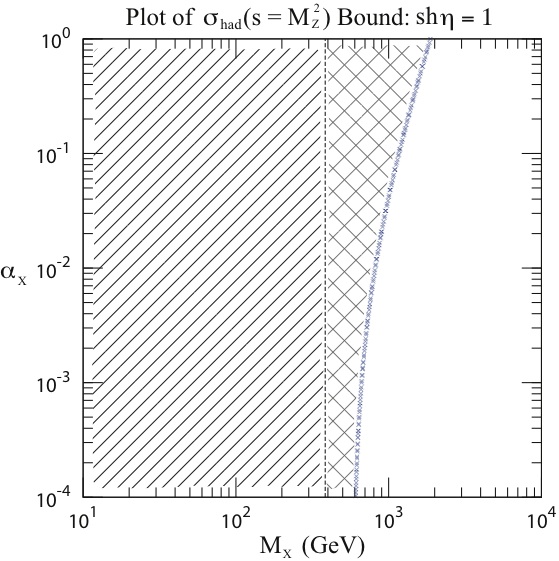}
 }
 \end{tabular}
 \caption{The constraint obtained from $\sigma_{had}$ evaluated for $\sqrt{s} = M_\ssZ$, as a bound in the $\alpha_\ssX - M_\ssX$ plane for various values of $\sh\,\eta$. Blue crosses (red squares) indicate parameters where predictions differ by 2$\,\sigma$ from experiment on the upper (lower) side. The hatched regions are excluded, while diagonal shading indicates a region excluded by global fits to oblique parameters.}
\label{sigmahad}
}

\subsubsection*{The Hadronic Cross Section at the $Z$ Pole}

Summing the above over all quarks lighter than $M_\ssZ$ and evaluating at $\sqrt{s} = M_\ssZ$ gives the leading correction to the resonant
cross section into hadrons, $\sigma_{had} \left( s = M_\ssZ^2 \right)$, which is well-measured to be $41.541 \pm 0.037$ nb \cite{pdg}. Requiring the deviation from the SM to be smaller than the 2$\sigma$ error gives the desired constraints. Figure \ref{sigmahad} shows a number of exclusion limits for the coupling $\alpha_\ssX$ vs the $X$-boson mass (for $X_{f\ssL} = X_{f\ssR} = (B - L)_f$, and $M_\ssX$ in the range of $10 - 10^3$ GeV), with each panel corresponding to a different choice for $\eta$.

These plots reflect several features seen in the analytic expressions for the couplings:

\begin{itemize}
\item For $M_\ssX \ll M_\ssZ$ and when $\eta$ is small enough, the mass dependence of the bound on $\alpha_\ssX$ completely drops out, leaving $\alpha_\ssX \lsim 10^{-2}$ in this limit. For larger $\eta$ small values of $\alpha_\ssX$ can still be ruled out because the contributions of mixing are already too large. This mixing also ensures that the region near $M_\ssX = M_\ssZ$ tends to give the strongest bounds.

\item The regime $M_\ssX \gg M_\ssZ$ similarly constrains only the combination $M_\ssX^2/\alpha_\ssX \gsim 800$ GeV (when $\eta$ is small).

\item For $\eta$ not too small and $M_\ssX$ smaller than $M_\ssZ$, figure \ref{sigmahad} shows a window of unconstrained couplings, for the same kinds of reasons discussed above for $\Gamma_{\ell^+ \ell^-}$.
\end{itemize}

\FIGURE[h!]{
\includegraphics[scale=1]{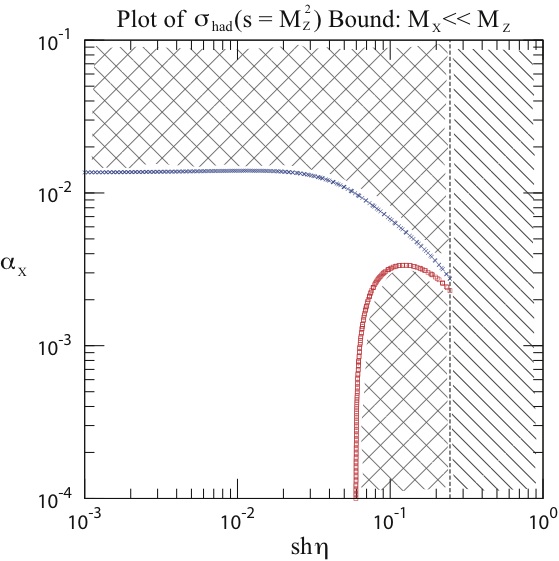}
\caption{Plot of the constraint from $\protect\sigma _{had}\left( s=M_{\ssZ}^2\right) $ in the region where $M_{\ssX}\ll M_\ssZ$. The parameters agreeing with the positive bound are marked with blue crosses, while those agreeing with the negative bound are marked with red squares. The hatched regions are excluded.}
\label{shadlowmcut}
}

Figure \ref{shadlowmcut} shows a sample slice of the constraint region in the $\alpha_\ssX$ vs $\sh\,\eta$ plane, in the limit $M_\ssX \ll M_\ssZ$. Once again, bounds are not plotted within regions of parameter space for which $z$ is not $\ll 1$. This plot shows that the smallest $\eta$ for which small $\alpha_\ssX$ can be ruled out is $\sh\,\eta \gsim 0.06$. Once $\eta$ is larger than this, mixing rules out the $X$ boson even with arbitrarily small gauge couplings.

\section{Constraints at intermediate energies}

Better constraints on lower-mass $X$ bosons can be obtained from low-energy scattering of muon neutrinos with electrons and nuclei. The purpose of this section is to quantify these bounds by identifying how the cross section depends on the parameters $g_\ssX$, $\eta$ and $M_\ssX$. We consider electron and nuclear scattering in turn.

\subsection{Neutrino-electron scattering}

The Feynman graphs relevant for $\nu_\mu e^-$ scattering are those of Fig.~\ref{schannel}, with three changes: ($i$) the gauge bosons are exchanged in the $t$-channel rather than $s$-channel; ($ii$) there is no photon-exchange graph and ($iii$) omission of right-handed neutrino polarizations.

Crossing to $t$-channel can be obtained by performing the following substitution
\begin{equation}
 \begin{array}{ccc}
 s\rightarrow t, & t\rightarrow u, & u\rightarrow s
 \end{array}
\end{equation}
among the Mandelstam variables in the invariant amplitude $\frac12 \sum \left\vert \mathcal{M} \right\vert^2$. With these replacements, the differential cross section for the process $\nu_\mu e^- \rightarrow \nu_\mu e^-$ is
\be
 \frac{\exd\sigma }{\exd t} \left( \nu_\mu  e^-
 \rightarrow \nu_\mu e^- \right) = - \frac{1}{8 \pi s^2}
 \left[ \left\vert A_{\ssL\ssL} (t) \right\vert^2
 s^2 + \left\vert A_{\ssR \ssL} (t) \right\vert^2
 (s+t)^2 \right] \,,
\ee
where
\begin{equation} \label{Aijnu}
 A_{\ssI \ssJ} (t) = e_\ssZ^2 \frac{ g_{e\ssI} g_{\nu \ssJ}}{
 t - M_\ssZ^2} + \frac{ k_{e\ssI} k_{\nu \ssJ}}{t - M_\ssX^2} \,.
\end{equation}

In the rest frame of the initial electron $s \simeq 2 \,m_e E_\nu$ and $t \simeq -2 y \,m_e E_\nu$, where $E_\nu$ is the incoming neutrino energy and $y$ is the fractional neutrino energy loss, $y := E^f_e/E_\nu$ where $E^f_e$ is the energy of the outgoing electron. (In such experiments \cite{CHARM2,radelbeyer}, $E_\nu,\,E^f_e \sim 1-10$ GeV so ratios of the form $m_e/ E_\nu$ and $m_e / E^f_e$ can be neglected.) In terms of these new variables the differential cross section is
\begin{equation} \label{nuediffcross}
 \frac{\exd\sigma }{\exd y} \left( \nu_\mu e^-
 \rightarrow \nu_\mu e^- \right) = \frac{m_e E_\nu}{4\pi}
 \left[ \Bigl\vert A_{\ssL\ssL} [t(y E_\nu)] \Bigr\vert^2
 + \Bigl\vert A_{\ssR \ssL} [t(y E_\nu)] \Bigr\vert^2 (1 - y)^2
 \right] \,.
\end{equation}

The cross section for anti-neutrino scattering is easily found from the above by interchanging $A_{\ssL \ssL} \leftrightarrow A_{\ssR \ssL}$.

\subsubsection*{Special case: Low-energy limit with $\eta = 0$}

One case of practical interest is when the boson masses, $M_\ssZ$ and $M_\ssX$, are much greater than the invariant energy exchange in the process of interest ({\em i.e.} $\sqrt{|t|} \ll M_\ssX, M_\ssZ)$. When this holds the amplitudes, $A_{\ssI \ssL}$, can be simplified to
\begin{eqnarray}
 A_{\ssI \ssL} &\simeq & - \frac{e_\ssZ^2 g_{e\ssI}
 g_{\nu \ssL}}{M_\ssZ^2} - \frac{k_{e \ssI}
 k_{\nu  \ssL} }{M_\ssX^2} \notag \\
 &=& - \frac{e_\ssZ^2 g_{\nu \ssL}}{M_\ssZ^2}
 \left[ g_{e \ssI} + \frac{M_\ssZ^2}{M_\ssX^2} \left(
 \frac{k_{\nu \ssL} k_{e \ssI}}{e^2_\ssZ g_{\nu \ssL}}
 \right) \right]  \,, \label{Aneuij}
\end{eqnarray}
allowing the effects of $X$-boson exchange be interpreted as an effective shift in the electron's electroweak couplings. For $E_\nu \simeq 1$ GeV and $y$ order unity this approximation remains good down to $M_\ssX \simeq 30$ MeV.

The resulting cross section is particularly simple in the case of no kinetic mixing, for which we can substitute the SM values $g_{e\ssI} = - \frac12 \delta_{\ssI\ssL} + s_\ssW^2$ and $g_{\nu \ssL} = \frac12$ and the $X$-boson couplings $k_{e\ssI} = g_\ssX X_{e \ssI}$ and $k_{\nu \ssJ} = g_\ssX X_{\nu \ssJ}$, and obtain
\begin{equation}
 A_{\ssI \ssL} \simeq -2 \sqrt{2} \,G_{\ssF}
 \left( -\frac12 \, \delta_{\ssI \ssL} + s_\ssW^2
 +\frac{g_\ssX^2 X_{e\ssI} X_{\nu \ssL}}{2\sqrt{2} \,
 G_{\ssF} M_\ssX^2} \right) \,,
\end{equation}
using the SM result $2\sqrt{2} \,G_{\ssF} \simeq e_\ssZ^2/2M_\ssZ^2$. We see that the $X$-boson contribution can be regarded as an additional contribution to $s_\ssW^2$ in this limit. This is convenient because it allows the simple use of constraints on $s_\ssW^2$ to directly constrain the ratio $g_\ssX^2/M_\ssX^2$.

The bounds are usually taken from the following ratio \cite{radelbeyer} of total cross sections,
\be
 R := \frac{\sigma(\nu_\mu e^- \to \nu_\mu e^-)}{
 \sigma (\ol\nu_\mu e^- \to \ol\nu_\mu e^-)} \,.
\ee
Given the differential cross section
\begin{equation}
 \frac{\exd\sigma }{\exd y} \left( \nu_\mu e^-
 \rightarrow \nu_\mu e^- \right) =\frac{2G_\ssF^2
 m_e E_\nu}{\pi } \left[ g_{e\ssL}^2 + g_{e\ssR}^2
 (1 - y)^2\right] \,,
\label{dsigneuelect}
\end{equation}
the total cross section becomes
\begin{equation}
 \sigma \left( \nu_\mu e^- \rightarrow
 \nu_\mu e^- \right) = \frac{2G_\ssF^2 m_e
 E_\nu}{\pi } \left( g_{e\ssL}^2 +\frac{g_{e\ssR}^2}{3}
 \right) \,,
\end{equation}
and so
\begin{equation}
 \sigma \left( \overline\nu_\mu e^-
 \rightarrow \overline\nu_\mu e^- \right)
 = \frac{2G_\ssF^2 m_e E_\nu }{\pi }
 \left(  \frac{g_{e\ssL}^2}{3} + g_{e\ssR}^2 \right) \,.
\end{equation}
Specializing to SM couplings the result depends only on $s_\ssW$:
\begin{equation}
 R = \frac{3 g_{e\ssL}^2 + g_{e\ssR}^2}{g_{e\ssL}^2
 +3g_{e \ssR}^2} = \frac{3 - 12 s_\ssW^2
 + 16s_\ssW^4}{1-4s_\ssW^2+16s_\ssW^4}
 = \frac{1+\kappa +\kappa ^2}{1-\kappa +\kappa ^2} \,,
\end{equation}
where $\kappa \equiv 1-4s_\ssW^2 \ll 1$.

Using the experimental limit \cite{CHARM2} $\Delta s_\ssW^2 = 0.0166$ ($2\sigma$ error) with $G_{\ssF} = 1.1664\times 10^{-5} \text{ GeV}^{-2}$ \cite{pdg} to constrain $\Delta s_\ssW^2 = {g_\ssX^2}/{2 \sqrt{2} \, G_{\ssF} M_\ssX^2}$ (assuming the choice $X_{e\ssI} X_{\nu \ssL} = 1$, as would be true for $X = B-L$ for example), gives \cite{gjneuelect}
\begin{equation}
\frac{M_\ssX}{g_\ssX}\gtrsim 4\ \text{TeV} \,.
\end{equation}

\FIGURE[h!]{
\includegraphics[scale=1]{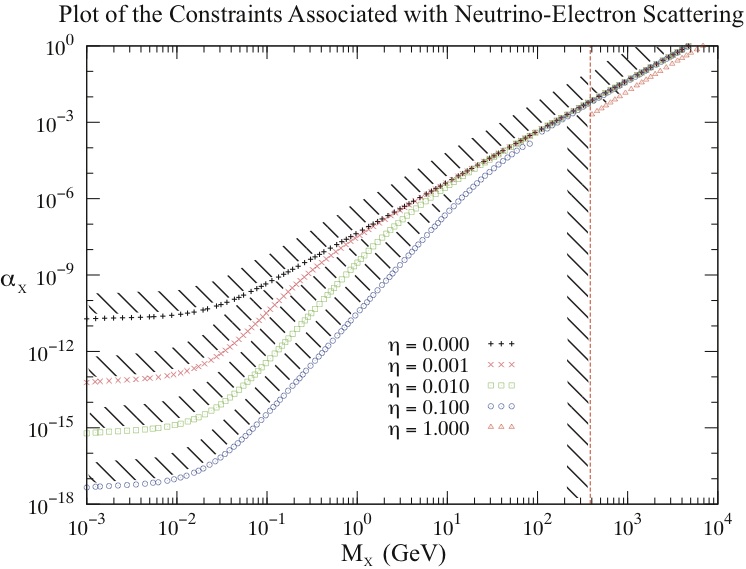}
\caption{Bound obtained on $\alpha_\ssX = g_\ssX^2/4\pi$ by limiting the influence of the $X$ boson on the $\nu-e^-$ cross section ratio $R$, obtained as a function of $M_\ssX$ for various values of $\sh\,\eta$. The vertical line indicates the region ruled out by electroweak oblique fits when $\eta = 1$.}
\label{nuelect}
}

\subsubsection*{General Case: $\protect\eta \neq 0$}

\medskip\noindent
More generally, the couplings $k_{f \ssI}$ also acquire contributions from $Z-X$ mixing even when $g_\ssX = 0$, as the above calculations show. In this case the more general bounds on $g_\ssX$, $\eta$ and $M_\ssX$ can be extracted by demanding that these contribute within the experimental limit $\Delta R$. Since the experimental limit is often quoted in terms of $s_\ssW^2$ \cite{CHARM2}, we translate using $\Delta R = \left\vert {\exd R_{SM}}/ {\exd s_\ssW^2 } \right\vert \Delta s_\ssW^2$. In obtaining $R$, we integrate over $y$ using $\sqrt{|t|}\ll M_\ssZ^2$, but {\em without} assuming that $\sqrt{|t|} \ll M_\ssX^2$. When evaluating $R$, we set $E_\nu$ to a nominal value of 1 GeV. Figure \ref{nuelect} shows the resulting bound in the $\alpha_\ssX - M_\ssX$ plane, for several choices for $\eta$ assuming $X_{e \ssI} X_{\nu \ssL} = 1$.

The resulting curves inspire a few comments:

\begin{itemize}

 \item For large $M_\ssX$ the bound is independent of $\eta$ due to the $M_\ssX / M_\ssZ$ suppression of the mixing in $\Delta g$ and $\Delta k$. This allows the direct $g_\ssX^2/M_\ssX^2$ term to dominate. The bounds in this regime are relatively strong, and compete with those found in direct searches (e.g., by CDF \cite{CDF} in the case of a SM-like $Z'$).
 \item For smaller $M_\ssX$, it is the terms in the couplings that are linear in $\eta$ that influence the deviation from the $\eta=0$ result. To see this, note that
     \be
     \Delta g_{\nu \ssL}=-\eta \, \frac{s_\ssW}{e_\ssZ} g_\ssX X_{\nu \ssL}+ \cO(\eta^2) \\,
     \ee
     and so there will be a term in $|A_{\ssI \ssL}|^2$ that is linear in $\eta$ with the parametric dependence $g_\ssX/M_\ssX^2$. When $g_\ssX \ll 0$, it is this term that is dominant compared to the $g_\ssX^2/M_\ssX^2$ term from $X$-boson exchange. However, when $g_\ssX\sim\eta$, this new term is no longer dominant and the bound regresses back to its original slope from the $\eta=0$ case at high masses.
 \item Once $1\leq M_\ssX \leq 10$ MeV and $M_\ssX \lsim \sqrt{|t|}$, the bound loses its dependence on $M_\ssX$ and levels out to some fixed value. This is expected from the form of eq.~\pref{Aijnu}.
 \item When $\sh \, \eta =1$, much of the parameter space is excluded due to the requirement that $z \ll 1$. Therefore, only a small region with $M_\ssX > 385$ GeV is bounded by electron-neutrino scattering in this case.
\end{itemize}

\subsection{Neutrino-nucleon scattering}

For the bounds from neutrino-nucleon scattering it is worth first recalling how the standard analysis is performed. In terms of the neutral-current quark couplings, the quark-level cross sections for neutral-current muon-neutrino scattering are
\begin{eqnarray} \label{neuxsects1}
 \sigma \left( \nu_\mu u \rightarrow \nu_\mu u\right)
 &=& \sigma_0 \left( g_{u\ssL}^2 + \frac{g_{u\ssR}^2}{3} \right)
 \,, \qquad
 \sigma \left( \nu_\mu d \rightarrow \nu_\mu d\right)
 = \sigma_0 \left( g_{d\ssL}^2 + \frac{g_{d\ssR}^2}{3} \right)  \\
 \sigma \left( \ol\nu_\mu u \rightarrow \ol\nu_\mu u \right)
 &=& \sigma_0 \left( \frac{g_{u\ssL}^2}{3} + g_{u \ssR}^2 \right)
 \,, \qquad
 \sigma \left( \ol\nu_\mu d \rightarrow \ol\nu_\mu d \right)
 = \sigma_0 \left( \frac{g_{d\ssL}^2}{3} + g_{d \ssR}^2 \right) \nn
\end{eqnarray}
while those for charged currents are
\be \label{neuxsects2}
 \sigma \left( \nu_\mu d \rightarrow \mu^- u\right)
 = \sigma_0 \quad \hbox{and} \quad
 \sigma \left( \ol\nu_\mu u \rightarrow \mu^+ d\right)
 =\frac{\sigma_0}{3} \,,
\ee
where $\sigma_0 := 2 N_c G_{\ssF}^2 m_e E_\nu/\pi$ and $N_{c}=3$.

These show that the quark neutral-current and charged-current cross sections are all proportional to one another. The resulting cross section for neutrino-nucleon scattering in the deep-inelastic limit is obtained by summing incoherently over the quark contributions, giving
\begin{eqnarray}
 \sigma \left( \nu_\mu N \rightarrow \nu_\mu X\right) &=&
 \varepsilon_\ssL^2 \, \sigma \left( \nu_\mu N \rightarrow
 \mu^- X\right) + \varepsilon_\ssR^2 \, \sigma \left( \ol\nu_\mu
 N\rightarrow \mu^+ X \right) \nn\\
 \sigma \left( \ol\nu_\mu N \rightarrow \ol\nu_\mu X \right)
 &=& \varepsilon_\ssL^2 \sigma \left( \ol\nu_\mu N
 \rightarrow \mu^+ X \right) + \varepsilon_\ssR^2
 \sigma \left( \nu_\mu N \rightarrow \mu^- X \right) \,,
\end{eqnarray}
where
\be
 \varepsilon_{\ssL \left( \ssR \right) }^2 :=
 g_{u \ssL \left( \ssR \right) }^2 +
 g_{d\ssL\left( \ssR\right) }^2 \,.
\ee

\FIGURE[ht]{
\begin{tabular}{cc}
\includegraphics[scale=0.68]{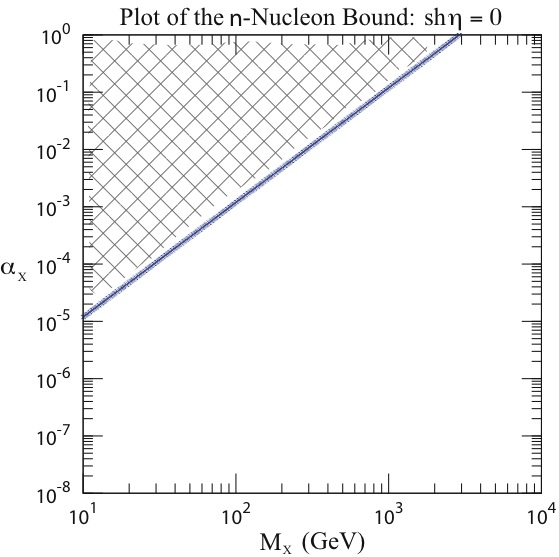} &
\includegraphics[scale=0.68]{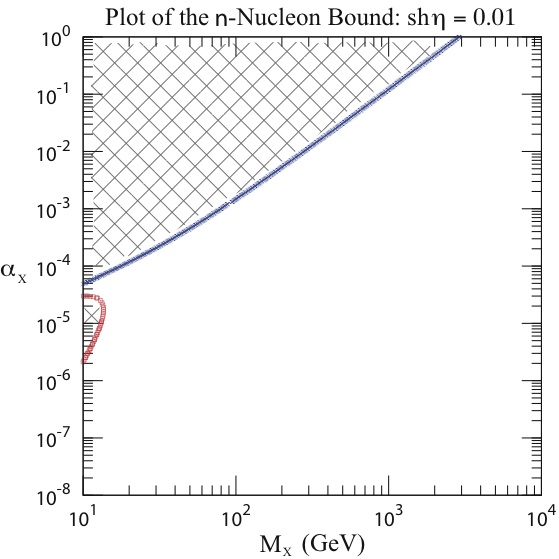} \\
\includegraphics[scale=0.68]{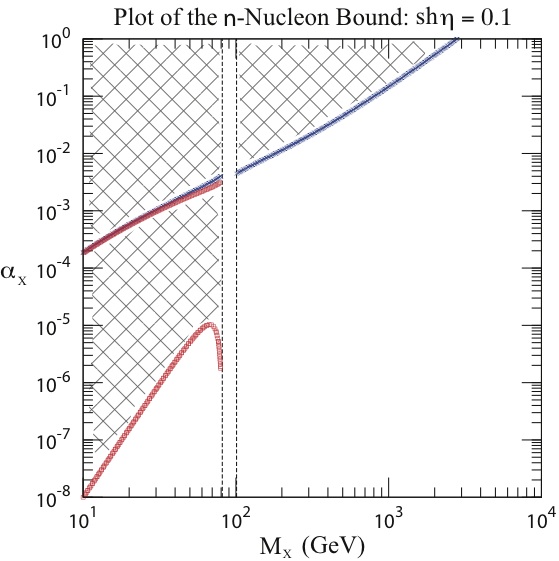} &
\includegraphics[scale=0.68]{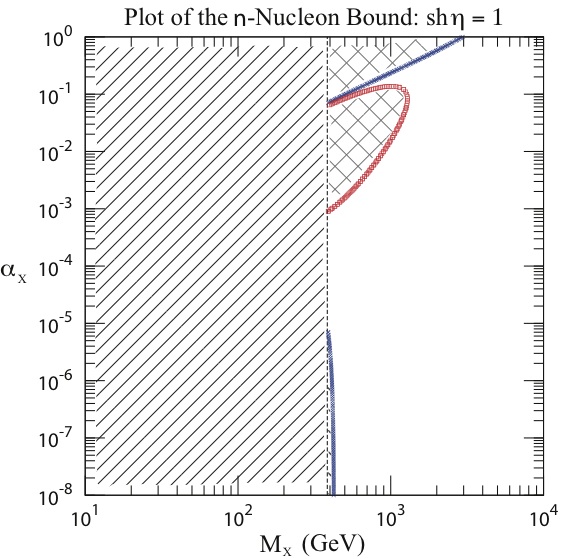} \\
\end{tabular}
\caption{Plot of the constraint from $R^{-}$ (neutrino-nucleon scattering) assuming $X = B-L$. Here, we plot the bound on $\protect\alpha_\ssX$ as a function of $M_\ssX$ for various values of $\protect\eta$. Blue squares (red crosses) indicate parameters whose predictions lie $2\,\sigma$ above (below) the central experimental value. The vertical line indicates the region excluded by precision oblique fits.}
\label{nuhad}
}

The experimental bounds come from the following ratios:
\begin{eqnarray}
 R^{\nu } &:= & \frac{\sigma \left( \nu_{\mu } N
 \rightarrow \nu_{\mu} X\right) }{\sigma \left( \nu_{\mu }
 N\rightarrow \mu ^{-} X\right) }
 = \varepsilon_\ssL^2 + r \, \varepsilon_\ssR^2 \nn\\
 R^{\bar{\nu}} &:=& \frac{\sigma \left( \ol{\nu}_{\mu }
 N\rightarrow \ol{\nu}_{\mu } X\right) }{\sigma
 \left( \ol{\nu}_{\mu } N \rightarrow \mu^{+} X\right) }
 =\varepsilon_{\ssL}^2 + \frac{\varepsilon_{\ssR}^2}{r}
\end{eqnarray}
where $r := \sigma \left( \bar{\nu}_{\mu }N \rightarrow \mu^{+} X\right) /\sigma \left( \nu_{\mu } N\rightarrow \mu^{-}X\right)$. Most useful is the Paschos-Wolfenstein ratio \cite{pw}, from which the comparatively uncertain ratio $r$ cancels:
\be \label{PWratio}
 R^{-} := \frac{R^{\nu } - rR^{\bar{\nu}}}{1-r}=\frac{\sigma
 \left( \nu _{\mu }N\rightarrow \nu _{\mu }X\right) -\sigma \left(
 \bar{\nu}_{\mu }N\rightarrow \bar{\nu}_{\mu }X\right) }{\sigma
 \left( \nu _{\mu }N\rightarrow \mu ^{-}X\right) -\sigma \left(
 \bar{\nu}_{\mu }N\rightarrow
 \mu ^{+}X\right) }
 = \varepsilon_{\ssL}^2-\varepsilon_{\ssR}^2 \,.
\ee
Experiments measure the following values \cite{nutev}
\begin{eqnarray} \label{nutevbound}
 \varepsilon _{\ssL}^2 &=&0.30005\pm 0.00137 \\
 \varepsilon _{\ssR}^2 &=&0.03076\pm 0.00110 \,.\nn
\end{eqnarray}

To constrain the $X$-boson coupling parameters we work in the regime with $\sqrt{-t} \ll M_\ssX$, for which the effects of $X$-boson mixing and exchange can both be rolled into a set of effective neutral-current couplings. The cross sections for quark-level scattering are then given by integrating eqs.~\pref{nuediffcross} using \pref{Aneuij}, leading to expressions identical with eqs.~\pref{neuxsects1} but with
\be
 g_{q \ssI} \to  g_{q \ssI}^{\rm eff} :=
 2 \left[ g_{q\ssI} g_{\nu \ssL} + \left( \frac{M_\ssZ^2}{M_\ssX^2}
 \right) \frac{k_{q \ssI} k_{\nu \ssL}}{e_\ssZ^2} \right] \,,
\ee
where $q = u, d$ and $I = L, R$. Using these in eq.~\pref{PWratio} gives constraints on $\varepsilon_{\ssI}^2 = \left( g_{u\ssI}^{\SM} + \Delta g_{u \ssI}^{\rm eff} \right)^2 + \left( g_{d\ssI}^{\SM} + \Delta g_{d \ssI}^{\rm eff} \right)^2$.

Figure \ref{nuhad} plots the constraint found by requiring $\Delta \varepsilon_{\ssL}^2 \leq 0.00137$, assuming that $X = B-L$. The plots are cut off at low mass where the condition $\left\vert t/M_{\ssX}^2 \right\vert \leq 0.01$ breaks down. Notice that for $\eta =0$ the bound is similar to that found for neutrino-electron scattering, with stronger bounds on $\alpha_\ssX$ at smaller $M_\ssX$. For nontrivial $\eta$ the strength of $X-Z$ mixing eventually provides the strongest constraint, leading to strong bounds even for small $g_\ssX$ at sufficiently low $M_\ssX$.

\section{Low-energy constraints}

We finally turn to constraints coming from lower-energy processes.

\subsection{Anomalous magnetic moments\label{amm-sec}}

The accuracy of anomalous magnetic moment (AMM) measurements \cite{pdg} produce a strong constraint on the parameters of an extra gauge boson. We consider the bound arising from both the electron and muon AMM on the $X$ gauge coupling as a function of the mass $M_\ssX$, for various values of the kinetic mixing parameter $\sh\,\eta$.

\FIGURE[h!]{
\begin{tabular}{cc}
\includegraphics[scale=0.64]{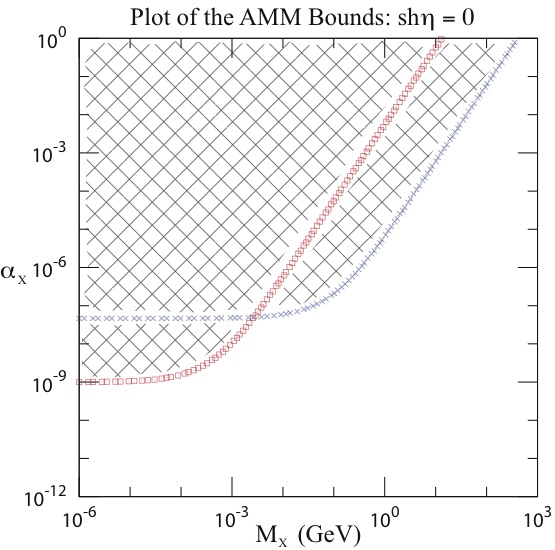} &
\includegraphics[scale=0.64]{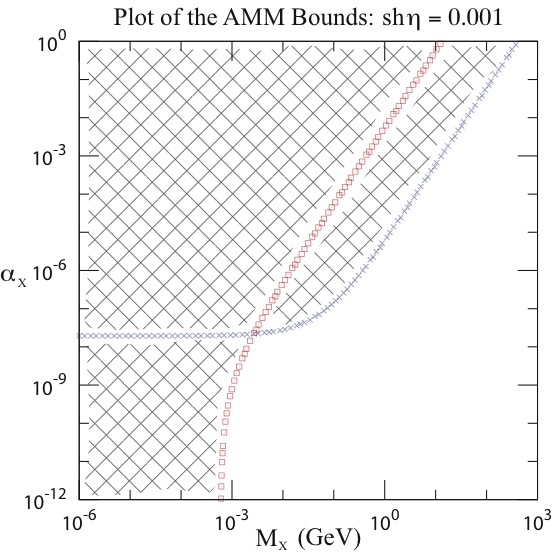} \\
\includegraphics[scale=0.64]{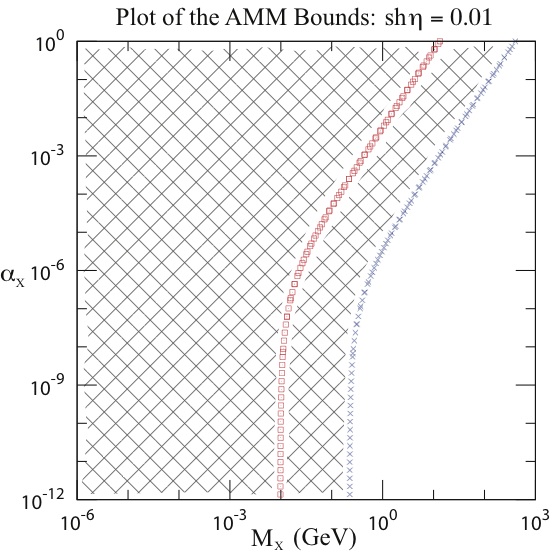} &
\includegraphics[scale=0.64]{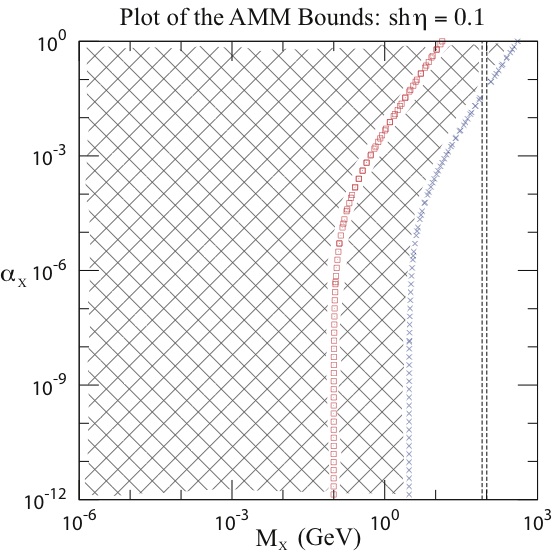} \\
\multicolumn{2}{c}{
\includegraphics[scale=0.64]{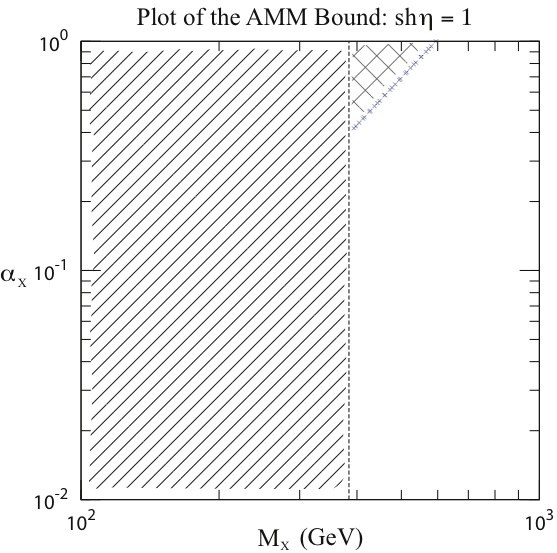}
}
\end{tabular}
\caption{Plots of the constraint on the gauge coupling $\alpha_\ssX$ arising from the electron and muon AMM as a function of $M_\ssX$, for various values of $\sh\,\eta$. The electron AMM bound is marked with blue crosses; the muon AMM bound is marked with red squares. The plot assumes a coupling $X_{\ell \ssL} = X_{\ell \ssR} = -1$, such as would be true if $X = B-L$. Hatched regions are excluded.}
\label{amm}
}

The correction to the AMM of a lepton, $\ell$, is given by \cite{ovanesyan}
\be
 \delta a_\ell = \frac{m_\ell^2}{4 \pi^2 M_\ssX^2}
 \int_0^1 \! dz \frac{k_{\ell \ssV}^2 z (1-z)^2 - k_{\ell \ssA}^2 \left[z (1-z) (3+z)+ 2 (1-z)^3 m_\ell^2 / M_\ssX^2 \right]}{z+(1-z)^2 m_\ell^2 / M_\ssX^2} \,,
\ee
where the vector and axial couplings to the $X$ boson are of the form
\bea
 k_{\ell \ssV} &:=&\frac{k_{\ell \ssL}+k_{\ell \ssR}}{2}= c_\xi \left[\ch\,\eta \, g_\ssX X_{\ell \ssV} - \sh\,\eta \, \frac{e}{c_\ssW} \left(-\frac14 +1\right)\right]-s_\xi e_\ssZ \left(-\frac14+s_\ssW^2 \right)  \\
 k_{\ell \ssA} &:=&\frac{k_{\ell \ssL}-k_{\ell \ssR}}{2}= c_\xi \left[\ch\,\eta \, g_\ssX X_{\ell \ssA} - \sh\,\eta \, \frac{e}{c_\ssW} \left(-\frac14 \right)\right]-s_\xi e_\ssZ \left(-\frac14\right) \nn \,.
\eea

There is, however, some subtlety in comparing this shift with experiment \cite{maximsec}: since the electron AMM, $\delta a_e$, is used to determine the fine-structure constant,  $\alpha$. The best bound on $X$ boson couplings therefore comes from the next most precise experiment that measures $\alpha$, and not the errors from the $(g-2)$ experiments themselves. Following \cite{maximsec} this leads to the constraints $\delta a_e < 1.59 \times 10^{-10}$ and $\delta a_\mu < 7.4 \times 10^{-9}$, which when compared with the above expression gives the bounds shown in Figure \ref{amm}. These plots reproduce the results found in \cite{ovanesyan} when $\sh\,\eta = 0$. In particular, the $M_\ssX$ values below which any gauge coupling is excluded are consistent with the bounds shown in \cite{maximsec}.

Since these bounds are often considered (e.g. in \cite{maximsec}, \cite{HiddenU1Bounds, toroschuster}) in the context of a constraint on kinetic mixing, we also plot the constraint on $\sh\,\eta$ as a function of the $X$ boson mass, for various values of the gauge coupling. This is shown in Figure \ref{alphaamm}.

\FIGURE[h!]{
\begin{tabular}{cc}
\includegraphics[scale=0.69]{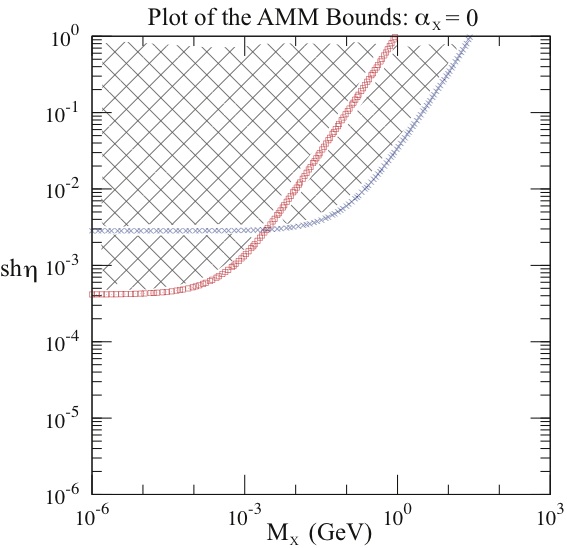} &
\includegraphics[scale=0.69]{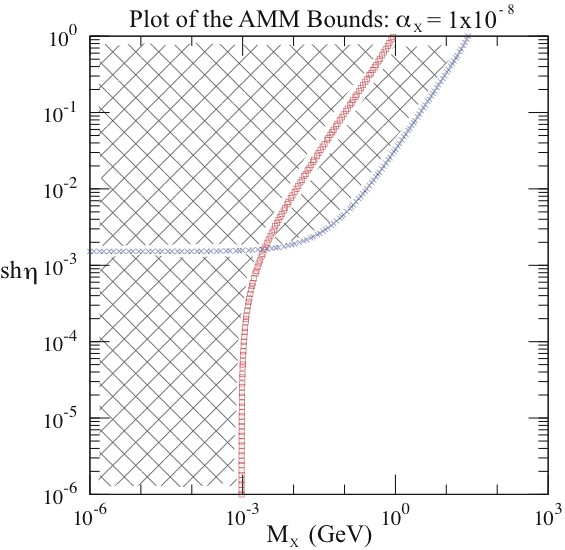} \\
\includegraphics[scale=0.69]{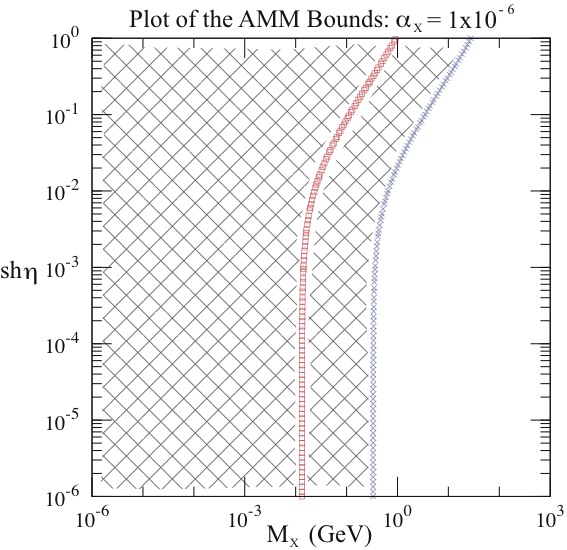} &
\includegraphics[scale=0.69]{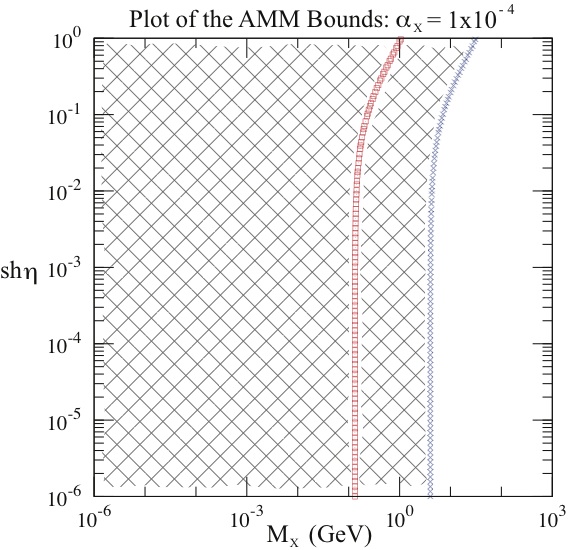}
\end{tabular}
\caption{Plots of the constraint on the kinetic mixing, $\sh\,\eta$, arising from the electron and muon AMM as a function of $M_\ssX$, for various values of the gauge coupling $\alpha_\ssX$. The electron AMM bound is marked with blue crosses; the muon AMM bound is marked with red squares. The plot assumes a coupling $X_{\ell \ssL} = X_{\ell \ssR} = -1$, such as would be true if $X = B-L$. Hatched regions are excluded.}
\label{alphaamm}
}

\subsection{Upsilon decay}

The bound we present here is an extension of the result found in \cite{toroschuster}. By looking at the decay rate of the $\Upsilon(3 s)$ $b\ol b$ bound state, researchers from the BABAR collaboration were able to place a bound on the occurrence of a particular channel involving a light pseudoscalar $A_0$ \cite{dimuon}:
\be
 e^+ + e^- \rightarrow \Upsilon(3 s) \rightarrow \gamma +
 A_0 \rightarrow \gamma + \mu^+ + \mu^- \,.
\ee
Their upper limit on the number of events
\be
 N = \sigma(e^+ + e^- \rightarrow \Upsilon(3 s)) \times \cL \times \mathrm{Br} (\Upsilon(3 s) \rightarrow \gamma + A_0) \times \mathrm{Br} (A_0 \rightarrow \mu^+ + \mu^-) \,,
\ee
places a bound on the quantity $Q:=\mathrm{Br} (\Upsilon(3 s) \rightarrow \gamma + A_0) \times \mathrm{Br} (A_0 \rightarrow \mu^+ + \mu^-)$.

However, the reaction of interest to us is
\be
 e^+ + e^- \rightarrow \gamma + X \rightarrow \gamma + \mu^+ + \mu^- \,,
\ee
which would have an identical signature. So the measured bound can also be reinterpreted as applying to the quantity
\be
 Q_\ssX := \frac{\sigma(e^+ + e^- \rightarrow \gamma + X)}{\sigma(e^+ + e^- \rightarrow \Upsilon(3 s))} \times \mathrm{Br} (X \rightarrow \mu^+ + \mu^-)
\ee
The experimental limit \cite{dimuon} $Q_\ssX < 3 \times 10^{-6}$ gives the plots found in Figure \ref{upsilon} over the range $2 m_\mu < M_\ssX < E_{cm} ( = 10.355$ GeV). This bound is quite strong, as it eliminates the entire region for $\sh\,\eta \gtrsim 0.002$ (as is shown in \cite{toroschuster}). For smaller $\sh\,\eta$, the bound is roughly constant when $M_\ssX \ll  E_{cm}$.

\FIGURE[h!]{
\begin{tabular}{cc}
\includegraphics[scale=0.7]{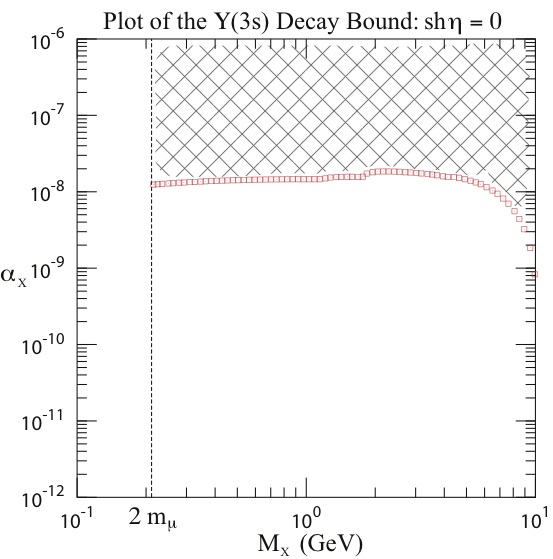} &
\includegraphics[scale=0.7]{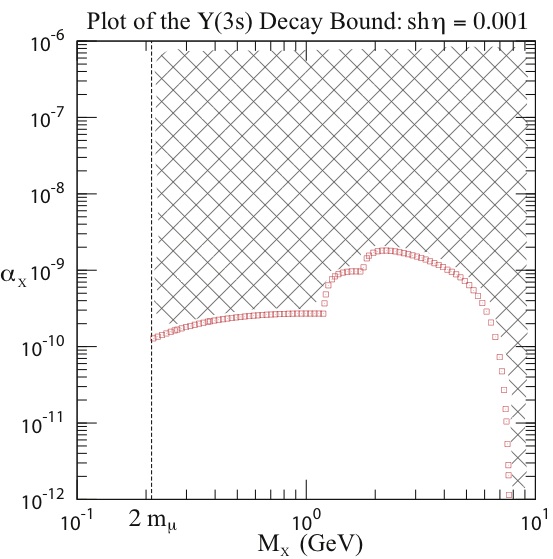}
\end{tabular}
\caption{Plots of the constraint on the gauge coupling $\alpha_\ssX$ arising from $\Upsilon (3s)$ decay as a function of $M_\ssX$, for $\sh\,\eta = 0, 0.001$. The plot assumes a coupling $X_{\ell \ssL} = X_{\ell \ssR} = -1$, such as would be true if $X = B-L$. Hatched regions are excluded.}
\label{upsilon}
}

As in the case of the AMM bounds, we also plot the constraint on $\sh\,\eta$ as a function of the $X$ boson mass for various values of the gauge coupling --- as shown in Figure \ref{alphaupsilon}.

\FIGURE[h!]{
\begin{tabular}{cc}
\includegraphics[scale=0.7]{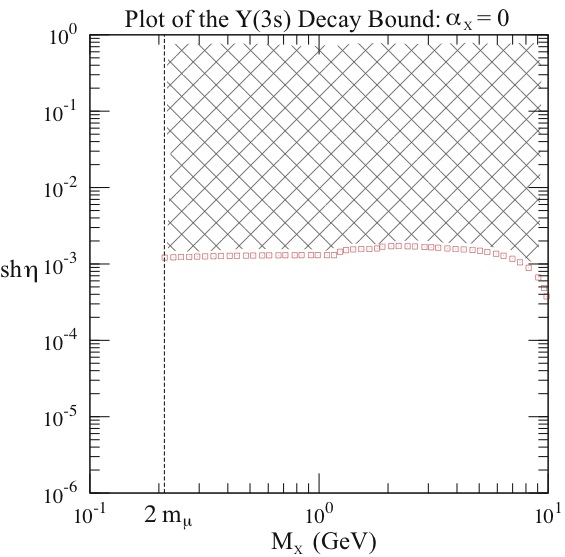} &
\includegraphics[scale=0.7]{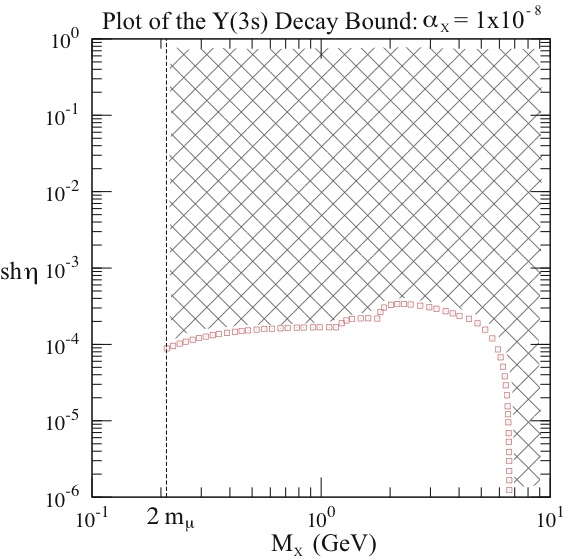}
\end{tabular}
\caption{Plots of the constraint on the kinetic mixing, $\sh\,\eta$, arising from $\Upsilon (3s)$ decay as a function of $M_\ssX$, for $\alpha_\ssX = 0,\, 1\times10^{-8}$. The plot assumes a coupling $X_{e \ssL} = X_{e \ssR} = -1$, such as would be true if $X = B-L$. Hatched regions are excluded.}
\label{alphaupsilon}
}

\subsection{Beam-dump experiments} \label{bdexpts}

In the MeV$-$GeV mass range, small $g_\ssX$ and $\eta$ are constrained by several beam dump experiments. These bounds are considered in detail in \cite{toroschuster}; we apply a simplified version of their analysis here.

In these experiments, a large number $N_e$ of electrons with initial energy $E$ are collided with a fixed target made of either aluminum or tungsten. Many of the resulting collision products are absorbed either by the target or by some secondary shielding. (Here, we use $t$ to denote the total thickness of both the target and the shielding.) The remaining products continue along an evacuated tube to the detector, located at some distance $D$ away from the target. For a summary of values for these parameters, see Table \ref{beamdumpvals}.

\TABLE[tbh]{
\begin{tabular}{|c|c|c|c|c|c|c|}
\hline
Experiment & Target & $N_e$ & Beam Energy & $t$ & $D$ \\ \hline \hline
E774 & W & $0.52\times 10^{10}$ & 275 GeV & 30 cm & 7.25 m  \\ \hline
E141 & W & $2\times10^{15}$ & 9 GeV & 12 cm & 35 m \\ \hline
E137 & Al & $1.87\times10^{20}$ & 20 GeV & 200 m & 400 m \\
\hline
\end{tabular}%
\caption{Parameter values for the E774, E141, and E137 beam dump experiments.}
\label{beamdumpvals}
}

The bound arises from the non-observation of $X$ decay products. The incoming electron emits an $X$ boson as bremsstrahlung during photon exchange with the nucleon ($N$): $e^- + N \rightarrow e^- +N+X$. The $X$ can then decay into either an $e^+ e^-$ or $\mu^+ \mu^-$ pair. However, a decay that occurs too soon is absorbed by the shield while a decay that occurs too late occurs past the detector. Therefore, the number of lepton anti-lepton pairs observed at the detector can be computed by multiplying the number of $X$ bosons produced, $N_\ssX$, by the probability for the $X$ to decay between $z=t$ to $z=D$:
\be
 N_{obs} = N_\ssX \int_t^D \! dz \, \left(\frac{1}{\ell_0}
 \, e^{-z/\ell_0}\right) \,.
\ee
Here, we write the lab frame decay length as $\ell_0:=\gamma c \tau$, where $\gamma = (1-v^2)^{-1/2}$ is the relativistic time-dilation factor and $\tau$ is the inverse of the $X$ rest-frame decay rate: $\tau:=1/\Gamma_\ssX$.

\FIGURE[h!]{
\begin{tabular}{cc}
\includegraphics[scale=0.73]{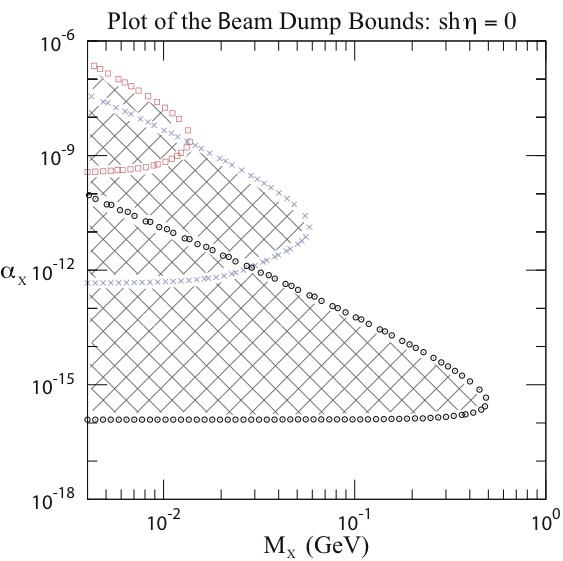} &
\includegraphics[scale=0.73]{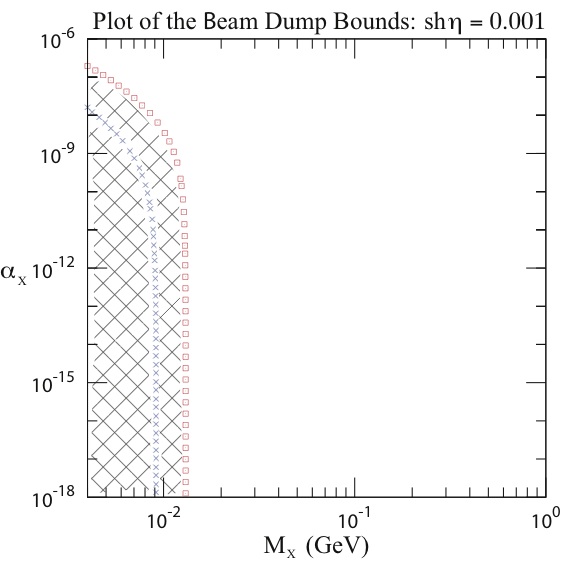}
\end{tabular}
\caption{The constraint arising from beam dump experiments on the coupling $\alpha_\ssX = g^2_\ssX/4\pi$ as a function of $M_\ssX$, for $\sh\,\eta=0,0.001$. The E774 bound is marked with red squares; the E141 bound is marked with blue crosses; the E137 bound is marked with black circles. The plot assumes a coupling $X_{\ell \ssL} = X_{\ell \ssR} = -1$, such as would be true if $X = B-L$. Hatched regions are excluded.}
\label{beamdump}
}

In estimating the number of $X$'s produced, we use the following result from \cite{toroschuster}:
\be
 N_\ssX \sim N_e \, \mu^2 \frac{\epsilon^2}{M_\ssX^2} \,,
\ee
where $\epsilon = \chi c_\ssW$ and $\mu^2 \simeq 2.5 \,\, \textrm{MeV}^2$ is an overall factor that contains information regarding the details of the nuclear interaction, and is shown in \cite{toroschuster} to be roughly constant for $M_\ssX$ between 1 and 100 MeV. There is, however, an obstacle in applying this result directly to our analysis: it was derived without including any coupling to $J_\ssX^\mu$. In order to introduce the $k_{e \ssL(\ssR)}$-dependence in this expression, we note from \cite{toroschuster} that the $\epsilon$-dependence above arises from the cross section $\sigma(e^- \gamma \rightarrow e^- X)$ under the assumption that the electron is massless. This means that the left- and right-handed helicity $X - e$ interactions contribute equally to the cross section, allowing the substitution
\be
 \epsilon^2 \rightarrow \frac{1}{4\pi\alpha} \left( \frac{k^2_{e\ssL}+k^2_{e\ssR}}{2}
 \right)\,,
\ee
with the normalization chosen so that the above expression reduces to $\chi^2 c_\ssW^2$ in the case where $X_{e\ssL(\ssR)}=0$, $\sh\,\eta \ll 1$ and $M_\ssX \ll M_\ssZ$.

All in all, we find that the number of $X$'s we expect to observe is given by
\be
 N_{obs} \sim \frac{N_e\,\mu^2}{M_\ssX^2} \left(
 \frac{ k^2_{e\ssL} + k^2_{e\ssR} }{8\pi\alpha} \right)
 \left( e^{-t/\ell_0} - e^{-D/\ell_0} \right) \,.
\ee
Applying the experimental exclusions \cite{toroschuster} $N_{obs} < 17$ events (E774), $N_{obs} < 1000$ events (E141), and $N_{obs} < 10$ events (E137) gives the bounds shown in Figure \ref{beamdump}.

\FIGURE[h!]{
\begin{tabular}{cc}
\includegraphics[scale=0.73]{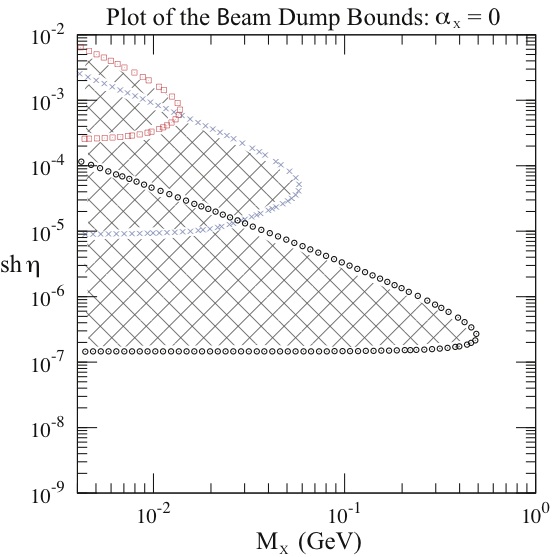} &
\includegraphics[scale=0.73]{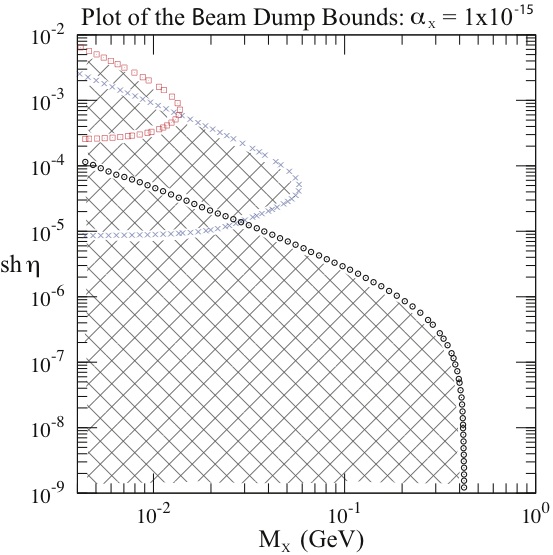} \\
\includegraphics[scale=0.73]{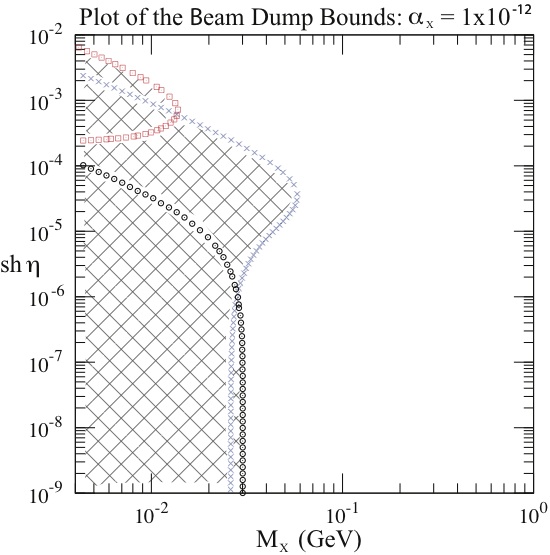} &
\includegraphics[scale=0.73]{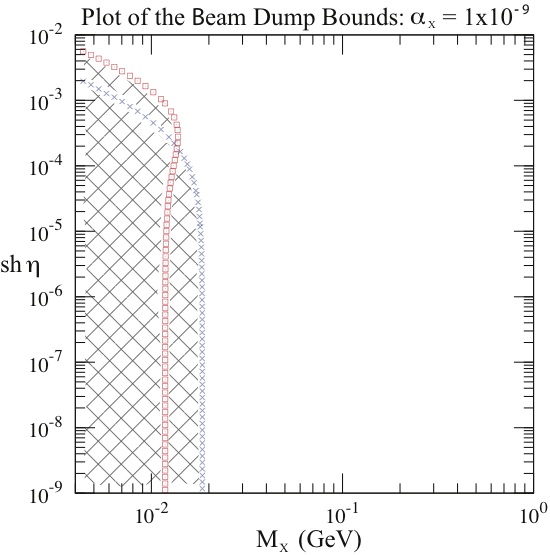}
\end{tabular}
\caption{The constraint arising from beam dump experiments on the kinetic mixing $\sh\,\eta$ as a function of $M_\ssX$, for $\alpha_\ssX = 0$, $1 \times 10^{-15}$, $1 \times 10^{-12}$, and $1 \times 10^{-9}$. The E774 bound is marked with red squares; the E141 bound is marked with blue crosses; the E137 bound is marked with black circles. The plot assumes a coupling $X_{\ell \ssL} = X_{\ell \ssR} = -1$, such as would be true if $X = B-L$. Hatched regions are excluded.}
\label{alphabeamdump}
}

The plot for $\sh\,\eta=0$ gives good agreement with a similar plot in \cite{ovanesyan} (in the region over which these results overlap). The lower bounds for each experiment are approximately flat because, in the region where $t \ll D \ll \ell_0$, the fraction of $X$'s that decay is just $D/\ell_0$, which gives
\be
 N_{obs} \sim \frac{N_e\,\mu^2}{M_\ssX^2} \left(
 \frac{ k^2_{e\ssL} + k^2_{e\ssR} }{8\pi\alpha} \right) \frac{D}{\ell_0} \,.
\ee
The leading $M_\ssX$-dependence then cancels since $\ell_0 \sim 1/M_\ssX^2$. The upper bound results from the situation where the $X$ bosons decay too quickly, and the decay products do not escape the shielding.

We have only included plots for the cases where $\sh\,\eta=0$ and 0.001 because the bounds become too weak to constrain any region of this parameter space whenever $\sh \, \eta > 0.007$.

An interesting feature of these bounds is that, at any given value of $\sh\,\eta$, the gauge coupling can be increased such that the bounds are evaded. This occurs because a stronger gauge coupling causes the $X$ bosons to decay within the shielding. Therefore, any bound on kinetic mixing which results from these experiments can weaken if the direct coupling of electrons to the $X$ is taken to be non-zero. To demonstrate this, consider the bounds shown in Figure \ref{alphabeamdump}, which plots the bound on kinetic mixing as a function of the $X$-boson mass, for various values of $\alpha_\ssX$. Note that, for $\alpha_\ssX \gsim 1\times 10^{-6}$, these bounds are satisfied for all values of $\sh\,\eta$ in the relevant mass range.

\subsection{Neutron-nucleus scattering\label{neutnuclsec}}

Low-energy neutron-nucleus scattering is important because most of the other low-energy bounds evaporate if the new boson doesn`t couple to leptons (such as if $X = B$). For neutron-nucleus scattering a bound is obtained by considering the effects of the new Yukawa-type potential that would arise from a non-zero vector coupling of the $X$ to neutrons. For light $X$ bosons this can be seen over the strong nuclear force because it has a longer range, and can affect the angular dependence of the differential cross section for elastic scattering, $d\sigma(n N \to n N)/d\Omega$. This bound is discussed in the context of a scalar boson in \cite{barbericson} and more generally in \cite{psizeanom}.

Following these authors we parameterize the differential cross section as
\be
 \frac{d\sigma}{d\Omega} = \frac{\sigma_0}{4 \pi} \left( 1+ \omega E cos\theta \right) \,,
\ee
where $\sigma_0$ and $\omega$ are to be taken from experiments. Then an interaction of the form
\be
 \Delta V_{nN}(r) = \left( \frac{g_n^2}{4\pi} \right)
 \frac{e^{-M_\ssX r}}{r}
\ee
leads to a correction to the expected value of $\omega$, which is measured experimentally in the energy range $E\sim 1$--$10$ keV for neutrons scattering with ${}^{208}$Pb.

Agreement with observations leads to the bound \cite{barbericson,psizeanom}
\be
 \frac{k_{n\ssV}^2}{4\pi M_\ssX^4} < 3.4\times10^{-11} \,,
\ee
where $k_{n\ssV} = k_{u\ssV} + 2 k_{d\ssV}$ with $k_{f\ssV} := \frac12 (k_{f\ssL}+k_{f\ssR})$, as above. Figure \ref{neutnucl} shows a plot of this bound for the nominal case $\sh\,\eta=0$.

For this combination of couplings, an interesting cancellation occurs. For small kinetic mixing, the correction $\Delta k_{f\ssL(\ssR)}$ has the form
\begin{eqnarray}
 \Delta k_{f\ssL(\ssR)} &= \eta \frac{e}{c_\ssW} (Q_f c_\ssW^2 - g^\SM_{f\ssL(\ssR)})+\eta s_\ssW e_\ssZ g^\SM_{f\ssL(\ssR)} +\cO(\eta^2) \nn\\
 &= \eta e c_\ssW Q_f + \cO(\eta^2) \,,
\end{eqnarray}
and so the leading correction in $\eta$ vanishes for any electrically neutral particle, like a neutron. This makes this bound relatively insensitive to changes in kinetic mixing, not varying appreciably over the range $0\leq \sh\,\eta \leq 1$. A similar cancellation occurs in the case of nucleosynthesis, considered in \S\ref{bbn}.

\FIGURE[tbh] {
\includegraphics[scale=1]{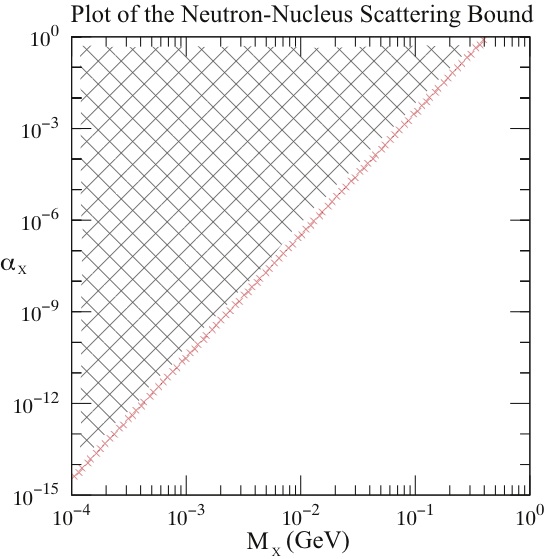}
\caption{Plot of the constraint on the gauge coupling $\alpha_\ssX$ due to neutron-nucleus scattering as a function of the $X$-boson mass $M_\ssX$. The hatched regions are excluded.}
\label{neutnucl} }

\subsection{Atomic parity violation\label{weakcharge}}

The Standard Model predicts a low-energy effective coupling between the electron axial current and the vector currents within a given nucleus. The so-called weak charge of a nucleus with $Z$ protons and $N$ neutrons is defined (up to an overall constant) as the coherent sum of the $Z$-boson vector couplings over the constituents of that nucleus \cite{bouchfayet}:
\be
 Q_\ssW(Z,N):=4 \left[ Z \left( 2 g_{u\ssV} + g_{d\ssV}\right) + N \left( g_{u\ssV} + 2 g_{d\ssV }\right) \right] \,.
\ee
where
\be
 g_{f\ssV f}:=\frac{g_{f\ssL}+g_{f\ssR}}{2}
  \quad \hbox{and} \quad
  g_{f\ssA }:=\frac{g_{f\ssL }-g_{f\ssR }}{2} \,.
\ee

In terms of these the leading parity-violating effective electron-nuclear interaction generated by $Z$ boson exchange is
\be
 \cL_{eff}=-\sqrt{2} G_\ssF g_{e\ssA } Q_\ssW \left(\ol e \gamma_\mu \gamma_5 e\right) \left(\ol \Psi \gamma^\mu \Psi \right) \,.
\ee
where $\Psi$ is the field describing the nucleus. $X$-boson exchange adds an additional term to this effective lagrangian of the form
\be
 \cL_{eff}^\ssX=-\frac{k_{e\ssA } Q_\ssX}{M_\ssX^2} \left(\ol e \gamma_\mu \gamma_5 e\right) \left(\ol \Psi \gamma^\mu \Psi \right) \,
\ee
where
\be
 Q_\ssX := Z \left( 2 k_{u\ssV } + k_{d\ssV }\right) + N \left( k_{u\ssV } + 2 k_{d\ssV }\right) \,.
\ee
Therefore, the total shift in the $Q_\ssW$ due to the $X$ boson is
\be \label{deltaqw}
 \Delta Q_\ssW = \left[\frac{g_{e\ssA }}{(-1/4)}Q_\ssW -Q_\ssW^\SM \right]- \frac{2 \sqrt{2}}{G_\ssF} \frac{k_{e\ssA } Q_\ssX}{M_\ssX^2}
\ee
where
\beqa
 Q_\ssW^\SM (Z,N) &=& 4 \left[ Z \left( 2 g_{u\ssV }^\SM + g_{d\ssV }^\SM \right) + N \left( g_{u\ssV }^\SM + 2 g_{d\ssV }^\SM \right) \right] \nn \\
 &=& Z \left(1- 4 s_\ssW^2 \right) -N \,.
\eeqa
Notice that the bracketed term in $\Delta Q_\ssW$ goes to $0$ as $\eta\rightarrow0$, whereas the second term does not as long as $k_{\ssA e}$ does not vanish in the same limit. The total effective lagrangian for this system can then be written as
\be
 \cL_{eff}+\cL_{eff}^\ssX = \frac{G_\ssF}{2 \sqrt{2}} \left( Q_\ssW^\SM + \Delta Q_\ssW \right) \left(\ol e \gamma_\mu \gamma_5 e\right) \left(\ol \Psi \gamma^\mu \Psi \right) \,.
\ee

It is expected that the second term in eq.~\pref{deltaqw} will be dominant, so it is useful to consider the form of $k_{\ssA e}/M_\ssX^2$ in the limit where $M_\ssX \ll M_\ssZ$ and $\eta \ll 1$:
\be
 \frac{k_{e\ssA }}{G_\ssF M_\ssX^2} = \frac{g_\ssX X_{e\ssA} \left(1+\frac12 \, {c_\ssW^2} \eta^2 \right)}{G_\ssF M_\ssX^2} - \sqrt{2} \; \frac{s_\ssW}{e_\ssZ} \eta \,.
\ee
Therefore, if $X_{\ssL e} = X_{\ssR e}$, then the constraint becomes significantly less stringent at low masses since, instead of bounding the ratio $g_\ssX^2/M_\ssX^2$, it is now the combination $g_\ssX \eta$ that is bounded. In order to emphasize the strength of this bound when $X_{\ssA e}\neq0$, we use the charge assignments as shown in Table \ref{apvcharges}.

\TABLE[tbh]{
\begin{tabular}{|c|c|}
\hline
SM Fermion & Charge $X$ \\ \hline
$u_{L},d_{L}$ & $0$ \\ \hline
$u_{R}$ & $-1/3$ \\ \hline
$d_{R}$ & $+1/3$ \\ \hline
$\nu _{L},e_{L}$ & $0$ \\ \hline
$e_{R}$ & $-1/3$ \\
\hline
\end{tabular}%
\caption{Charge assignments for the ``right-handed'' $U\left( 1\right)$.}
\label{apvcharges}
}

If the $X$ boson is light enough the above effective interaction eventually becomes inaccurate in describing the electron-nucleus interactions. In this case, rather than pursuing a detailed analysis of the microscopic lagrangian, we follow ref.~\cite{bouchfayet} and introduce a corrective factor $K(M_\ssX)$ to account for the non-locality caused by the small mass of the $X$ boson. This modifies our expression for $\Delta Q_\ssW$ as follows:
\be
 \Delta Q_\ssW = \left[ \frac{g_{e\ssA }}{(-1/4)}Q_\ssW -Q_\ssW^\SM \right] - \frac{2 \sqrt{2}}{G_\ssF} \frac{k_{e\ssA } Q_\ssX}{M_\ssX^2} K(M_\ssX) \,.
\ee
In \cite{bouchfayet} a table is given for $K$ for various values of $M_\ssX$ in the range $0.1$ MeV $< M_\ssX < 100$ MeV. In order to render the graphs shown here, we have interpolated values of $K$ by doing a least squares fit to the values in \cite{bouchfayet}.

\FIGURE[h!] {
\includegraphics[scale=1]{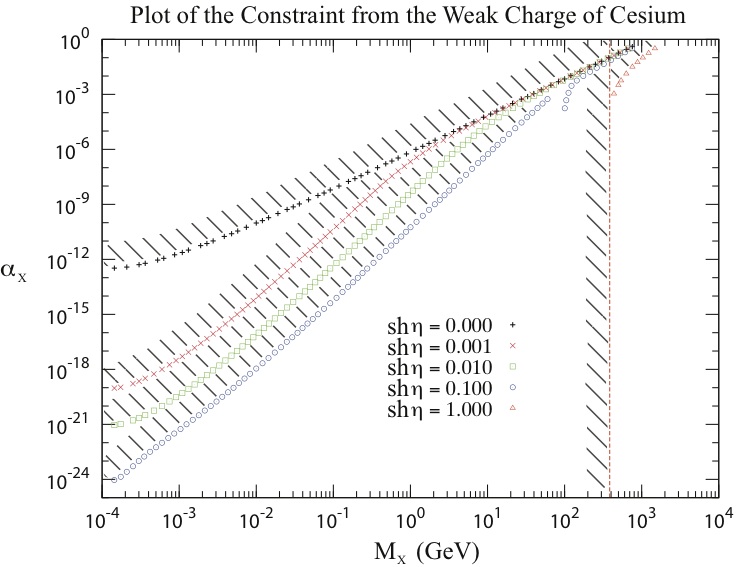}
\caption{Plot of the constraint on the gauge coupling $\alpha_\ssX$ due to the weak charge of cesium as a function of the $X$-boson mass $M_\ssX$, for various values of $\sh\,\eta$. The hatched regions are excluded.}
\label{apv} }

As with the neutrino-electron scattering bounds, the slope of the bound changes for $\eta\neq0$ due to the production of a new dominant term through cancellation with the modified $Z$-fermion coupling. Once again, we exclude the region below $385$ GeV for the $\sh\,\eta=1$ plot in order to avoid conflict with the electroweak oblique fits that require $z\ll1$.

Since this bound relies crucially on there being an axial vector coupling to the electron, we did not include it when compiling the summary of bounds given in figures in \S1.

\subsection{Primordial nucleosynthesis\label{bbn}}

We close with the study of constraints coming from cosmology, which for the mass range of interest in this paper consists dominantly of Big Bang Nucleosynthesis.

Any $X$ bosons light enough to be present in the primordial soup at temperatures below $T \sim 1$ MeV can destroy the success of Big Bang Nucleosynthesis (BBN) if they make up a sufficiently large fraction ($\lsim 10\%$) of the universal energy density, leading to potentially strong constraints. In particular, such a boson poses a problem if it is in thermal equilibrium at these temperatures.

Quantitatively, measurements of primordial nuclear abundances forbid the existence of the number of additional neutrino species (beyond the usual 3 of the SM) to be \cite{4hebbn} $\delta N_{\nu } \leq 1.44$ (at $95\%$ C.L.). But since each boson in equilibrium counts $\frac87$ times more strongly in the equilibrium abundance, and since a massive $X$ boson carries 3 independent spin states, the corresponding bound on the number, $N_\ssX$, of new species of spin-1 particles in equilibrium at BBN is
\be
 N_\ssX \leq 0.84 \,.
\ee
Even just one additional massive spin-1 boson into relativistic equilibrium is excluded at the $95\%$ confidence level.

In a universe containing only the $X$ boson and ordinary SM particles at energies of order 1 MeV, this leads to two kinds of constraints: either the $X$ boson's couplings are weak enough that it does not ever reach equilibrium; or if the $X$ boson is in equilibrium it must be heavy enough ($\gsim 1$ MeV) to have a Boltzmann-suppressed abundance. Figure \ref{bbnweta} sketches the regions in the coupling-mass plane that are excluded by these conditions. The vertical line corresponds to the situation where abundance is suppressed by Boltzmann factors.

\FIGURE[ht]{
\begin{tabular}{cc}
\includegraphics[scale=0.73]{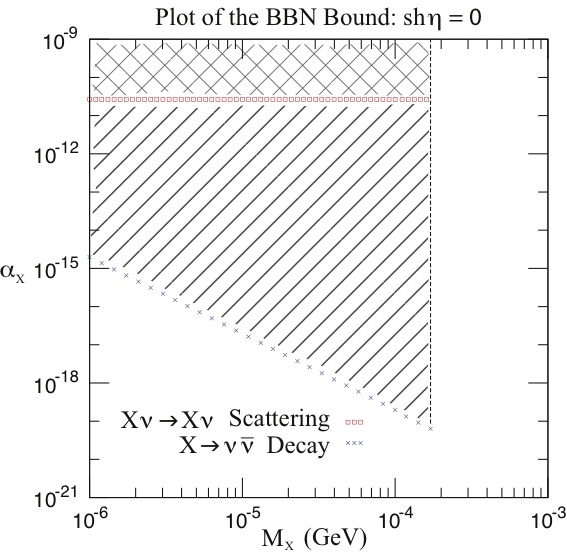} &
\includegraphics[scale=0.73]{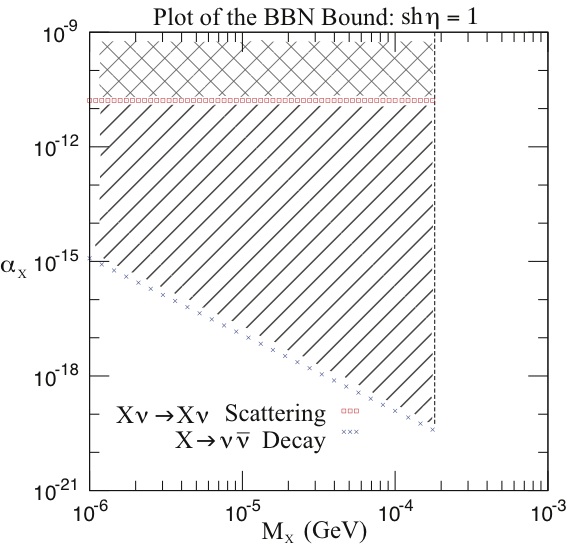}
\end{tabular}
\caption{Constraint on the gauge coupling of the $X$ due to its effect on nucleosynthesis, as a function of the $X$-boson mass. The red squares indicate the bound due to $X\nu \raro X\nu$ scattering; the blue crosses indicate the bound due to $X \raro \nu \ol\nu$ decay. }
\label{bbnweta}
}

The constraints on couplings arise only for sufficiently light particles, and express the condition that the couplings be weak enough to avoid equilibrium, at least up until the freeze-out temperature $T_\ssF$. There are two curves of this type drawn, which differ by whether it is collision or decay processes that are the dominant equilibration mechanisms. Qualitatively, the requirement that reactions like $X \nu \leftrightarrow X \nu$ not equilibrate the $X$ bosons leads to a constraint on the couplings that is $M_\ssX$-independent in the limit where $M_\ssX \ll T_\ssF$, because then the size of both the reaction rate and Hubble scale is set by the temperature. The same is not true for decay reactions, $X \leftrightarrow \ol\nu \, \nu$, since the rate for this also depends on the $X$-boson mass.

A few other comments are appropriate for Figure \ref{bbnweta}. First, because they are outside the main scope of this study, the bounds shown are derived assuming that $M_\ssX \ll T$ (rather than being evaluated numerically as a function of $M_\ssX$) and so are drawn only up to the mass range within 0.5 MeV of the freeze-out temperature. Second, the resulting expressions depend only weakly on $\eta$, showing little difference over the range $0<\sh\,\eta<1$. As discussed in earlier sections, this is a consequence of the neutrino's electrical neutrality, which ensures that the leading small-$\eta$ limit of the kinetic mixing first arises at $\cO(\eta^2)$ rather than $\cO(\eta)$.

\section*{Acknowledgments}

We would like to thank Brian Batell, Joseph Conlon, Rouven Essig, Sven Krippendorf, David Poland, Maxim Pospelov, Philip Schuster, Natalia Toro and Michael Trott for helpful
discussions. AM and FQ thank McMaster University and Perimeter Institute for hospitality. CB and AM thank the Abdus Salam International Centre for Theoretical Physics (ICTP) for its kind hospitality while part of this work was done. The work of AM was supported by the EU through the Seventh Framework Programme and Cambridge University. CB and MW's research was supported in part by funds from the Natural Sciences and Engineering Research Council (NSERC) of Canada. Research at the Perimeter Institute is supported in part by the Government of Canada through Industry Canada, and by the Province of Ontario through the Ministry of Research and Information (MRI).

\section*{Note Added}

Since posting, we have learned of a beam dump analysis \cite{newBeamDumpBB} that has enlarged\footnote{We thank Johannes Bl\"umlein for bringing this to our attention} the exclusion regions discussed in \S\ref{bdexpts}.

\appendix

\section{Diagonalizing the gauge action}

This appendix provides the details of the diagonalization of the gauge boson kinetic and mass mixings. The starting point is eq.~\pref{startLeq},
\begin{equation} \label{AppstartLeq}
 \cL = -\frac14 \hat{\bf V}_{\mu\nu}^\ssT \hat K
 \hat{\bf V}^{\mu\nu} - \frac12 \hat{\bf V}_\mu^\ssT
 \hat M \hat{\bf V}^\mu + \hat{\bf J}_\mu^\ssT
 \hat{\bf V}^\mu \,,
\end{equation}
with $\hat K$ and $\hat M$ given in eqs.~\pref{startKMeq}.

\subsubsection*{Diagonalization}

We begin by performing the usual weak-mixing rotation to diagonalize the mass term:
\begin{equation}
 \hat{\bf V} = R_1 \check{\bf V} :=
 \begin{bmatrix}
 \hat{c}_\ssW & \hat{s}_\ssW
 & 0 \\
 -\hat{s}_\ssW & \hat{c}_\ssW
 & 0 \\
 0 & 0 & 1
 \end{bmatrix}
 \begin{bmatrix}
 \check{Z} \\
 \check{A} \\
 \check{X}%
\end{bmatrix}%
\end{equation}
where
\be
 \hat{c}_\ssW := \cos \hat\theta_\ssW :=
 \frac{g_2}{\sqrt{g_1^2 + g_2^2}}
 \quad \hbox{and} \quad
  \hat{s}_\ssW := \sin\hat \theta_\ssW
 := \frac{g_1}{\sqrt{g_1^2 + g_2^2}} \,.
\ee
The lagrangian then becomes
\begin{equation}
 \cL = - \frac14 \check{\bf V}_{\mu\nu}^\ssT
 \check K \check{\bf V}^{\mu\nu} - \frac12
 \check{\bf V}_\mu^\ssT \check M \check{\bf V}^\mu
 +  \check{\bf J}_\mu^\ssT \check{\bf V}^\mu  \,,
\end{equation}
with new matrices
\begin{equation}
 \check{K} = R_1^\ssT \hat K R_1 =
 \begin{bmatrix}
 1 & 0 & \chi \hat s_\ssW \\
 0 & 1 & -\chi \hat c_\ssW \\
 \chi \hat s_\ssW & -\chi \hat c_\ssW & 1%
 \end{bmatrix}
 \quad \hbox{and} \quad
 \check{M} = R_1^\ssT \hat M R_1 =
 \begin{bmatrix}
 m_\ssZ^2 & 0 & 0 \\
 0 & 0 & 0 \\
 0 & 0 & m_\ssX^2%
 \end{bmatrix}%
\end{equation}
where $m_\ssZ^2 := \frac14 \left( g_1^2 + g_2^2 \right) v^2$. Under the same transformation the currents become
\begin{eqnarray}
 \check{\bf J}_\mu &=& R_1^\ssT \hat{\bf J}_\mu =
 \begin{bmatrix}
 \hat J_\mu^3 \,\hat c_\ssW - \hat J_\mu^\ssY \,\hat s_\ssW \\
 \hat J_\mu^3 \,\hat s_\ssW + \hat J_\mu^\ssY \,\hat c_\ssW \\
 \hat{J}_\mu^\ssX
 \end{bmatrix}  \\
 &=& \sum_f
 \begin{bmatrix}
 \hat ie_\ssZ \ol{f} \gamma_\mu \left[ T_{3f} \gamma_\ssL
 - Q_f \hat s_\ssW^2 \right] f \\
 ie \, \ol{f} \gamma_\mu Q_f f \\
 ig_\ssX \ol{f} \gamma_\mu \left[ X_{f\ssL} \gamma_\ssL
 + X_{f \ssR} \gamma_\ssR \right] f
 \end{bmatrix} :=
 \begin{bmatrix}
 \check J_\mu^\ssZ \\
 \check J_\mu^\ssA \\
 \check J_\mu^\ssX
 \end{bmatrix} \,, \notag
\end{eqnarray}
which defines $\hat e_\ssZ := e/(\hat s_\ssW \hat c_\ssW)$ and uses the standard SM relations $g_2 \hat s_\ssW = g_1 \hat c_\ssW := e$ and $Q_{f} = T_{3f} + Y_{f\ssL} = Y_{f\ssR}$.

The kinetic term is diagonalized by letting
\begin{equation}
 \check{\bf V} := L \tilde{\bf V} :=
 \begin{bmatrix}
 1 & 0 & - \hat s_\ssW \, \sh \, \eta  \\
 0 & 1 & \hat c_\ssW \, \sh \, \eta \\
 0 & 0 & \ch \, \eta
 \end{bmatrix}%
 \begin{bmatrix}
 \tilde Z \\
 \tilde A \\
 \tilde X
 \end{bmatrix}
\end{equation}
with
\be
 \sh \, \eta := \sinh \eta := \frac{\chi}{\sqrt{1 - \chi^2}}
 \quad \hbox{and} \quad
 \ch \, \eta := \cosh \eta := \frac{1}{\sqrt{1 - \chi^2}} \,.
\ee
This gives, by construction
\begin{equation}
 \tilde K = L^\ssT \check K L =
 \begin{bmatrix}
 1 & 0 & 0 \\
 0 & 1 & 0 \\
 0 & 0 & 1%
 \end{bmatrix}%
\end{equation}
and
\begin{equation}
 \tilde M = L^\ssT \check M L=
 \begin{bmatrix}
 m_\ssZ^2 & 0 & - m_\ssZ^2 \hat s_\ssW \sh \, \eta \\
 0 & 0 & 0 \\
 - m_\ssZ^2 \hat s_\ssW \sh \, \eta & 0 & m_\ssX^2 \ch^2 \, \eta
 + m_\ssZ^2 \hat s_\ssW^2 \sh^2 \, \eta
 \end{bmatrix}\,,
\end{equation}
while the currents become
\begin{equation}
 \tilde{\bf J}_\mu := L^\ssT \check{\bf J}_\mu =
 \begin{bmatrix}
 \check J_\mu^\ssZ \\
 \check J_\mu^\ssA \\
 - \check J_\mu^\ssZ \hat s_\ssW \sh \, \eta
 + \check J_\mu^\ssA \hat c_\ssW \sh \, \eta
 + \check J_\mu^\ssX \ch \, \eta
 \end{bmatrix} \,.
\end{equation}
(Notice that $L$ and $R_1$ satisfy $L R_1 = R_1 L$, so it is immaterial whether we first diagonalize the SM mass or the kinetic terms.)

Finally, the mass matrix is diagonalized by letting
\begin{equation}
 \tilde{\bf V} = R_2 {\bf V} :=
 \begin{bmatrix}
 c_\xi & 0 & -s_\xi \\
 0 & 1 & 0 \\
 s_\xi & 0 & c_\xi%
 \end{bmatrix}%
 \begin{bmatrix}
 Z \\
 A \\
 X
 \end{bmatrix}%
\end{equation}
where $c_\xi := \cos \xi$ and $s_\xi := \sin \xi$ with the angle $\xi $ given by
\begin{equation}
 \tan 2\xi =\frac{-2 \hat s_\ssW \sh \eta }{1-\hat s_\ssW^2 \sh^2 \eta-r_\ssX^2 \ch^2\eta} \,,
 \label{Apptan2alpha}
\end{equation}
where we define for convenience
\be
 r_\ssX := \frac{m_\ssX}{m_\ssZ} \,.
\ee

The diagonalized lagrangian then is
\begin{equation}
 \cL = -\frac14 \, {\bf V}_{\mu\nu}^\ssT
 {\bf V}^{\mu\nu} - \frac{M_\ssZ^2}{2} \, Z_\mu Z^\mu
 -\frac{M_\ssX^2}{2} \, X_\mu X^\mu +
  {\bf J}_\mu^\ssT {\bf V}^\mu  \,,
\end{equation}
where the physical masses are
\beqa
 M_\ssX^2 &=& \frac{m_\ssZ^2}{2}
 \left( 1 + \hat s_\ssW^2 \sh^2 \eta + r_\ssX^2 \ch^2 \eta
 + \vartheta_\ssX \sqrt{ \left( 1+ \hat s_\ssW^2 \sh^2 \eta
 + r_\ssX^2 \ch^2 \eta \right)^2
 - 4 r_\ssX^2 \ch^2 \eta} \right)  \\
 M_\ssZ^2 &=& \frac{m_\ssZ^2}{2}
 \left( 1 + \hat s_\ssW^2 \sh^2 \eta + r_\ssX^2 \ch^2 \eta
 - \vartheta_\ssX \sqrt{ \left( 1+ \hat s_\ssW^2 \sh^2 \eta
 + r_\ssX^2 \ch^2 \eta \right)^2
 - 4 r_\ssX^2 \ch^2 \eta} \right)
\label{Appmasseigs}
\eeqa
with $\vartheta_{\ssX}$ defined such that $M_\ssZ \rightarrow m_\ssZ$ and $M_\ssX \rightarrow m_\ssX$ as $\eta \rightarrow 0$:
\be
 \vartheta_{\ssX}  :=
 \left\{ \begin{array}{c} +1 \quad \hbox{if} \quad r_{\ssX} > 1 \\
 -1 \quad\hbox{if} \quad r_{\ssX} < 1 \end{array} \right. \,.
\ee

The currents in the physical basis are similarly read off as
\begin{equation}
 {\bf J}_\mu =
 \begin{bmatrix}
 \check J_\mu^\ssZ c_\xi +
 \left( -\check J_\mu^\ssZ \hat s_\ssW \sh \, \eta
 + \check J_\mu^\ssA \hat c_\ssW \sh \, \eta
 + \check J_\mu^\ssX \ch \, \eta \right) s_\xi \\
 \check J_\mu^\ssA \\
 - \check J_\mu^\ssZ s_\xi
 + \left( - \check J_\mu^\ssZ \hat s_\ssW \sh \, \eta
 + \check J_\mu^\ssA \hat c_\ssW \sh \, \eta
 + \check J_\mu^\ssX \ch \, \eta \right) c_\xi
 \end{bmatrix} :=
 \begin{bmatrix}
 J_\mu^\ssZ \\
 J_\mu^\ssA \\
 J_\mu^\ssX
 \end{bmatrix} \,.
\end{equation}

Since we are eventually interested in obtaining bounds in terms of the physical masses $M_\ssZ$ and $M_\ssX$, it is useful to invert these mass equations to find the input parameters $m^2_\ssZ$ and $m^2_\ssX$ as a function of the physical masses and $\eta$. This gives
\beqa
 m_\ssX^2 &=& \frac{M_\ssZ^2}{2 \ch^2\eta}\left(1+R_\ssX^2
 +\vartheta_\ssX \sqrt{\left(1+R_\ssX^2\right)^2-4 \left(1+ \hat s_\ssW^2 \sh^2\eta\right) R_\ssX^2}\right) \\
 m_\ssZ^2 &=& \frac{M_\ssZ^2}{2 \left(1+ \hat s_\ssW^2 \sh^2\eta\right)}\left(1+R_\ssX^2
 -\vartheta_\ssX \sqrt{\left(1+R_\ssX^2\right)^2-4 \left(1+ \hat s_\ssW^2 \sh^2\eta\right) R_\ssX^2}\right)
\eeqa
where $R_\ssX$ is used to denote the ratio of the physical masses:
\be \label{AppRXeqn}
R_\ssX:=\frac{M_\ssX}{M_\ssZ} \,.
\ee
Also, the sign $\vartheta_\ssX$ is now $+1$ for $R_\ssX>1$ and $-1$ for $R_\ssX<1$.

Given this inversion, the angle $\xi$ can now be written as a function of $R_\ssX$ and $\eta$ only:
\begin{equation}
 \tan 2\xi(R_\ssX,\eta) =- \left(\frac{2 \hat s_\ssW \sh \eta }{1-\hat s_\ssW^2 \sh^2\eta - r_\ssX^2(R_\ssX,\eta) \ch^2\eta}\right) \,,
\end{equation}
where
\be
r_\ssX^2(R_\ssX,\eta) = \frac{\left(1+ \hat s_\ssW^2 \sh^2\eta\right)
\left(1+R_\ssX^2+\vartheta_\ssX \sqrt{\left(1+R_\ssX^2\right)^2
-4 \left(1+ \hat s_\ssW^2 \sh^2\eta\right) R_\ssX^2}\right)}
{\ch^2\eta\left(1+R_\ssX^2-\vartheta_\ssX \sqrt{\left(1+R_\ssX^2\right)^2
-4 \left(1+ \hat s_\ssW^2 \sh^2\eta\right) R_\ssX^2}\right)} \,.
\ee

\subsection*{Physical couplings} \label{Appphyscouplings}

We are now in a position to read off the physical implications of the $X$ boson. That is, we may write
\begin{equation}
 \cL = \cL_\SM + \delta \cL_\SM +
 \cL_\ssX \,, \label{AppLeff1}
\end{equation}
where the modification to the SM self-couplings are given by
\be \label{AppLeff2}
 \delta \cL_\SM = - \frac{z}{2} \, m_\ssZ^2 Z_\mu Z^\mu
 + i \hat e_\ssZ \sum_f \left[ \ol{f} \gamma^\mu
 \left( \delta g_{f\ssL} \gamma_{\ssL} + \delta g_{f \ssR}
 \gamma_{\ssR}\right) f \right] Z_\mu \,,
\ee
with \cite{bigfit}
\be
 z (R_\ssX,\eta) := \frac{M_\ssZ^2 - m_\ssZ^2}{m_\ssZ^2}
 = \frac{\hat s_\ssW^2 \sh^2\eta - \Delta_\ssX + \vartheta_\ssX \sqrt{\Delta_\ssX^2 - R_\ssX^2 \hat s_\ssW^2 \sh^2\eta }}
 {1 + \Delta_\ssX - \vartheta_\ssX \sqrt{\Delta_\ssX^2 - R_\ssX^2 \hat s_\ssW^2 \sh^2\eta }}  \,, \label{Appzeqn}
\ee
where
\be \label{AppDeltaeqn}
 \Delta_\ssX:=\frac{1}{2}(R_\ssX^2-1) \,.
\ee
(Note that the $\eta \rightarrow 0$ limit of $z$ is easily verified by implementing the identity $\Delta_\ssX=\vartheta_\ssX \sqrt{\Delta_\ssX^2}$.)

Given the form of $z$, one might worry that, for some choice of the parameters $M_\ssX$ and $\sh\,\eta$, $z$ would yield a complex value. However, any such choice does not correspond to a choice of real values for the original parameters of the lagrangian, $m_\ssX$, $m_\ssZ$, and $\chi$. This happens because sufficiently large kinetic mixing tends to preclude the existence of mass eigenvalues, $M_\ssX$ and $M_\ssZ$, that are too close to one another. This is why this region of parameter space is excluded from the plots of \S \ref{e+e-annihil}.

The fermion couplings are similarly
\be
 \delta g_{f\ssL(\ssR)} = \left(c_\xi-1\right) \hat g_{f\ssL(\ssR)}+ s_\xi \left(\sh\,\eta \, \hat s_\ssW (Q_f \hat c_\ssW^2 -\hat g_{f\ssL(\ssR)}) + \ch \eta \, \frac{g_\ssX}{\hat e_\ssZ} X_{f\ssL(\ssR)} \right) \,.
\ee

The terms explicitly involving the $X$ boson are
\begin{eqnarray}
 \cL_\ssX &=& - \frac14 \, X_{\mu \nu } X^{\mu \nu } -
 \frac{M_\ssX^2}{2} \, X_\mu X^\mu  \\
 && \qquad + i \sum_f \ol{f} \gamma_\mu \left( k_{f\ssL} \gamma_{\ssL}
 + k_{f\ssR} \gamma_{\ssR} \right) f  X^\mu \,, \notag
\end{eqnarray}
with
\be
 k_{f\ssL(\ssR)} = c_\xi \left(\ch\,\eta \, g_\ssX X_{f\ssL(\ssR)} + \sh\,\eta \, \frac{e}{\hat c_\ssW} (Q_f \hat c_\ssW^2-\hat g_{f\ssL(\ssR)})\right)
 -s_\xi \hat e_\ssZ \hat g_{f\ssL(\ssR)} \,.
\ee
Notice that in this basis $X_{\mu }$ does not couple directly to the electroweak gauge bosons at tree-level, but has acquired modified fermion couplings due to the mixing.

\subsubsection*{Oblique parameters}

The only remaining step is to eliminate parameters like $\hat s_\ssW$ and $m_\ssZ$ in the lagrangian in favour of a physically defined weak mixing angle, $s_\ssW$, and mass $M_\ssZ$. This process reveals the physical combination of new-physics parameters that is relevant to observables, and thereby provides a derivation \cite{bigfit} of the $X$-boson contributions to the oblique electroweak parameters \cite{oblique}.

We have already seen how to do this for the $Z$ mass, for which
\begin{equation}
 m_\ssZ \simeq M_\ssZ \left( 1 - \frac{z}{2} \right) \,.
\end{equation}
For the weak mixing angle it is convenient to define $s_\ssW$ so that the Fermi constant, $G_\ssF$, measured in muon decay is given by the SM formula,
\begin{equation}
 \frac{G_\ssF}{\sqrt{2}} := \frac{e^2}{8 s_\ssW^2 c_\ssW^2 M_\ssZ^2} \,.
 \label{AppG_fdef}
\end{equation}
But this can be compared with the tree-level calculation of the Fermi constant obtained in our model from $W$-exchange,
\begin{equation}
 \frac{G_\ssF}{\sqrt{2}} = \frac{g_2^2}{8 m_\ssW^2}
 = \frac{e^2}{8 \hat s_\ssW^2 \hat c_\ssW^2 m_\ssZ^2} \,,
\label{AppG_f}
\end{equation}
to infer
\begin{equation}
 \hat s_\ssW^2 \hat c_\ssW^2 = s_\ssW^2 c_\ssW^2
 \left( 1+z\right) \,,
\end{equation}
which, to linear order in $z$, implies that
\begin{equation}
 \hat s_\ssW^2 = s_\ssW^2 \left[
 1 + \frac{z \, c_\ssW^2}{c_\ssW^2 - s_\ssW^2}
 \right] \,.
\end{equation}

Eliminating $\hat s_\ssW$ in favour of $s_\ssW$ in the fermionic weak interactions introduces a further shift in these couplings, leading to our final form for the neutral-current lagrangian:
\begin{eqnarray}
 \cL_{\NC} &=&  \frac{ie}{\hat s_\ssW \hat c_\ssW}
 \sum_f \left[ \ol{f} \gamma^\mu \left( T_{3f} \gamma_{\ssL}
 -Q_f \hat s_\ssW^2 \right) f \right. \nn\\
 &&\qquad\qquad\qquad \left. +
 \ol{f} \gamma^\mu \left( \delta g_{f\ssL} \gamma_{\ssL}
 +\delta g_{f \ssR} \gamma_{\ssR} \right) f \right] Z_\mu \notag \\
 &\simeq& \frac{ie}{s_\ssW c_\ssW} \left(
 1 - \frac{z}{2} \right) \sum_f \left\{ \ol{f} \gamma^\mu
 \left[ T_{3f} \gamma_{\ssL} - Q_f s_\ssW^2 \left(
 1 + \frac{z \, c_\ssW^2}{c_\ssW^2 - s_\ssW} \right)
 \right] f \right. \nn\\
 && \qquad\qquad \left. + \phantom{\frac12} \overline{f} \gamma^\mu
 \left( \delta g_{f\ssL} \gamma_{\ssL}
 +\delta g_{f\ssR} \gamma_{\ssR}\right)
 f \right\} Z_\mu  \notag \\
 &:=& i e_\ssZ \sum_f \overline{f}
 \gamma^\mu \left[ \left( g_{f\ssL}^\SM + \Delta g_{f\ssL}
 \right) \gamma_\ssL + \left( g_{f\ssR}^\SM + \Delta g_{f\ssR}
  \right) \gamma_\ssR \right] f \, Z_\mu \,, \label{AppdLNC}
\end{eqnarray}
where $e_\ssZ := e/s_\ssW c_\ssW$ and
\begin{equation} \label{AppDgfLR}
  \Delta g_{f \ssL(\ssR)} = - \frac{z}{2} g_{f \ssL(\ssR)}^{\SM} - z \left(\frac{s_\ssW^2 c_\ssW^2}{c_\ssW^2-s_\ssW^2}\right) Q_f
  +\delta g_{f\ssL(\ssR)} \,,
\end{equation}
where (as usual) $g_{f\ssL}^\SM := T_{3f} - Q_f s_\ssW^2$ and $g_{f\ssR}^\SM := - Q_f s_\ssW^2$. It is assumed throughout that the corrections $z$, $\delta g_{f\ssL(\ssR)}$, and
\be \label{AppdefDkfLR}
\Delta k_{f\ssL(\ssR)}:=k_{f\ssL(\ssR)}-g_\ssX X_{f\ssL(\ssR)}
\ee
are small, so that any expression can be linearized in these variables. In particular, this means that one can replace hatted electroweak parameters (i.e. $\hat s_\ssW$, etc...) with unhatted ones in our previous expressions to give:
\beqa
 z (R_\ssX,\eta) &=& \frac{s_\ssW^2 \sh^2\eta - \Delta_\ssX + \vartheta_\ssX \sqrt{\Delta_\ssX^2 - R_\ssX^2 s_\ssW^2 \sh^2\eta }}
 {1 + \Delta_\ssX - \vartheta_\ssX \sqrt{\Delta_\ssX^2 - R_\ssX^2 s_\ssW^2 \sh^2\eta }} \\
 \delta g_{f\ssL(\ssR)} &=& \left(c_\xi-1\right) g_{f\ssL(\ssR)}^\SM+ s_\xi \left(\sh\,\eta \, s_\ssW (Q_f c_\ssW^2 -g_{f\ssL(\ssR)}^\SM) + \ch \eta \, \frac{g_\ssX}{e_\ssZ} X_{f\ssL(\ssR)} \right) \\
 \Delta k_{f\ssL(\ssR)} &=& \left(c_\xi \ch\,\eta -1\right) g_\ssX X_{f\ssL(\ssR)} + c_\xi \sh\,\eta \, \frac{e}{c_\ssW} (Q_f c_\ssW^2-g_{f\ssL(\ssR)}^\SM)-s_\xi e_\ssZ g_{f\ssL(\ssR)}^\SM \,. \label{AppDkfLR}
\eeqa

Alternatively, one can use the relationship between $z$ and $\eta$ to determine the contribution to the oblique parameters \cite{oblique} $S = U = 0$ and $\alpha T = - z$,
where (as usual) $\alpha := e^2/4\pi$. In this case, $\Delta g_{f \ssL(\ssR)}$ can be written as in \cite{bigfit}
\begin{equation}
  \Delta g_{f\ssL(\ssR)} = \frac{\alpha T}{2} \, g_{f\ssL(\ssR)}^{\SM}
  +\alpha T \left( \frac{s_\ssW^2 c_\ssW^2}{c_\ssW^2
  -s_\ssW^2} \right) Q_f + \delta g_{f\ssL(\ssR)} \,.
\end{equation}

\end{document}